\documentclass{aastex63}
\usepackage{appendix}

\usepackage{longtable}

\usepackage{booktabs}

\usepackage[caption=false]{subfig}
\usepackage{graphicx}

\usepackage{longtable}
\usepackage{multirow}
\usepackage{graphicx}
\usepackage{booktabs}
\usepackage{amsmath,amssymb}
\usepackage[T1]{fontenc}
\newcommand{\rowskip}{\noalign{\vspace{1pt}} \hline  \noalign{\vspace{1pt}} }
\graphicspath{{./}{figures/}}

\begin{document}


\title{The Magnetar Model and the Ejecta--Circumstellar-Matter Interaction Model for 31 Luminous, Rapidly Evolving Optical Transients}

\correspondingauthor{Shan-Qin Wang; Zi-Gao Dai; En-Wei Liang}
\email{shanqinwang@gxu.edu.cn;daizg@ustc.edu.cn;lew@gxu.edu.cn}

\author{Shan-Qin Wang}
\affiliation{Guangxi Key Laboratory for Relativistic Astrophysics,
School of Physical Science and Technology, Guangxi University, Nanning 530004,
China}

\author{Wen-Pei Gan}
\affiliation{Guangxi Key Laboratory for Relativistic Astrophysics,
School of Physical Science and Technology, Guangxi University, Nanning 530004,
China}

\author{Long Li}
\affiliation{School of Astronomy and Space Science, Nanjing
University, Nanjing 210093, China}
\affiliation{Key Laboratory of Modern Astronomy and Astrophysics
(Nanjing University), Ministry of Education, China}

\author{Alexei V. Filippenko}
\affiliation{Department of Astronomy, University of California, Berkeley, CA 94720-3411, USA}
\affiliation{Miller Institute for Basic Research in Science, University of
California, Berkeley, CA 94720, USA}

\author{Ling-Jun Wang}
\affiliation{Astroparticle Physics, Institute of High Energy Physics,
Chinese Academy of Sciences, Beijing 100049, China}

\author{Zi-Gao Dai}
\affiliation{Department of Astronomy, School of Physical Sciences, University of Science and Technology of China, Hefei 230026, Anhui, China}

\author{En-Wei Liang}
\affiliation{Guangxi Key Laboratory for Relativistic Astrophysics,
School of Physical Science and Technology, Guangxi University, Nanning 530004,
China}

\begin{abstract}
We study {31} luminous, rapidly evolving optical transients (REOTs),
and use {the magnetar model and the ejecta--circumstellar-matter (CSM)
interaction (CSI) model} to fit their {multiband} light curves.
We find that {28 events can be fitted by the magnetar model. and
the derived masses ($\sim0.05$--5.58\,M$_\odot$) are consistent with that of} stripped
and ultrastripped core-collapse supernovae (CCSNe). {On the other hand,  the CSI model can fit 25 events;
the derived ejecta masses are $\sim 3$--30\,M$_\odot$, consistent with that of
SNe~Ibn. The results, together with the fact that many luminous REOTs have been confirmed to be
luminous SNe~IIb/Ib/c or Ibn/IIn, suggest that at least a fraction of luminous REOTs
spectroscopically unclassified might be rapidly evolving SNe~Ib/Ic or Ibn.
Three events in the sample cannot be explained by the two models we use.
We expect that future intense photometry, spectroscopic classifications,
and systematic light-curve modeling of luminous REOTs will shed more light on their nature.}
\end{abstract}


\section{Introduction}

\label{sec:Intro}

Over the past {two} decades, {the Dark Energy Survey (DES),
the Pan-STARRS1 (PS1), the Supernova Legacy Survey (SNLS),
the Hyper Suprime-Cam (HSC), the Zwicky Transient Facility (ZTF), and other},
optical telescopes have discovered
many rapidly evolving optical transients (REOTs; e.g.,
\citealt{Cenko2012,Drout2014,Arc2016,Tanaka2016,Ho2018,Pren2018,Per2019,Pur2018,
Rest2018,Rodney2018,Ho2021}). While the rise and decline timescales
($\lesssim 10$\,d) of REOTs are shorter than those of normal supernovae (SNe),
their peak-luminosity distribution spans a wide range\citep{Drout2014,Per2019},
from $\sim10^{42}$\,erg\,s$^{-1}$ (which is
roughly equal to the peak luminosities of low-luminosity SNe {that
can be explained by the $^{56}$Ni model \citep{Arn1982}}) to
$\gtrsim 10^{44}$\,erg\,s$^{-1}$ (which reaches the luminosity threshold
of superluminous SNe (SLSNe; \citealt{Qui2011,Gal2012}).

The physical origin of REOTs is still elusive, and many scenarios have been
proposed to account for their properties.
\citet{Kasliwal2010} suggested that SN~2010X might be the accretion-induced collapse (AIC)
of an O/Ne/Mg white dwarf or a ``.Ia" explosion produced by the
thermonuclear detonation of the helium shell on a white dwarf \citep{Shen2010}.
The shock-breakout model \citep{Ofek2010} has also been invoked to
explain the light curves of REOTs.
However, the {$^{56}$Ni-powered core-collapse SN (CCSN)}, AIC, and helium-detonation models
cannot explain luminous REOTs, since the required mass of $^{56}$Ni {
and other radioactive elements} is larger than the inferred ejecta mass, or
the ratio of the mass of the radioactive elements (including $^{56}$Ni) to the
mass of ejecta is too large to be consistent with theoretical
expectations (see, e.g., \citealt{Drout2014,Pur2018});
alternative models must thus be considered.

\citet{Cenko2012} concluded that the short-lived, luminous
transient PTF10iya might be a tidal disruption event (TDE) produced by
a solar-type star that was disrupted by a $\sim10^7$\,M$_\odot$ black hole.
\citet{Per2019} argued that AT2018cow might be a TDE as well
(see also the discussion of \citealt{Liu18}), {while
\cite{Xiang2021} suggested that it might be a SN~Ibn/IIn (see also \citealt{Ho2021})
and demonstrated that the $^{56}$Ni plus the ejecta--circumstellar-matter (CSM)
interaction {(CSI)} model is best for explaining the bolometric light curve
of AT2018cow}.
\citet{Kashiyama2015} proposed that a fallback disk around a stellar-mass
black hole produced by a massive star could power a luminous REOT.
\citet{Yu2015} employed the merger-nova model \citep{Yu2013,Metpiro2014}
to suggest that a newly born magnetar from the remnant of the merger of a
neutron star (NS) binary can significantly enhance the luminosity of the
neutron-rich ejecta to fit the light curves, temperature evolution, and
photospheric radius evolution of three luminous REOTs (PS1-11bbq,
PS1-11qr, and PS1-12bv) discovered by PS1. \citet{Brooks2017}
suggested that the explosions of long-lived ($\sim10^{5}$\,yr) massive
remnants of a He white dwarf and a C/O or O/Ne white dwarf can account for
the light curves of luminous REOTs.

\citet{Hoto2017} used the magnetar-powered model
\citep{Kas2010,Woos2010} for the nascent magnetars produced by
ultrastripped SN explosions to explain
the {multiband} light curves of four REOTs reported by \citet{Arc2016}.
\citet{Rest2018} concluded that the light curve of the luminous REOT KSN2015K cannot be
explained by radioactivity or a central engine (magnetar or black
hole), but it can be powered by the interaction between the SN ejecta and
CSM from pre-SN mass loss.

{Although a major fraction of REOTs lack spectroscopic
classifications, we can get additional indirect information from
similar spectroscopically classified events.}
{Studies of individual optical transients have demonstrated that},
except for one kilonova (SSS17a/AT~2017gfo; e.g.,
\citealt{Arcavi17,Drout17,Kasen17,Pian17,Shappee17,Smartt17})
which is thus far a unique object, some
REOTs are SNe~Ib (e.g., SN~2002bj, \citealt{Poznanski2010}; {SN~2019dge,
\citealt{Yao2020}}), {SNe~Ic (e.g., SN~2005ek, \citealt{Drout2013}; iPTF14gqr,
\citealt{De2018}; SN~2020oi, \citealt{Rho2021}),
SNe~Ic-BL (e.g., iPTF16asu, \citealt{Whit2017}; SN~2018gep, \citealt{Pritchard2021})},
SNe~Ibn (e.g., SN~1999cq, \citealt{Matheson2010}; iPTF13beo, \citealt{Gorbikov2014};
LSQ13ccw, \citealt{Pastorello2015}; SN~2015U, \citealt{Shiv2016};
{PS15dpn, \citealt{Sma2016}; SN~2018bcc, \citealt{Karamehmetoglu2019}}),
and SNe~IIn (e.g., PTF09uj, \citealt{Ofek2010}). {The 5 (PTF12ldy, iPTF14aki,
iPTF15akq, iPTF15ul, SN~2015U) of 7 SNe~Ibn spectroscopical confirmed by
\cite{Hosseinzadeh2017} are also rapidly evolving events.
Recently, \cite{Ho2021} reported 42 REOTs observed by ZTF,
and 10 events were classified as SNe~II/IIb/Ib or Ibn/IIn (descriptions of the spectral
features of different types of SNe are given by \citealt{Filippenko1997} and \citealt{Gal2017});
see their Table 2\footnote{{Table 2 of \cite{Ho2021} listed 22
spectroscopically confirmed SNe, 12 of which had been classified by previous studies or reports.}}.
Additionally, almost all SNe~Ibn (including the rapid evolving ones) are rather luminous
(see Table 4 of \citealt{Hosseinzadeh2017} and Table 14 of \citealt{Ho2021});
for comparison, there are only two luminous, rapidly evolving SNe~Ic-BL (iPTF16asu and SN~2018gep).}
{The resemblance between the luminous, rapidly evolving SNe and the luminous REOTs, and
the fact that the host galaxies of the latter are star-forming galaxies \citep{Wiseman2020},
suggest that most luminous REOTs might be CCSNe} which must arise from the explosions of massive stars.
{We can say that the rapid evolving SNe are luminous REOTs with spectral classifications,
and speculate that some luminous REOTs lacking spectral classifications might be rapid evolving SNe~Ibn/IIn/II
or Ic/Ic-BL.}

{Currently, the prevailing models accounting for the light curves of luminous SNe and SLSNe
that cannot be explained by the $^{56}$Ni model are the magnetar model and the {CSI} model.
Both models are used to fit the light curves of (super)luminous SNe~I, while the CSI model is
the most reasonable one for explaining the light curves of (super)luminous SNe~Ibn/IIn.}
Assuming that the majority of luminous REOTs {in our sample are
luminous SNe~Ibn/IIn/Ic, it is interesting to test whether the magnetar and
CSI models can fit their light curves and to constrain the best-fit parameters
of the models.}

In this paper, we collect 31 luminous REOTs discovered by DES,
PS1, {SNLS, and HSC}, {investigating} the
{energy sources powering their multiband light curves}
{and constraining} the model parameters.
{We describe our sample in Section \ref{sec:sample}.
 In Section \ref{sec:fit}, we use the magnetar model and the CSI model to
fit the multiband light curves and derive the best-fit parameters of the models.
We discuss our results and draw the conclusions in Sections \ref{sec:dis} and \ref{sec:con}, respectively.
Throughout the paper, we assume $\Omega_m = 0.3$, $\Omega_\Lambda = 0.7$,
and H$_0 = 68.3$\,km\,s$^{-1}$\,Mpc$^{-1}$. The values of the Milky Way
extinction ($E_\mathrm{B-V}$) of all events are from \cite{Schlafly2011}.}

\section{The Sample of REOTs}

\label{sec:sample}

The luminous REOTs studied here were discovered by PS1 {in 2010--2013, DES in 2013--2016,
SNLS in 2017, and SNLS in 2004--2006}. Their detailed observational properties are given by
\citet{Drout2014}{, \cite{Arc2016},} \citet{Pur2018}, and \cite{Tampo2020}.
We select events that were observed in at least three filters, and exclude events whose
absolute peak magnitude{s are} dimmer than $-18$\,mag.
{Moreover, we exclude transients that exhibit rebrightening in at least two bands
and that show severe undulations around the peaks of the light curves.}

Using these {four} criteria, we {compile 19} DES REOTs ({DES13C3uig, DES13E2lpk, DES13X1hav}, DES13X3gms,
{DES13X3npb, DES13X3nyg}, DES15C3lpq, {DES15C3lzm, DES15C3nat, DES15C3opk, DES15C3opp},
DES15E2nqh, DES15S1fll, DES15X3mxf, DES16C1cbd, DES16C3gin, DES16E2pv, DES16X3cxn, and DES16X3ega),
{5 HSC REOTs (HSC17auls, HSC17bbaz,  HSC17bhyl, HSC17btum \footnote{The  brightess value of the $g$-band absolute magnitude of HSC17btum is at least
$\sim -17.73$\,mag. Pre-peak and post-peak photometry were obtained, and the $g$-band peak was missed; however,
it can be expected that the $g$-band peak magnitude might be $< -18$\,mag.}, HSC17dadp)},
{5} PS1 REOTs (PS1-10bjp, PS1-11bbq, PS1-11qr, PS1-12bv, and PS1-13duy),
and {2 SNLS REOTs (SNLS04D4ec and SNLS06D1hc)}.

Based on the tables of \citet{Drout2014}, {\cite{Arc2016}}, \citet{Pur2018},
{and \cite{Tampo2020}}, we list the relevant observational information for
{the 31} REOTs in Table \ref{tab:info}, from which we can obtain some clues about the nature
of these REOTs. First, the star-formation rates of the host galaxies are large, supporting
the CCSN scenario, though the degenerate-star model cannot be excluded.
Second, the TDE model involving supermassive black holes (SMBHs) cannot be plausible
to account for these luminous REOTs since the explosion-site offsets from their host-galaxy
nuclei are rather large; TDEs induced by tidal disruption of stars by SMBHs should generally
happen in the centers of the hosts. TDEs associated with intermediate-mass black holes (IMBHs)
can produce nonnuclear TDEs by capturing and disrupting stars. According to the calculation
of \citet{Strubbe2009}, however, \citet{Drout2014} pointed out that the timescale of IMBH
TDEs is $\lesssim 1$\,day, which is an order of magnitude shorter than the timescale of
the REOTs studied here. Therefore, the IMBH TDE model is also disfavored in explaining them.

\section{Evolution of the REOT Bolometric Luminosities, Temperatures, and Radii}

To derive the bolometric light curves, temperature evolution,
and radius evolution of {the} REOTs {they presented},
\citet{Drout2014}, {\cite{Arc2016}}, \citet{Pur2018},
{and \cite{Tampo2020}} {used} the
blackbody equation $F_{\lambda}(R_{\rm ph},T) = (2{\pi}hc^2/{\lambda}^5)
(e^{\frac{hc}{{\lambda}k_{\rm b}T(t)}}-1)^{-1}(R_{\rm ph}^2/D_L^2)$
to fit {their} spectral energy distributions (SEDs).
They showed that the SEDs of most of the REOTs
can be well fitted by the blackbody model, while some SEDs of
some REOTs (slightly) deviate from the blackbody model.

\begin{figure}[tbph]
\begin{center}
\includegraphics[width=0.18\textwidth,angle=0]{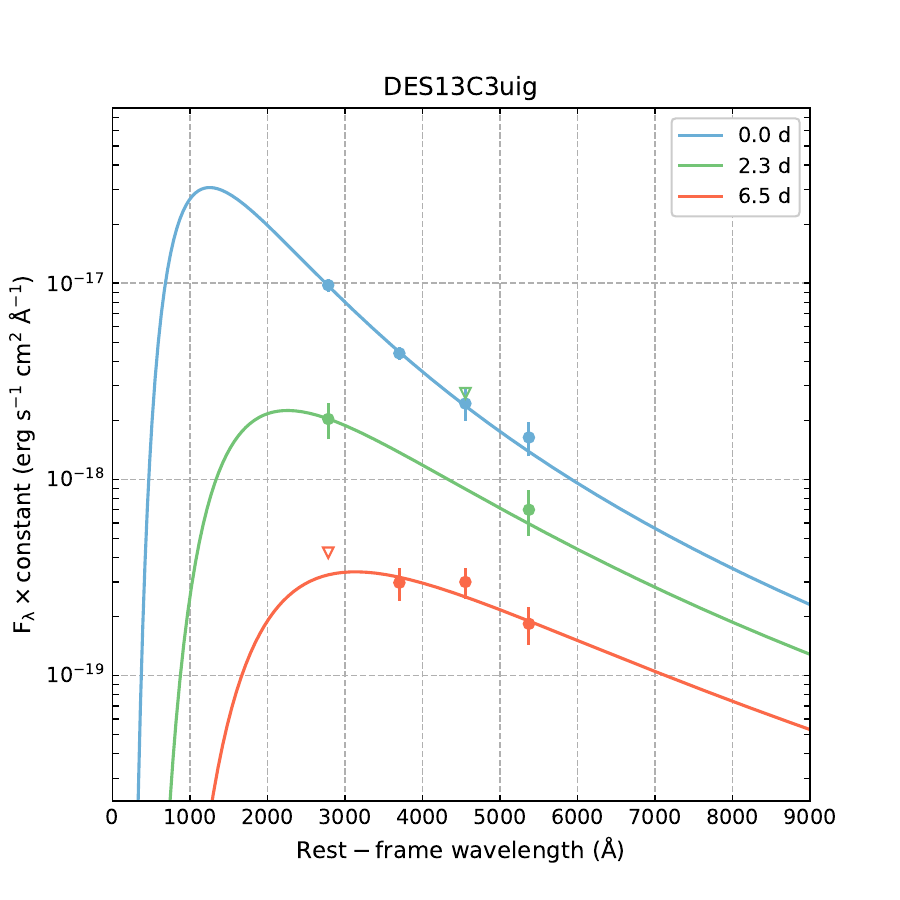}
\includegraphics[width=0.18\textwidth,angle=0]{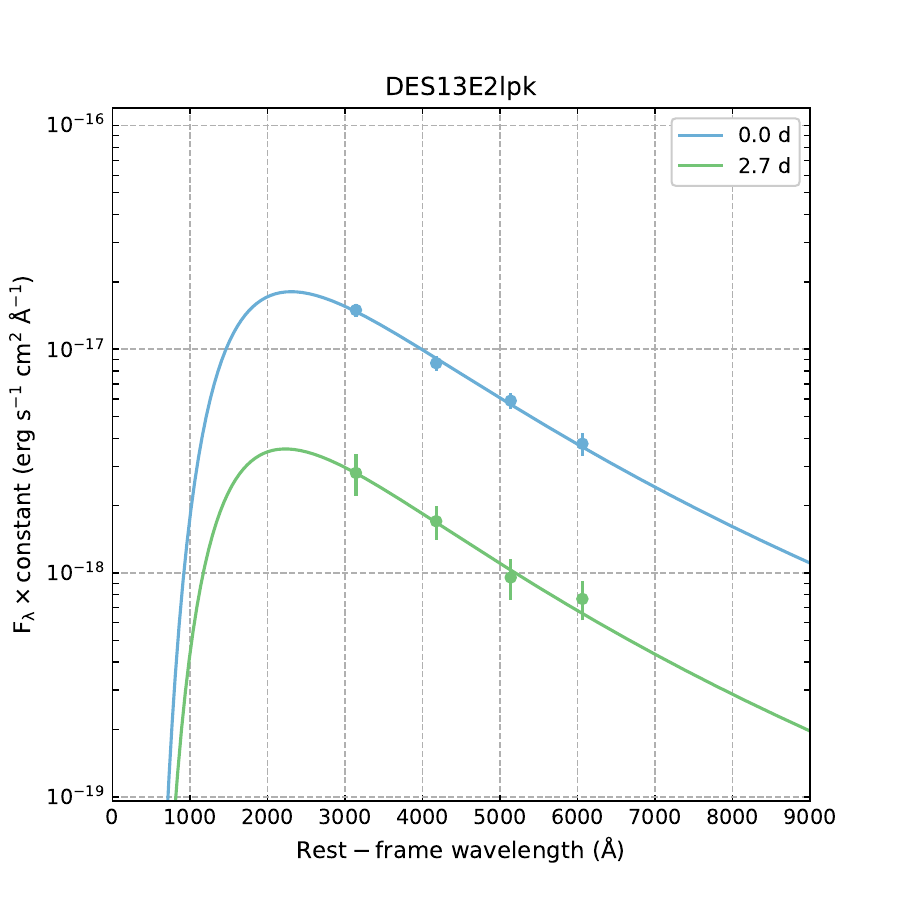}
\includegraphics[width=0.18\textwidth,angle=0]{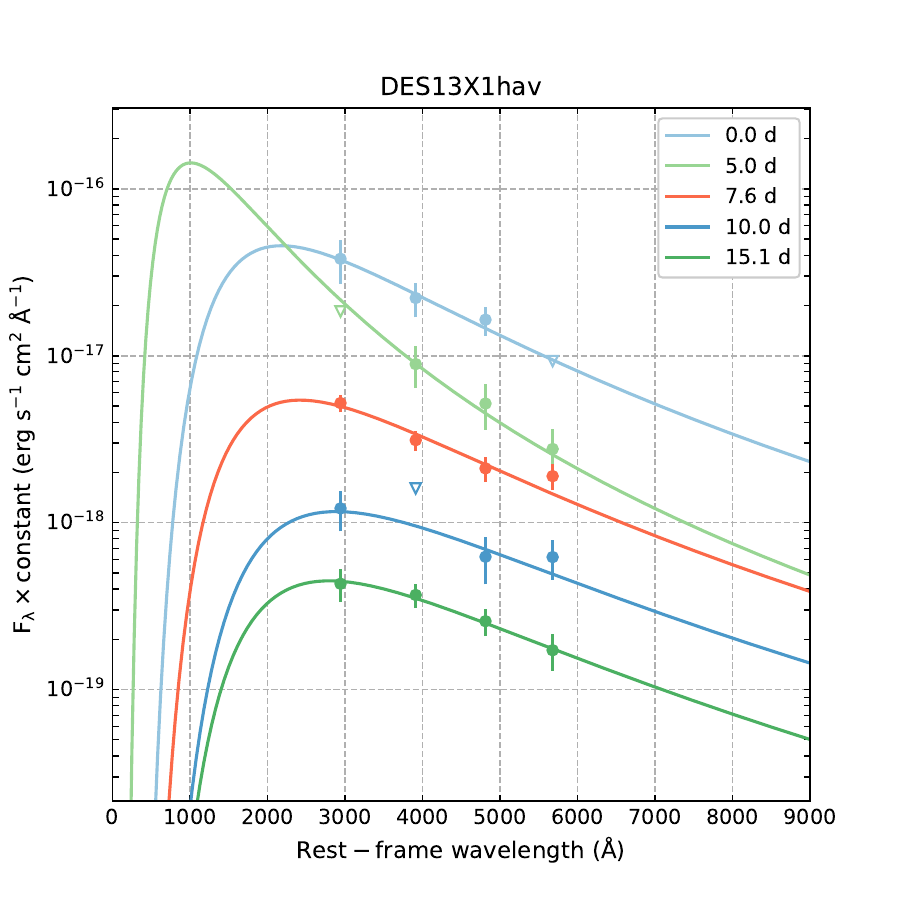}
\includegraphics[width=0.18\textwidth,angle=0]{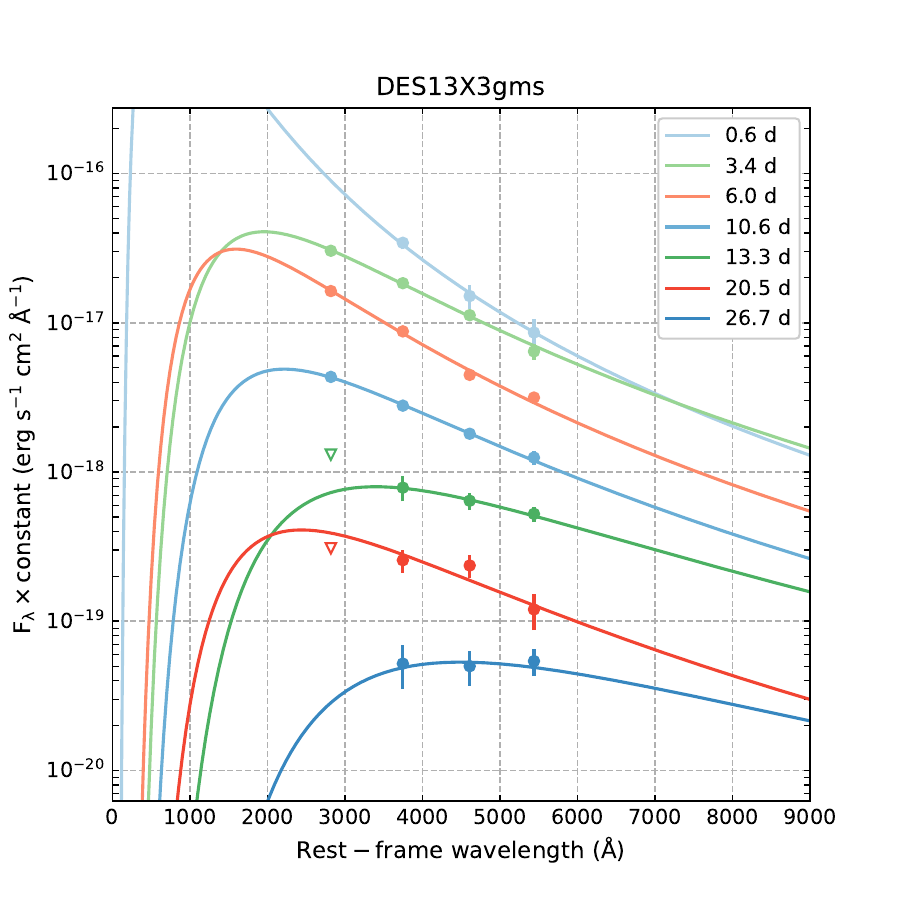}
\includegraphics[width=0.18\textwidth,angle=0]{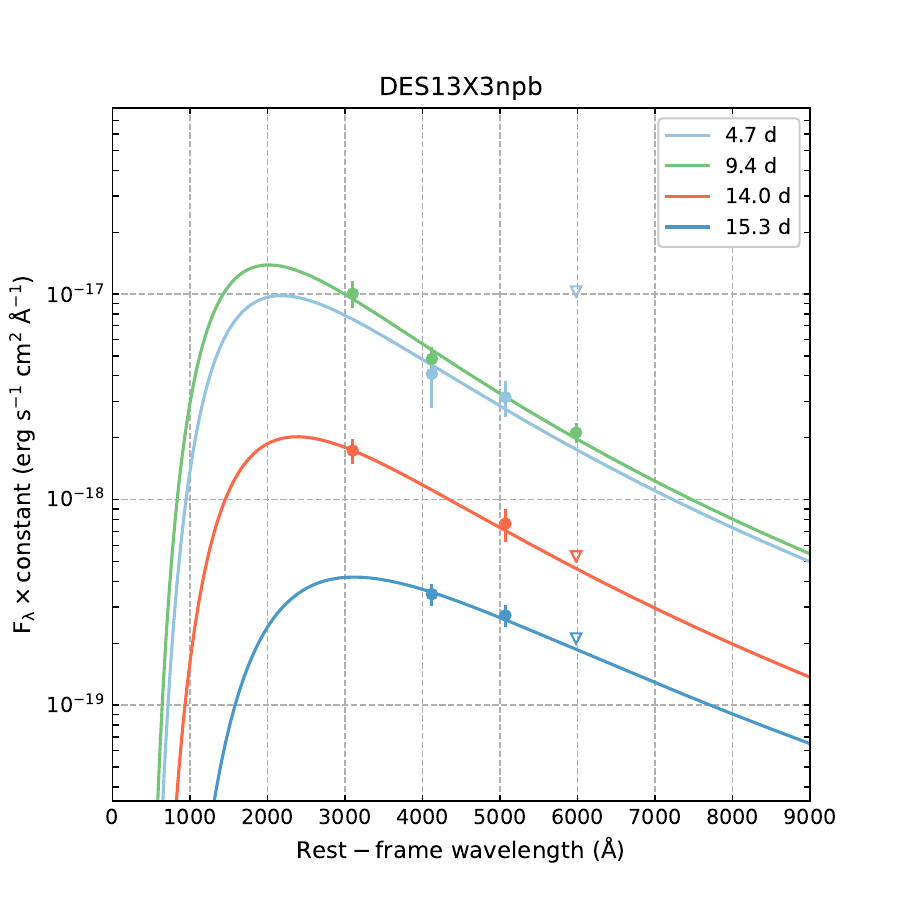}\\
\includegraphics[width=0.18\textwidth,angle=0]{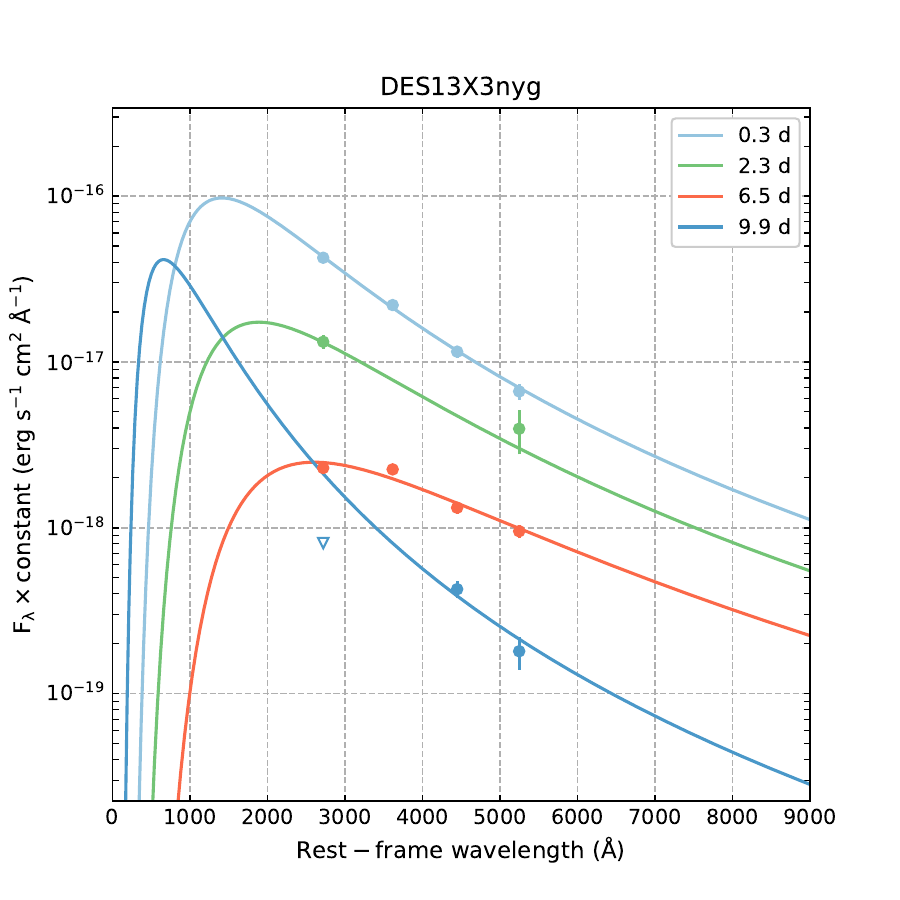}
\includegraphics[width=0.18\textwidth,angle=0]{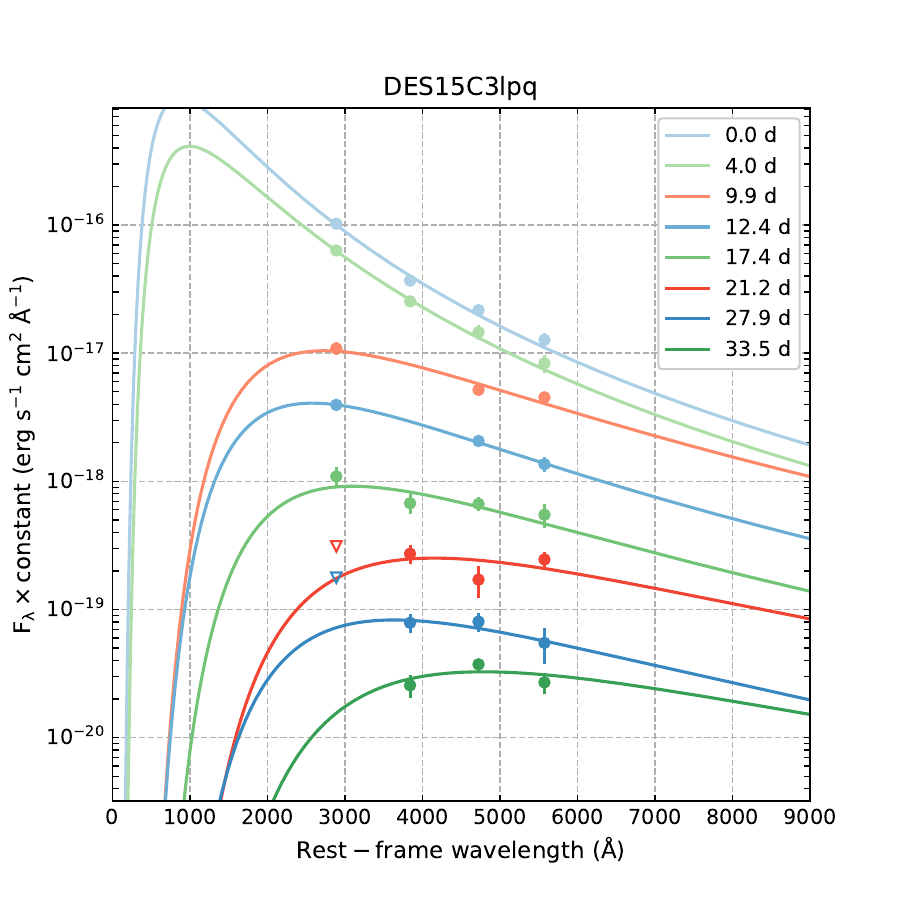}
\includegraphics[width=0.18\textwidth,angle=0]{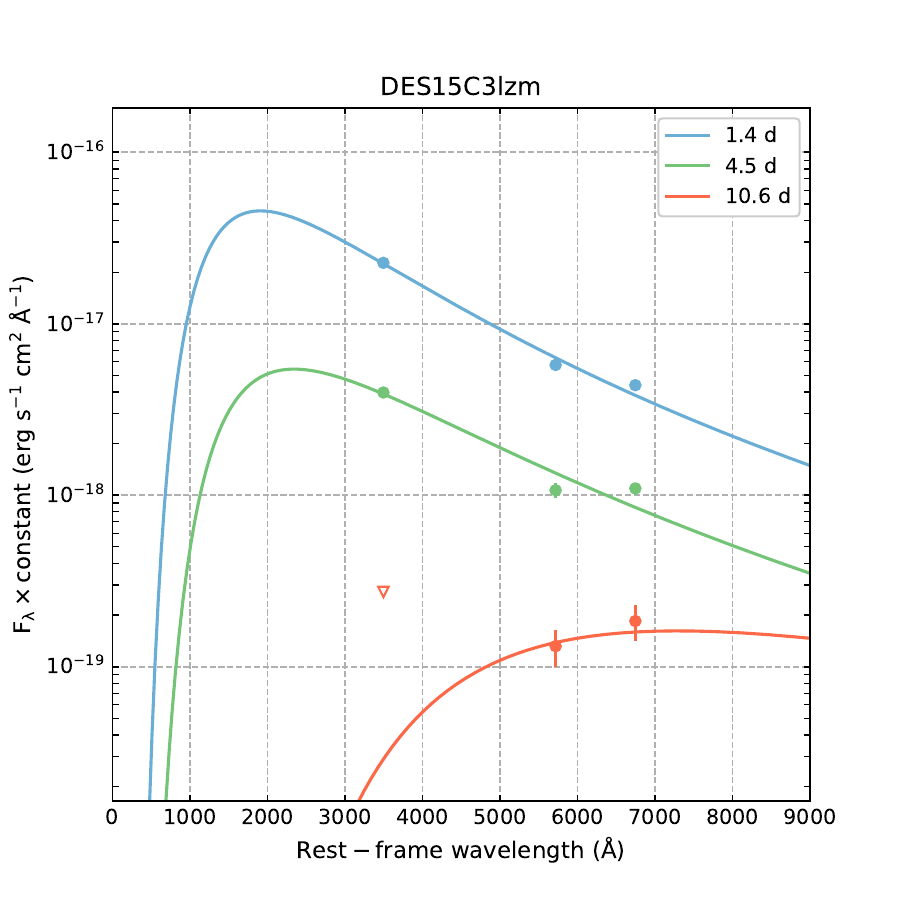}
\includegraphics[width=0.18\textwidth,angle=0]{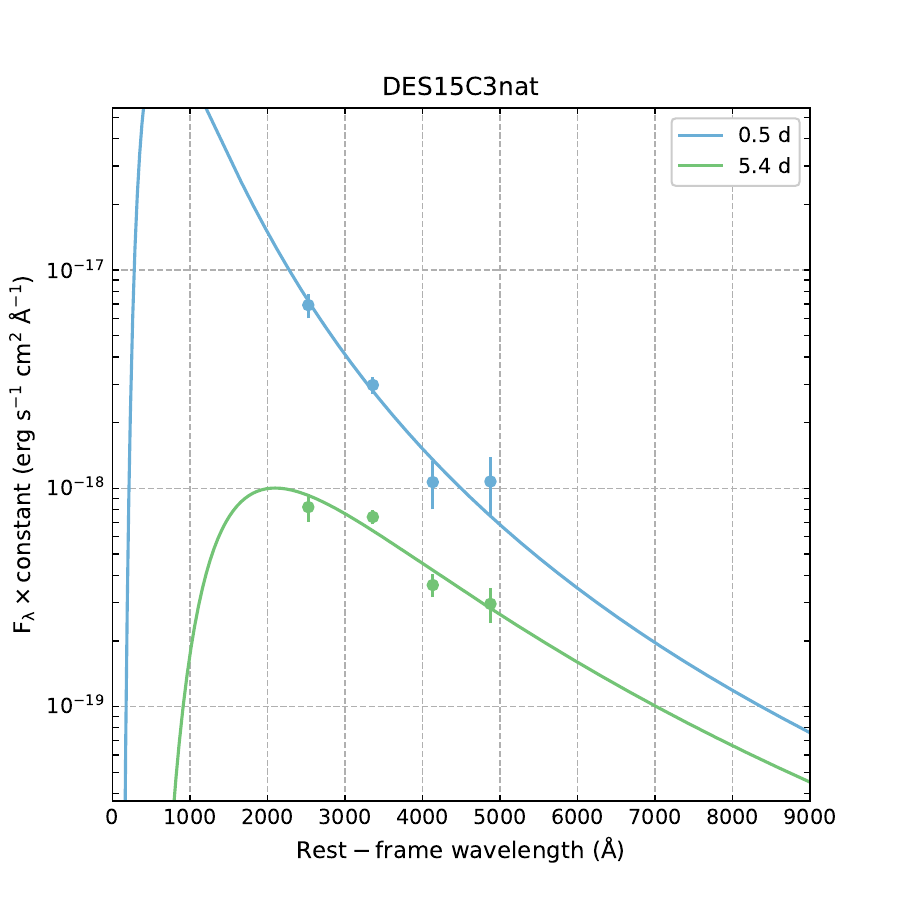}
\includegraphics[width=0.18\textwidth,angle=0]{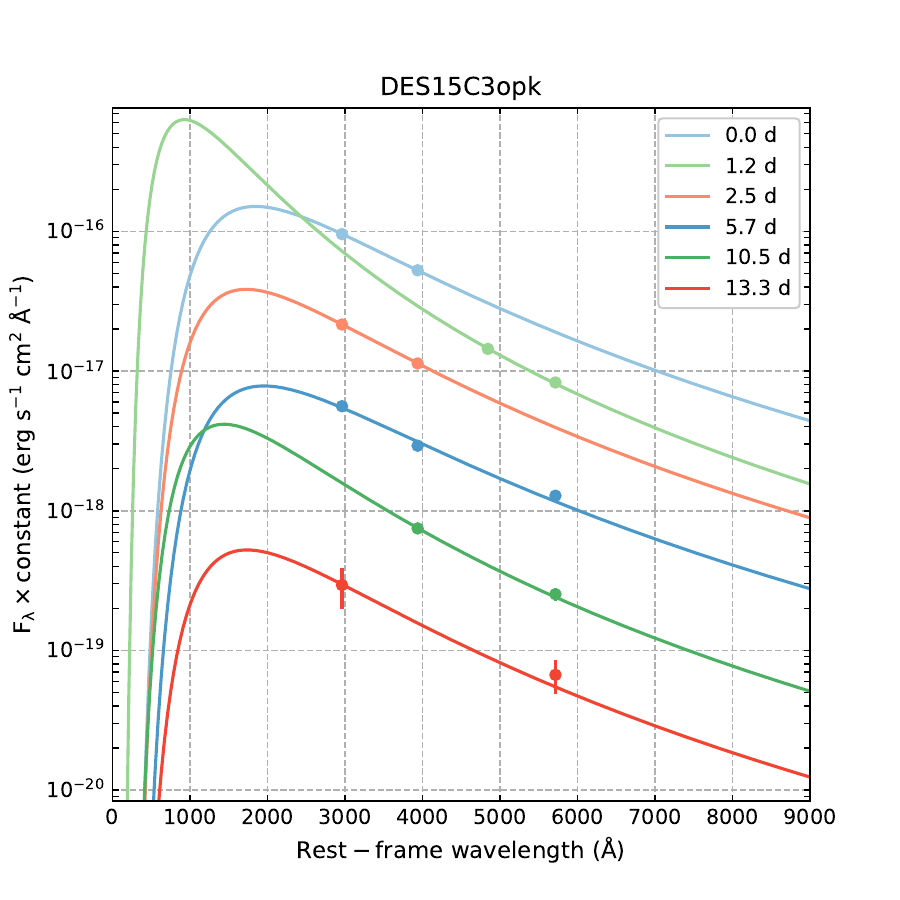}\\
\includegraphics[width=0.18\textwidth,angle=0]{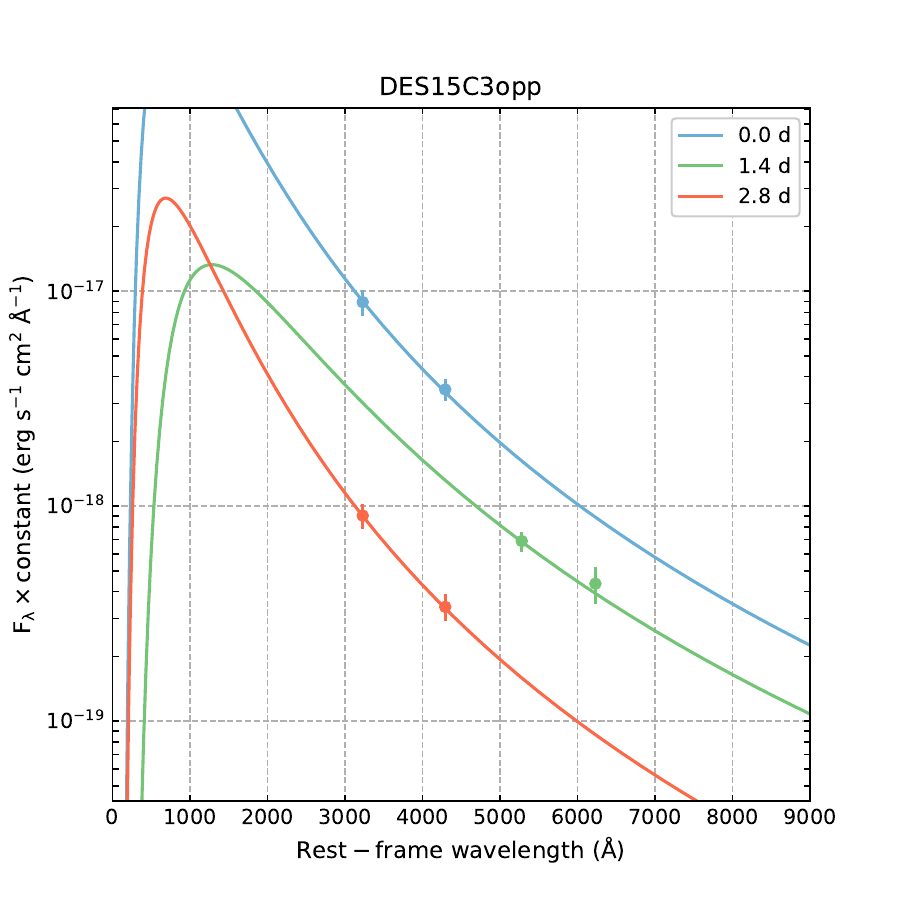}
\includegraphics[width=0.18\textwidth,angle=0]{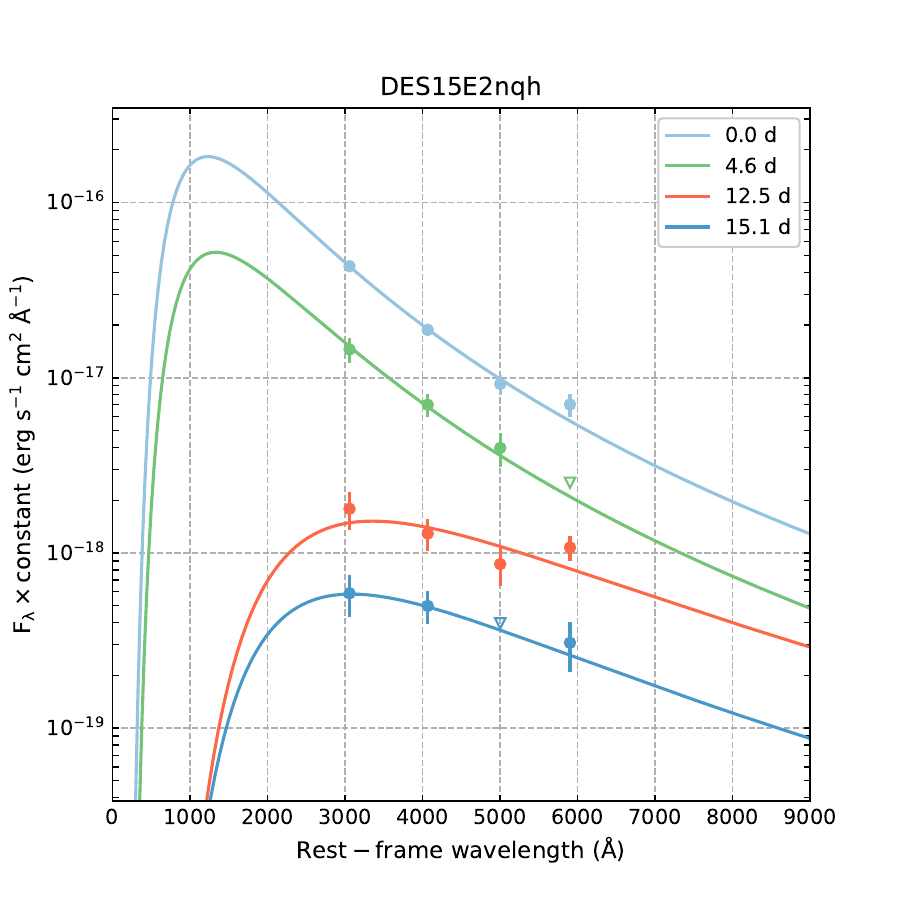}
\includegraphics[width=0.18\textwidth,angle=0]{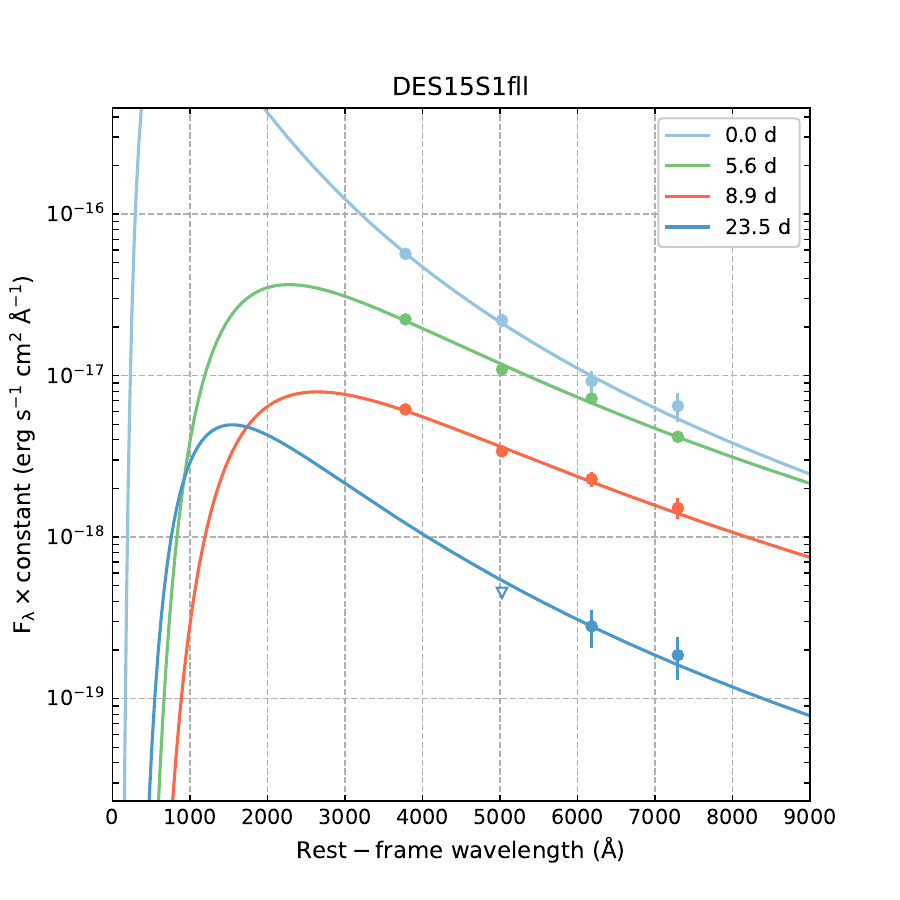}
\includegraphics[width=0.18\textwidth,angle=0]{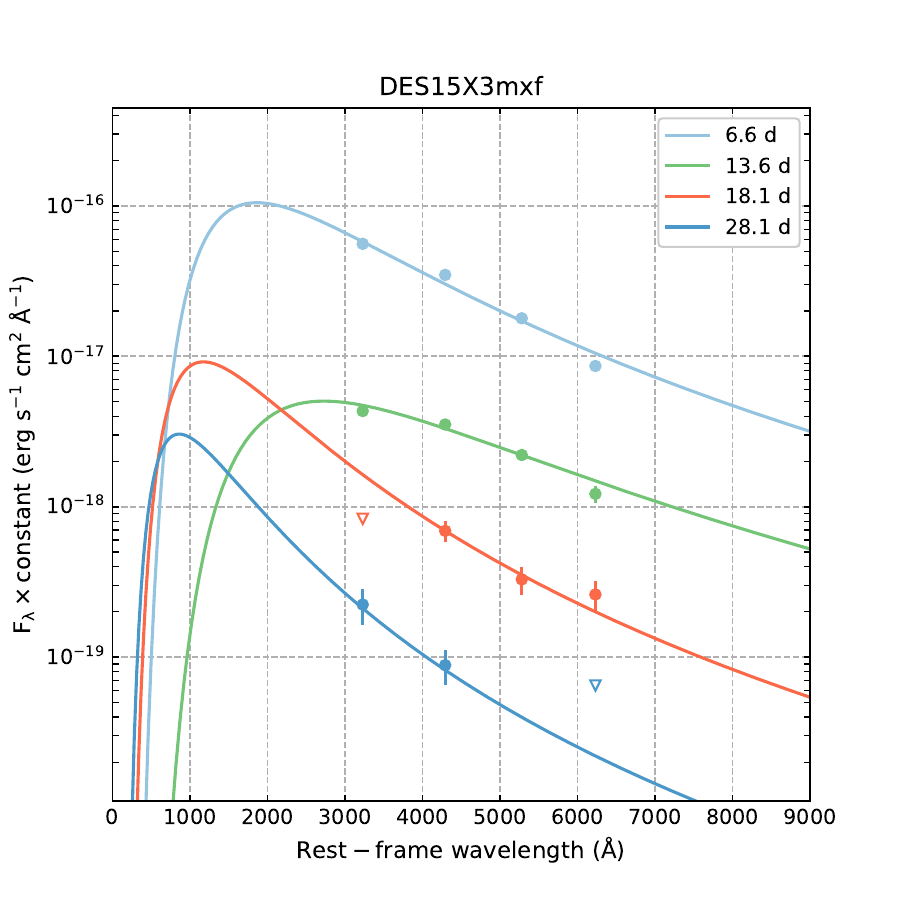}
\includegraphics[width=0.18\textwidth,angle=0]{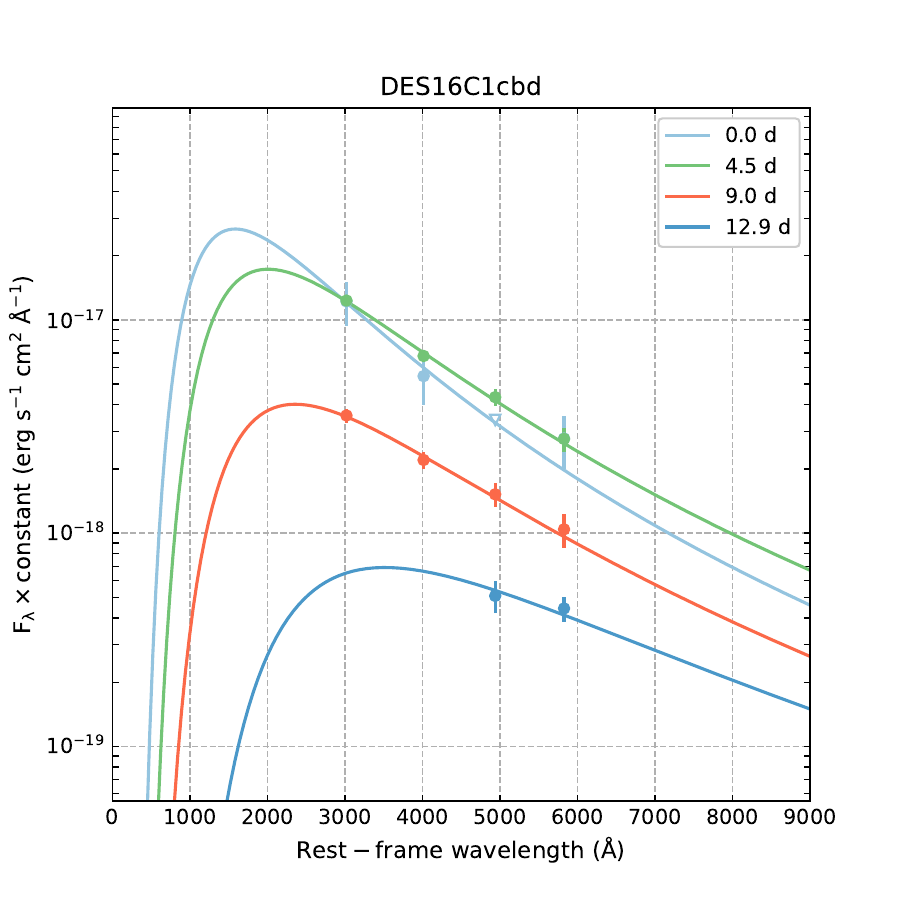}
\includegraphics[width=0.18\textwidth,angle=0]{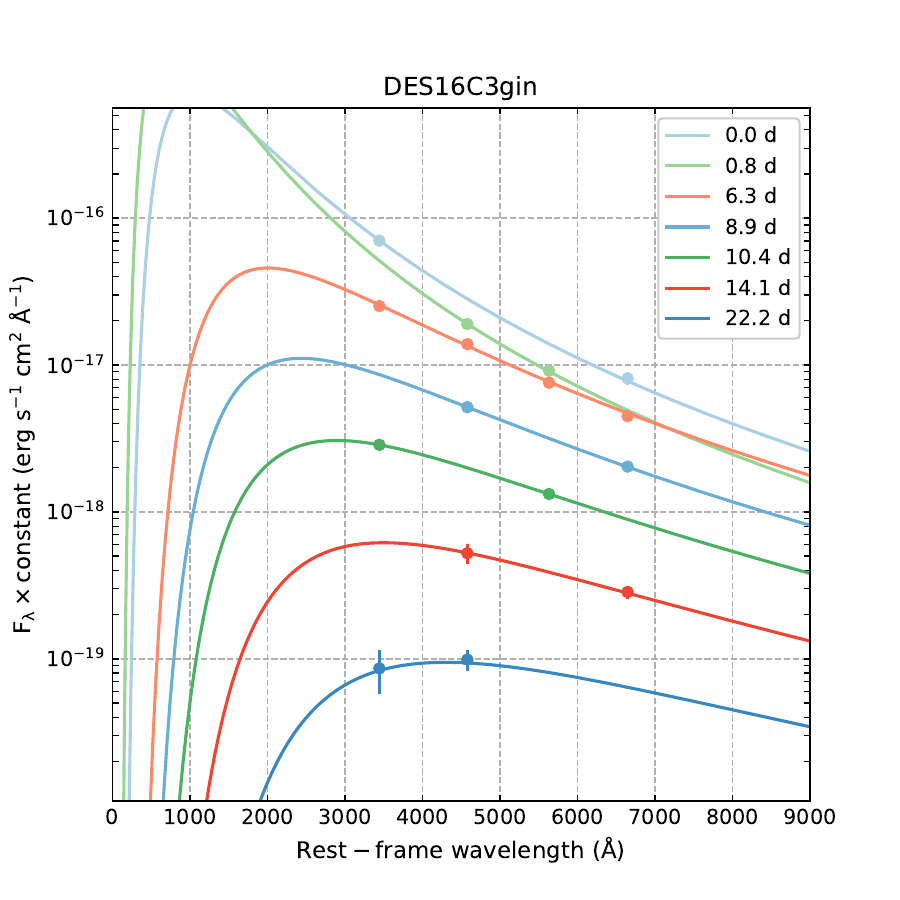}
\includegraphics[width=0.18\textwidth,angle=0]{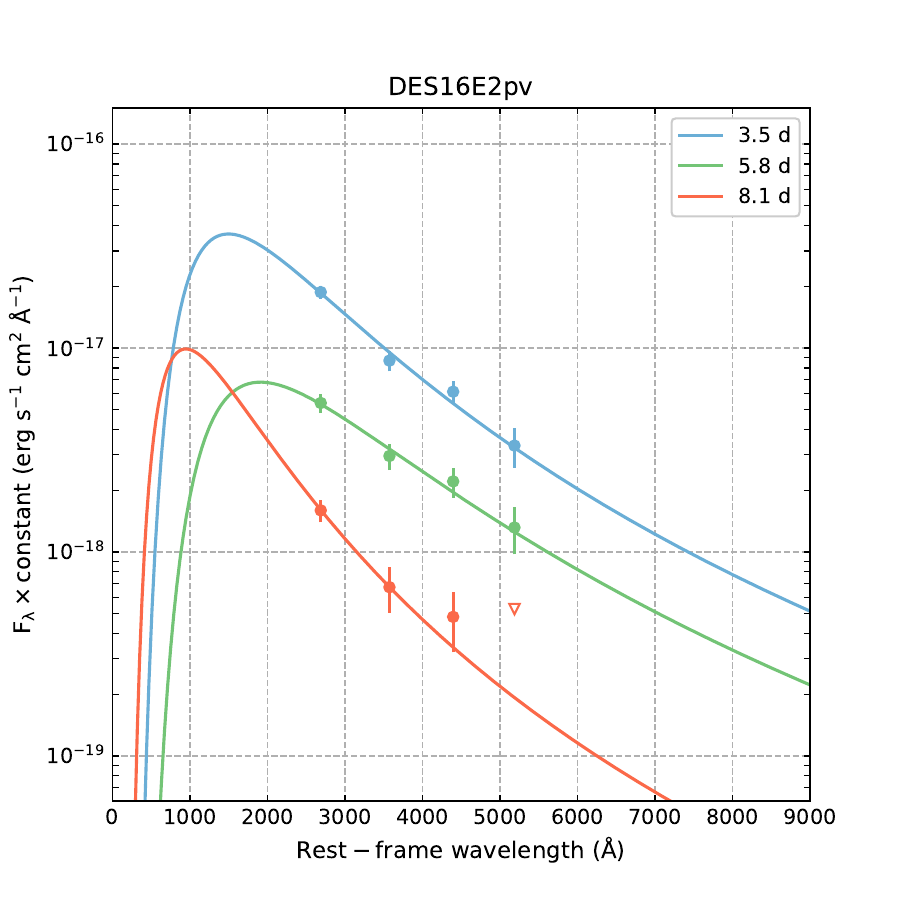}
\includegraphics[width=0.18\textwidth,angle=0]{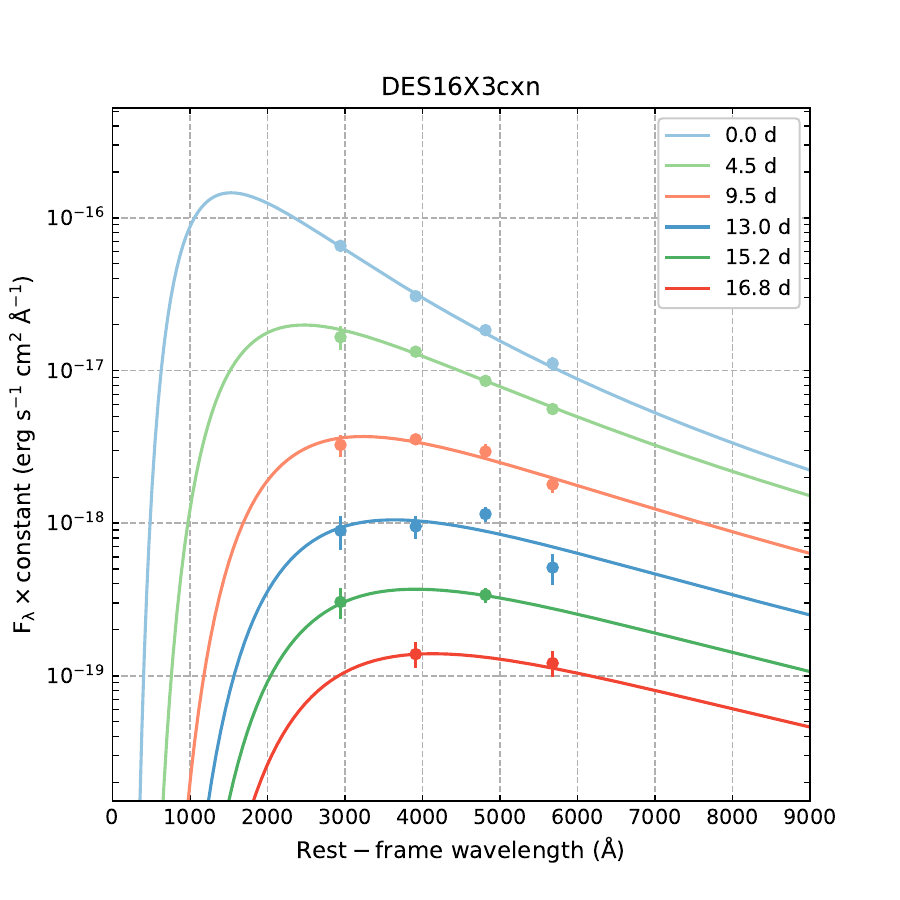}
\includegraphics[width=0.18\textwidth,angle=0]{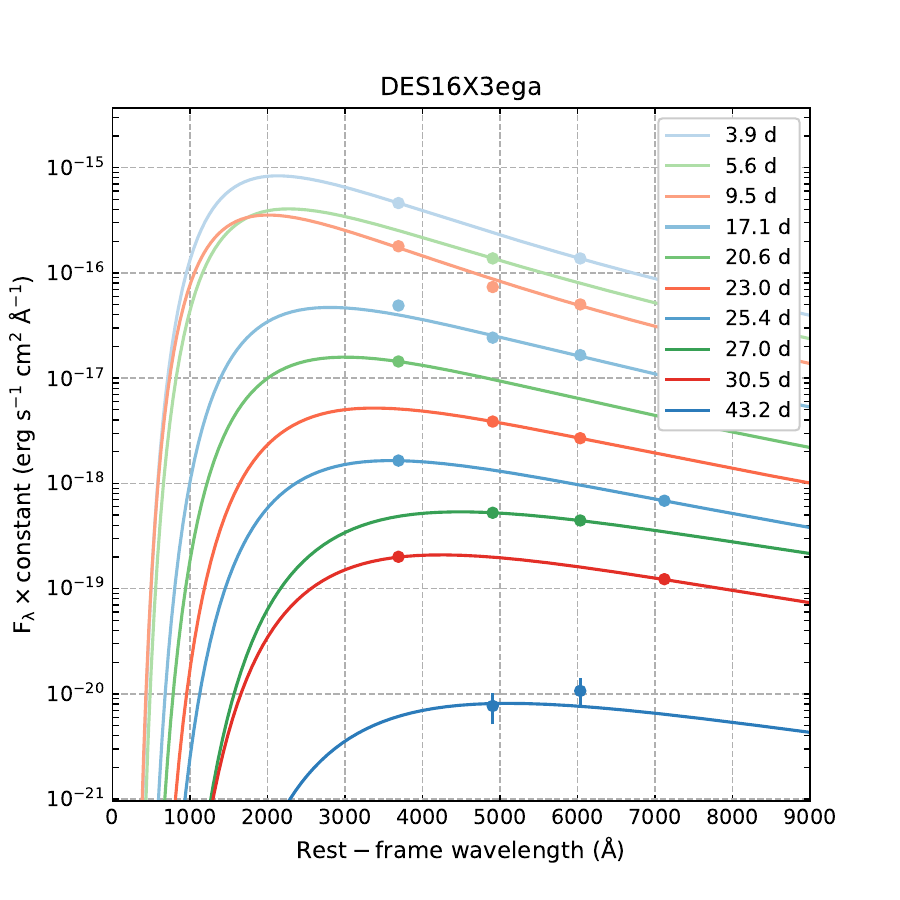}
\includegraphics[width=0.18\textwidth,angle=0]{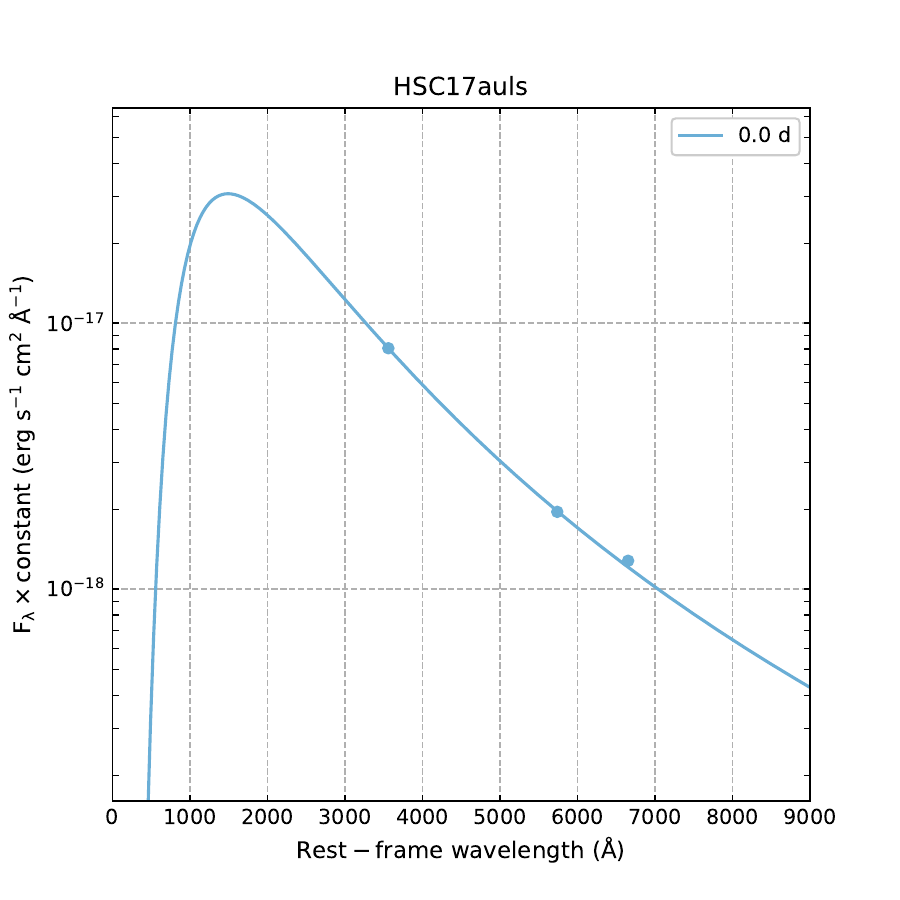}
\includegraphics[width=0.18\textwidth,angle=0]{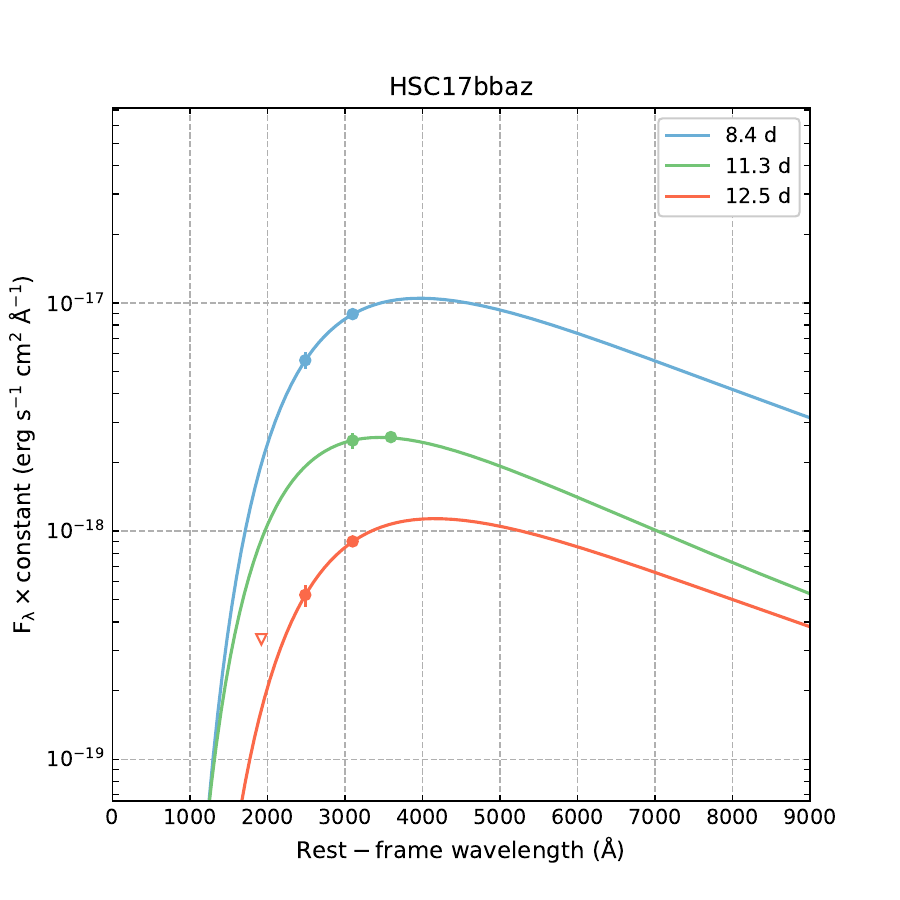}
\includegraphics[width=0.18\textwidth,angle=0]{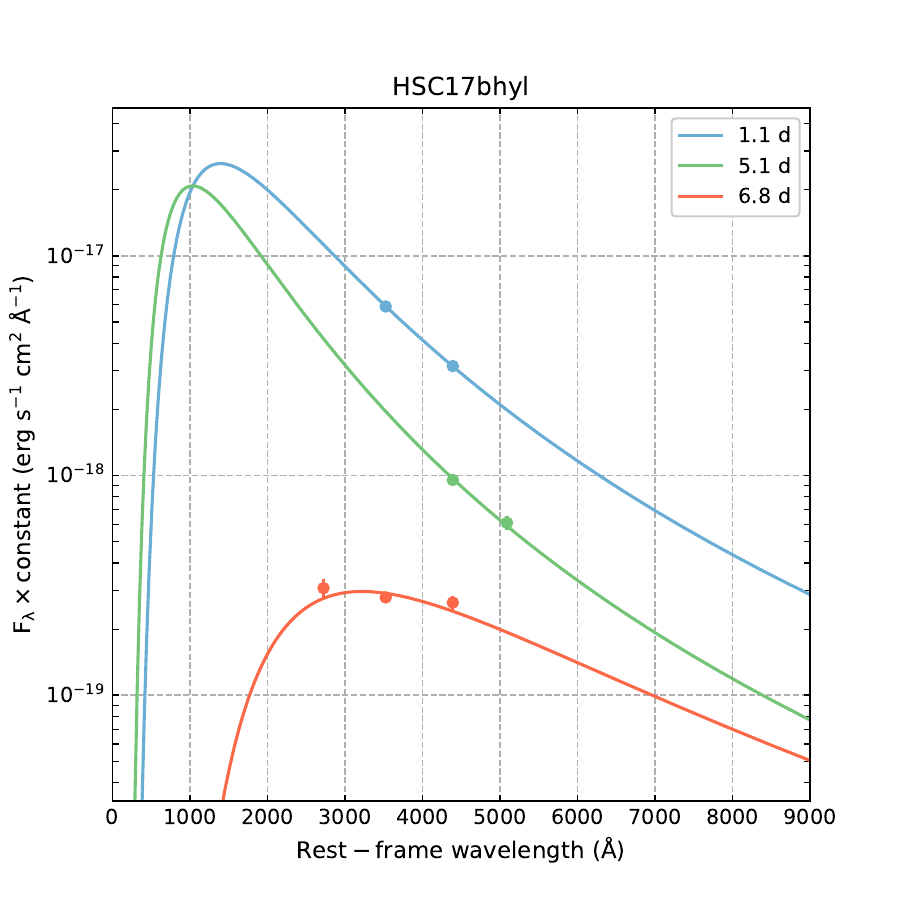}
\includegraphics[width=0.18\textwidth,angle=0]{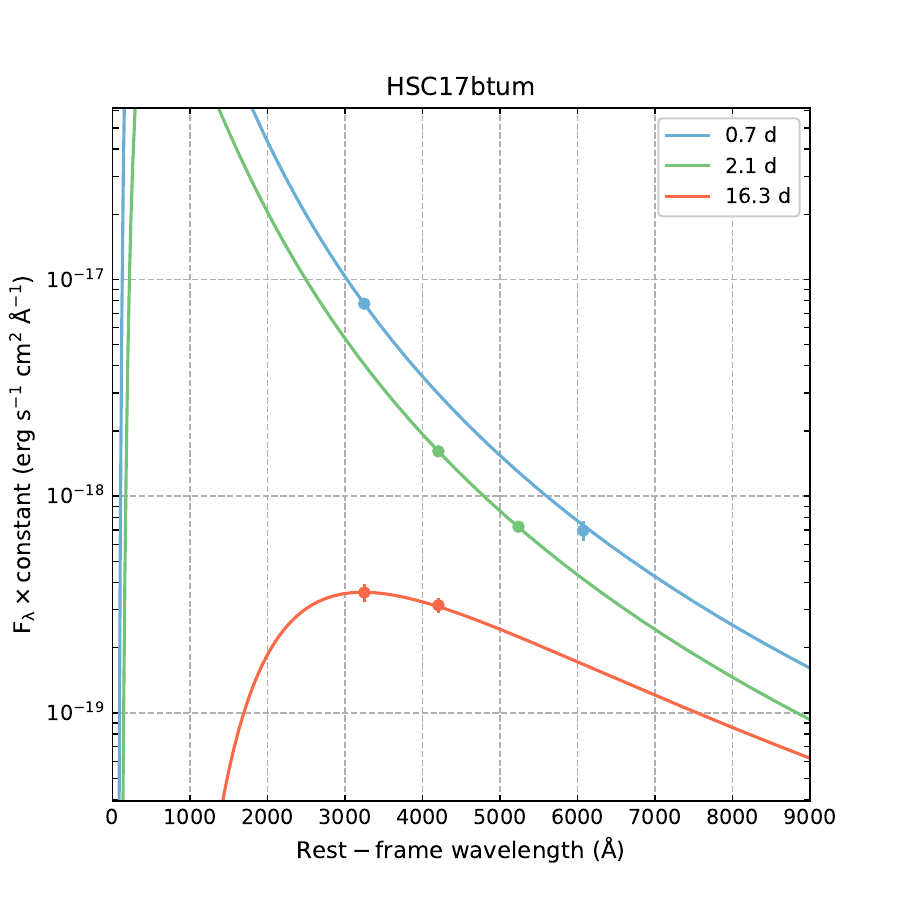}
\includegraphics[width=0.18\textwidth,angle=0]{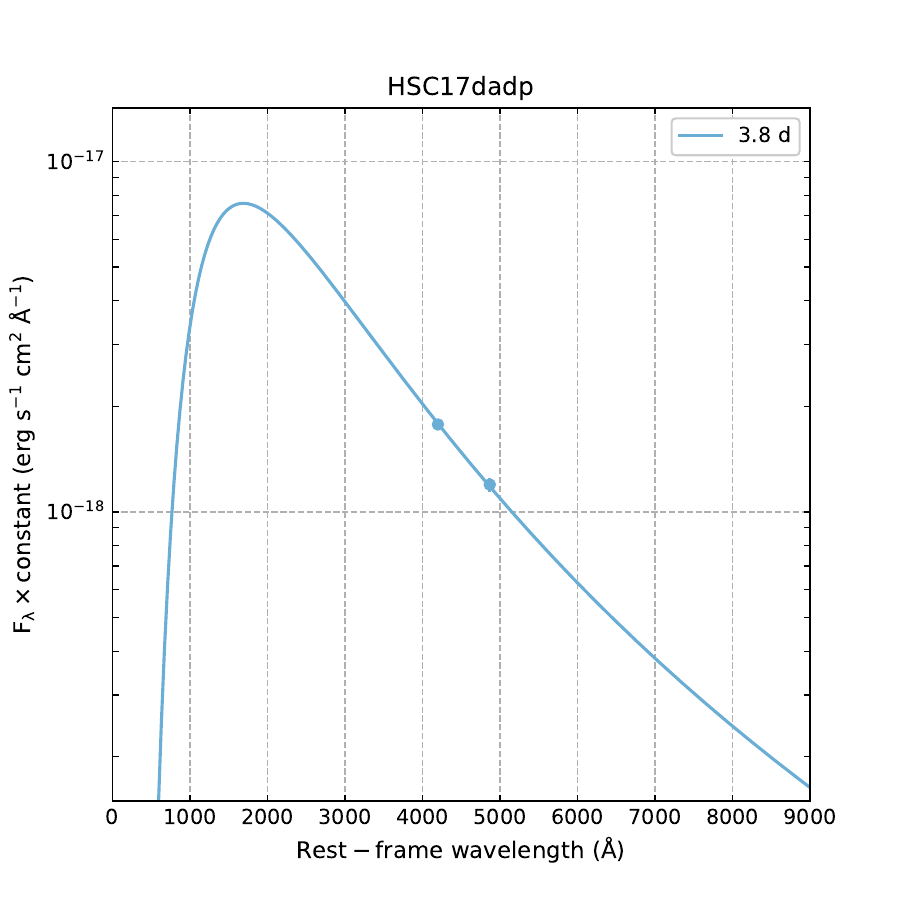}
\includegraphics[width=0.18\textwidth,angle=0]{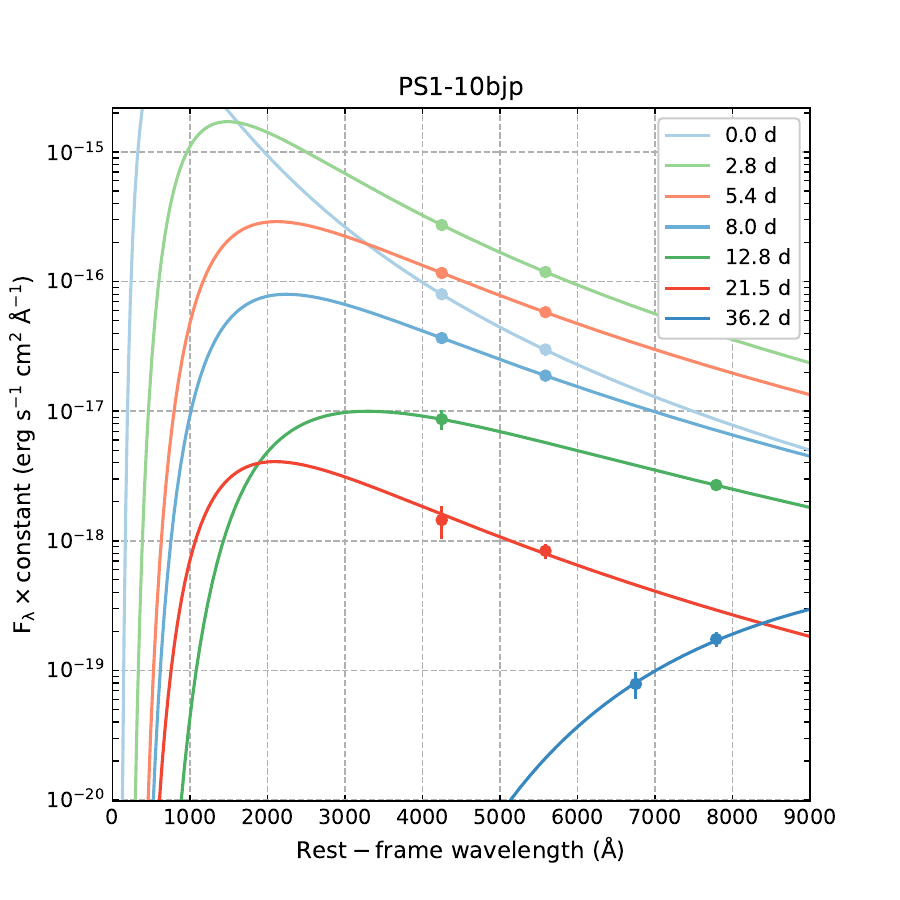}
\includegraphics[width=0.18\textwidth,angle=0]{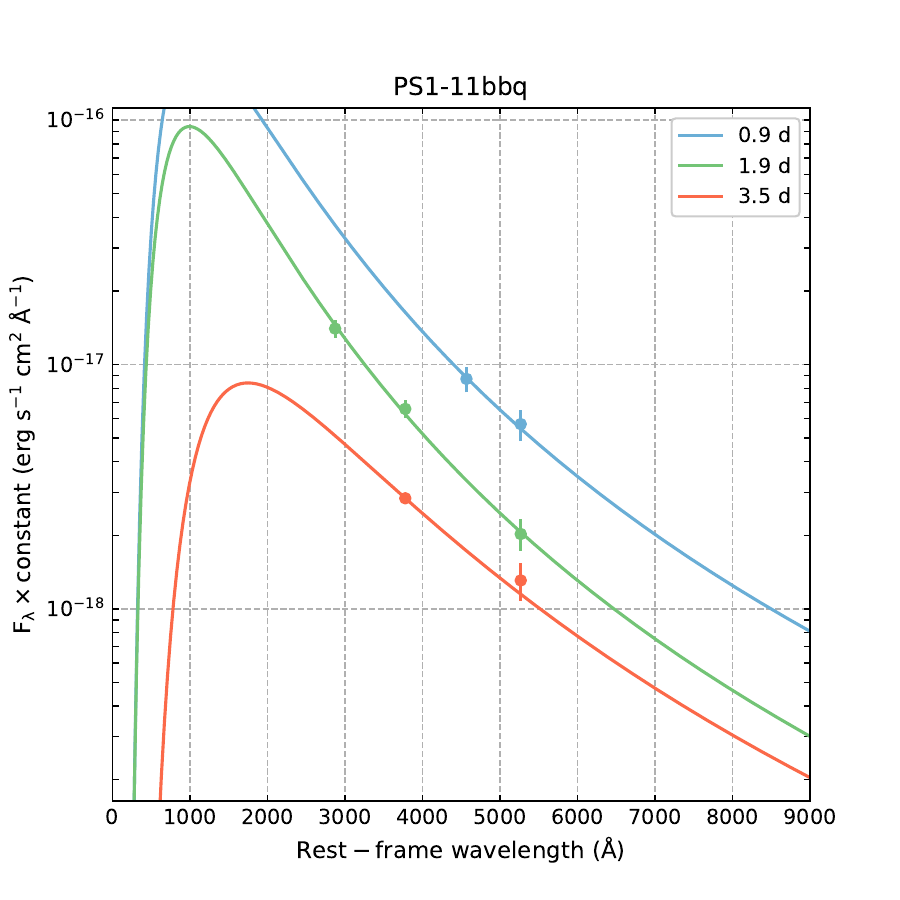}
\includegraphics[width=0.18\textwidth,angle=0]{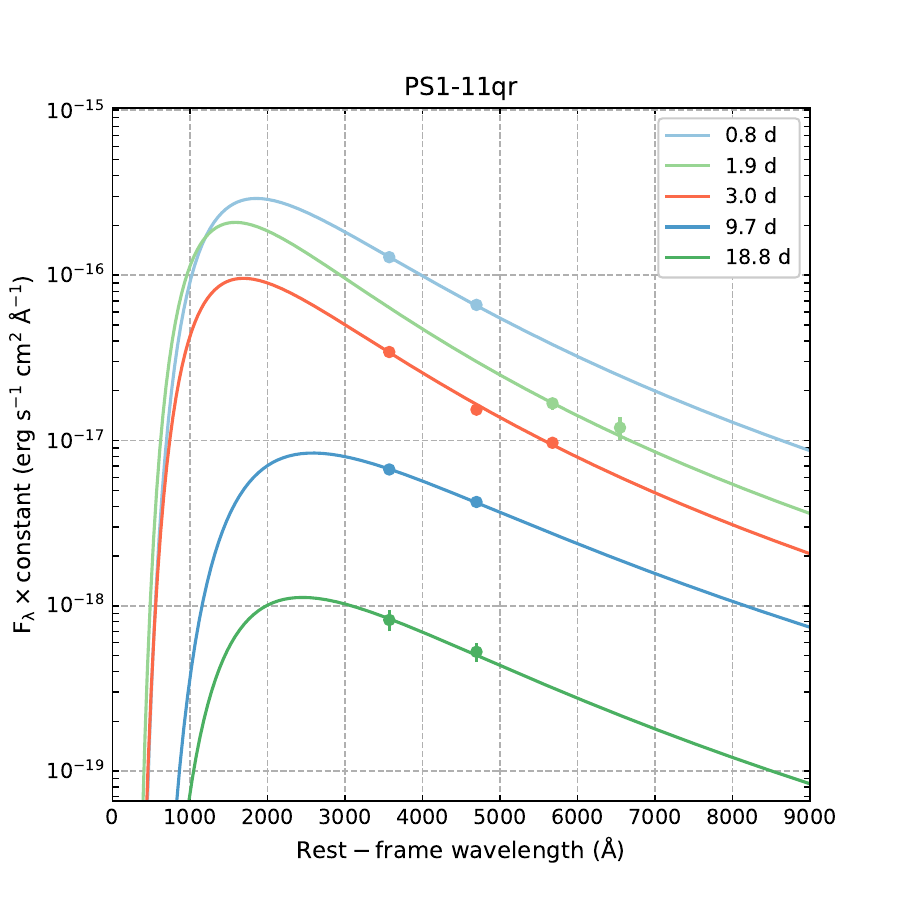}
\includegraphics[width=0.18\textwidth,angle=0]{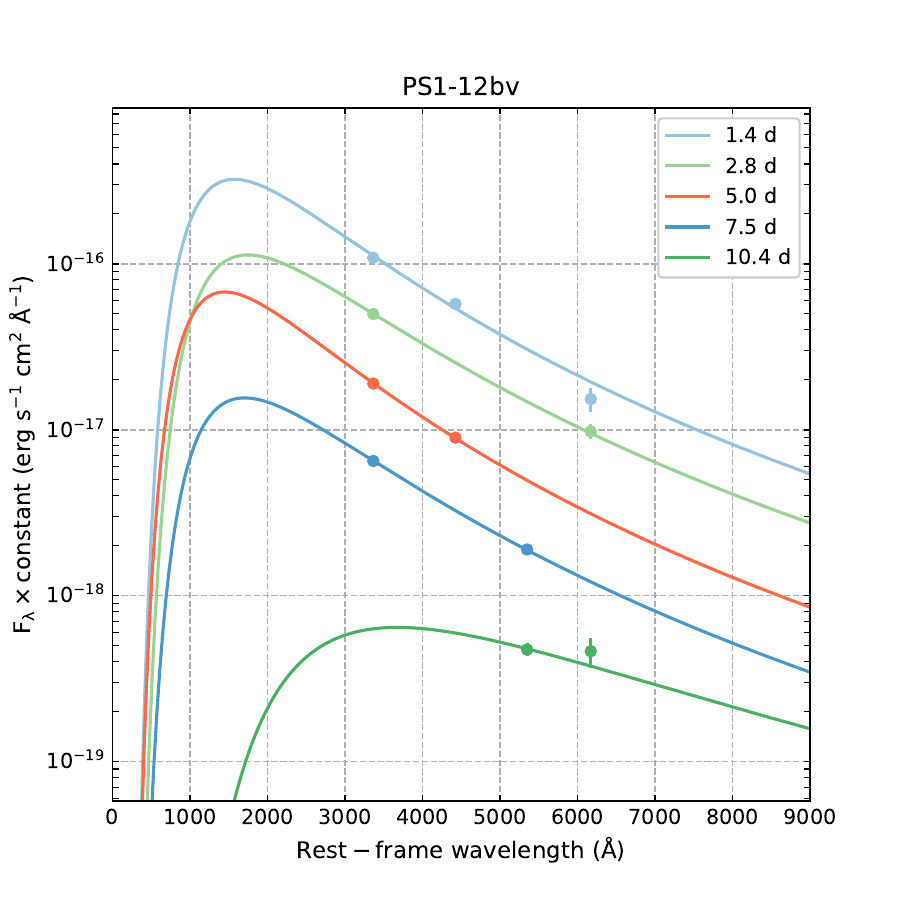}
\includegraphics[width=0.18\textwidth,angle=0]{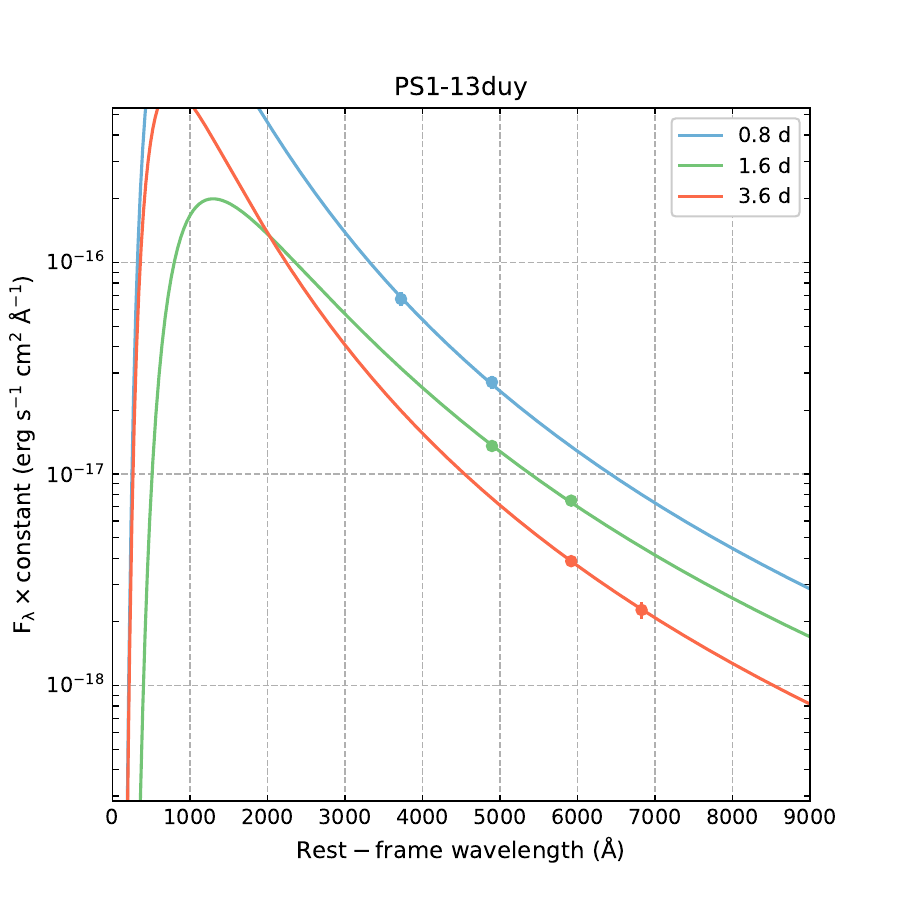}
\includegraphics[width=0.18\textwidth,angle=0]{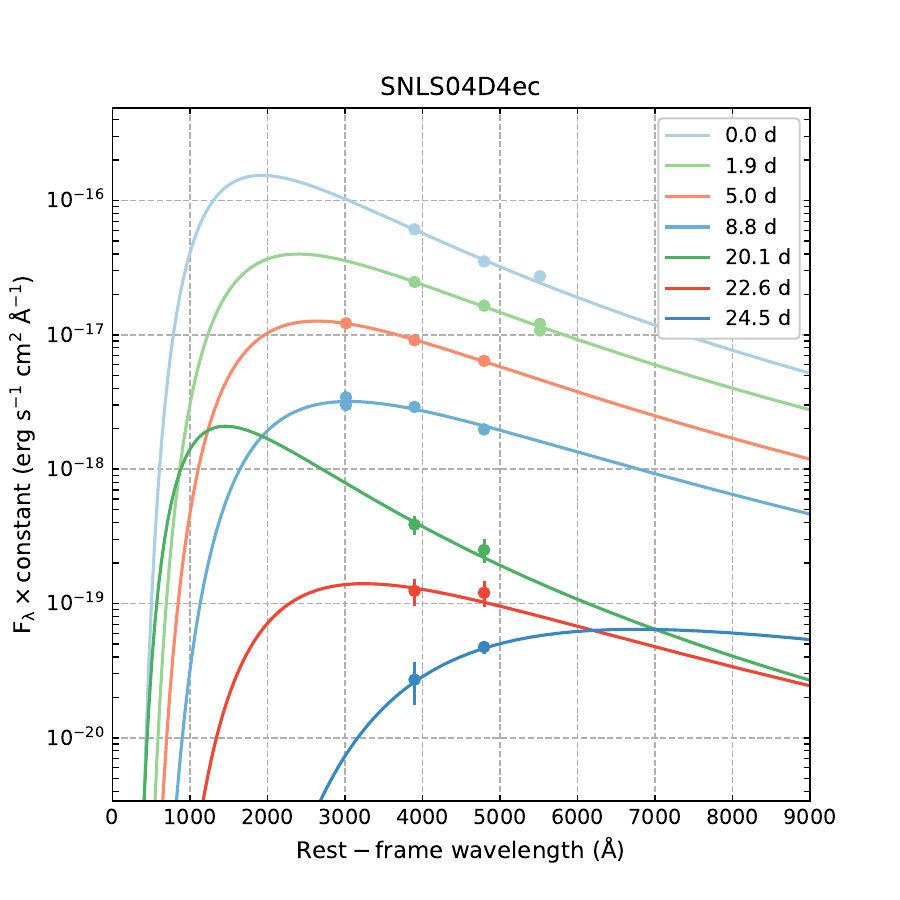}
\includegraphics[width=0.18\textwidth,angle=0]{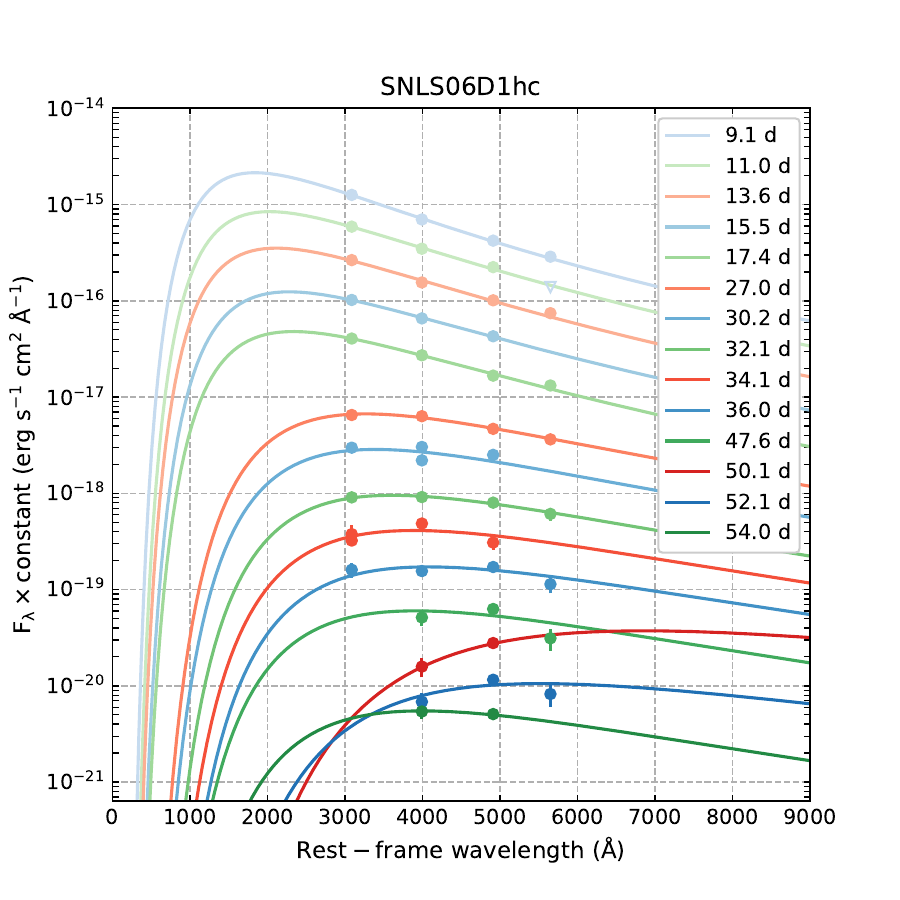}
\end{center}
\caption{{Fits to the multiepoch SEDs of 31 luminous REOTs using the blackbody model. The
points are the photometry, while the solid lines are the model SEDs.}}
\label{fig:bbfit}
\end{figure}

{Since the derived physical properties of the REOTs of \citet{Drout2014} and
\citet{Pur2018} have not previously been presented, we repeat the blackbody fits for the
SEDs of the 31 REOTs in Fig. \ref{fig:bbfit}.}
To get the best-fit parameters and the parameter ranges, we adopt {the Markov Chain Monte Carlo
(MCMC) method which is }performed by using the {\it emcee} Python package
\citep{Foreman-Mackey2013}. Once the MCMC is done, the best-fit values of the parameters are obtained by measuring
the medians of the posterior samples. The uncertainties are computed to $3\sigma$ confidence.
{Our fits also show that the SEDs of almost all events can be fitted by the blackbody model.
The best-fit values of the radii and the temperature, as well as the derived
blackbody luminosities of the events, are shown in Table \ref{tab:SED_Param_blackbody}.}

\begin{figure}[tbph]
\begin{center}
\includegraphics[width=0.32\textwidth,angle=0]{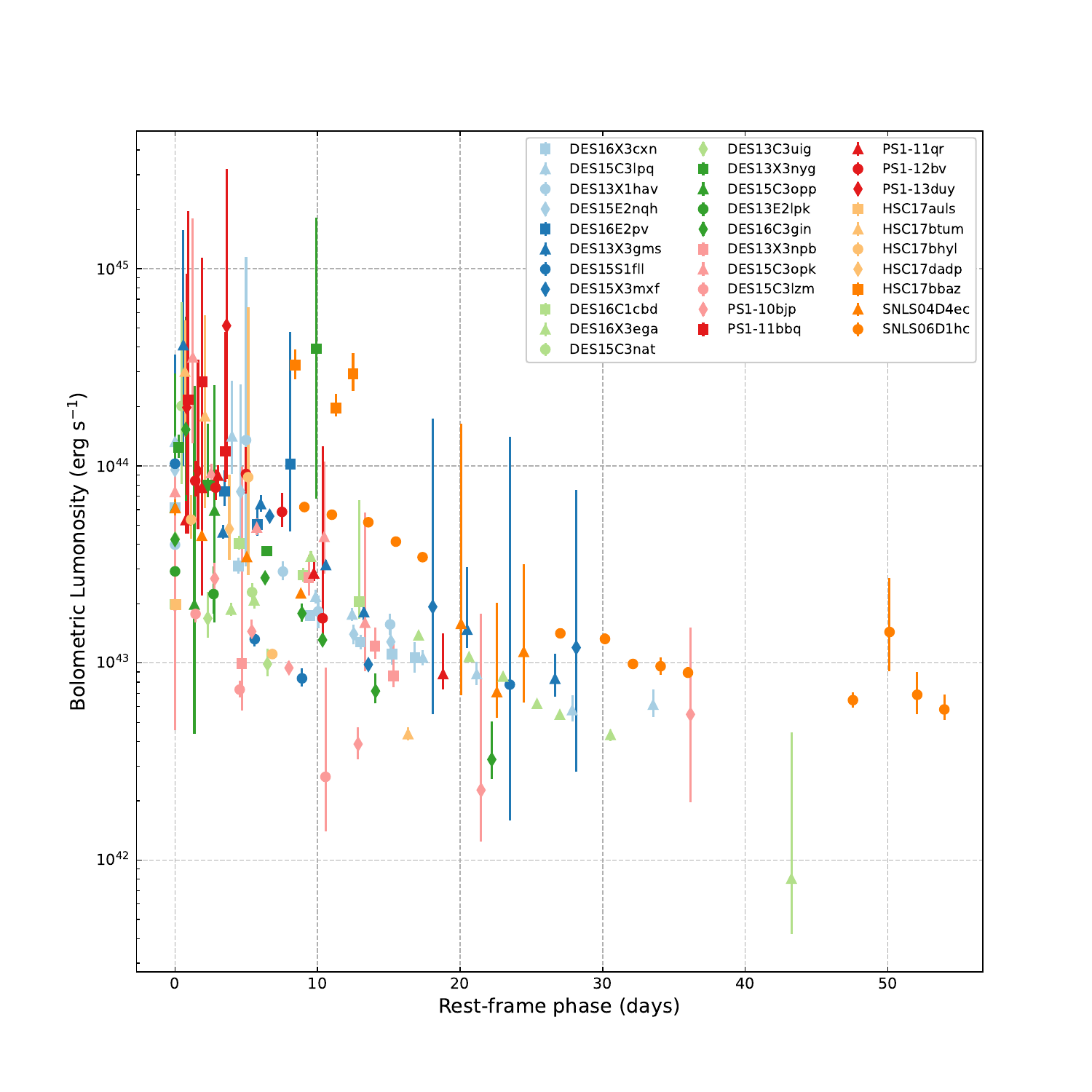}
\includegraphics[width=0.32\textwidth,angle=0]{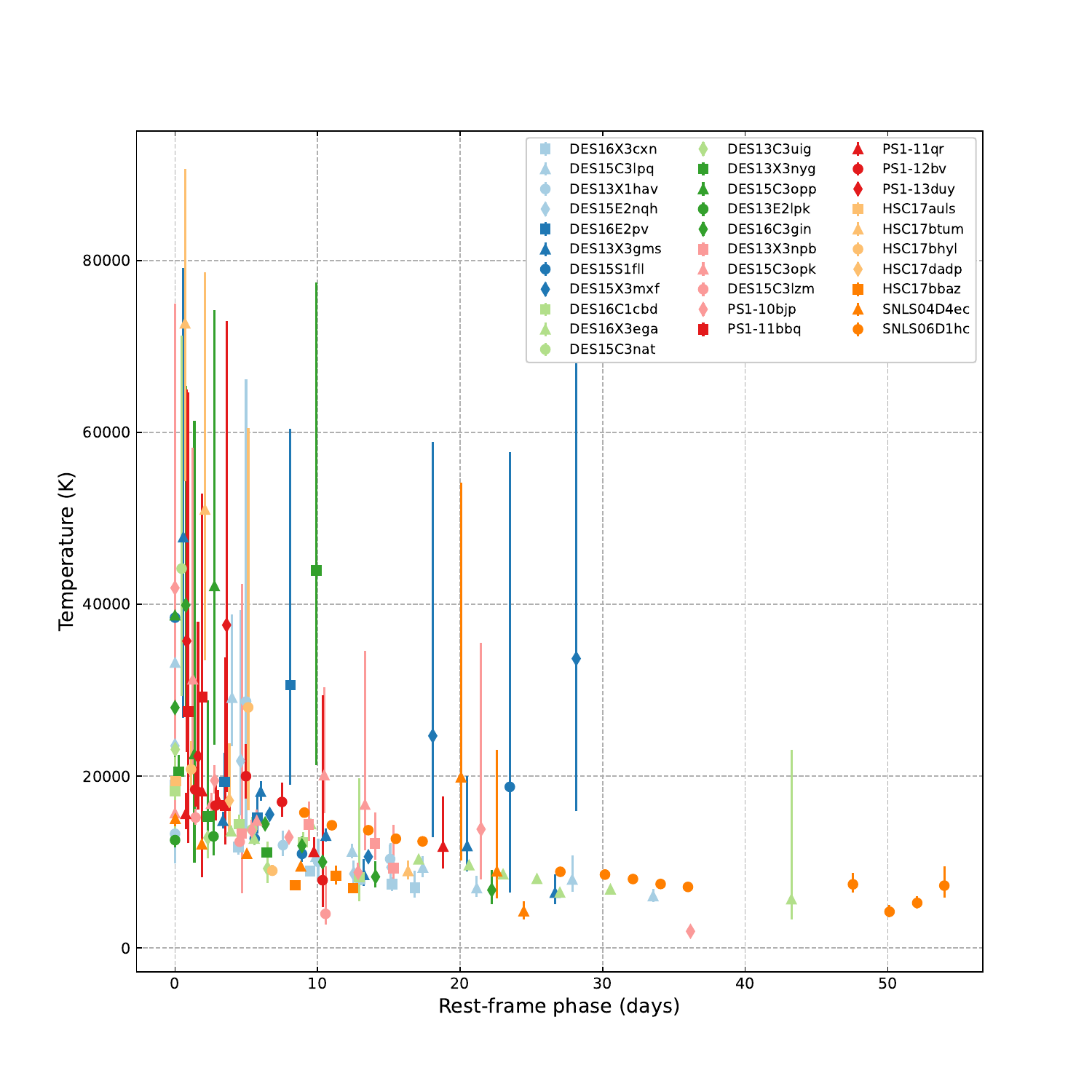}
\includegraphics[width=0.32\textwidth,angle=0]{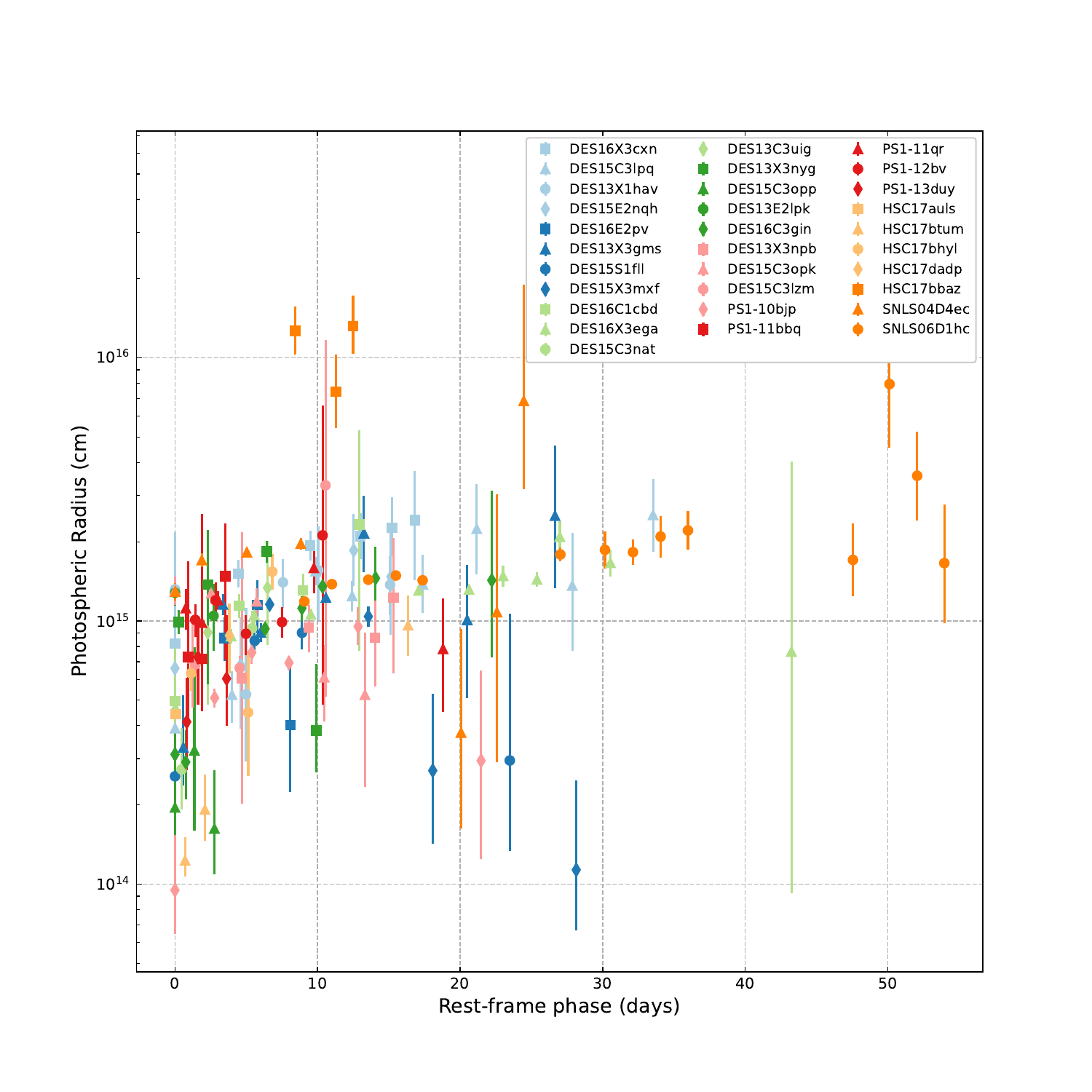}
\end{center}
\caption{{The derived bolometric light curves, the temperature evolution, and the radius
evolution of the 31 luminous REOTs using the blackbody model.}}
\label{fig:evo}
\end{figure}

To compare the {photospheric properties of the luminous REOTs},
we {plot} the {their blackbody} light curves, the temperature
evolution, {and the radius evolution} in Fig. \ref{fig:evo}.
{It should be pointed out that the zero-point of time of every event
is set to be their first detection, while the first point of every event in Fig. \ref{fig:evo}
corresponds to the epoch of the first SED. For some events, the multiband
observations were performed a few days after the first detections.}

{Fig. \ref{fig:evo} shows that}, while the shapes of the events are homogeneous, the peak
luminosities span from $\sim\,10^{43}$\,erg\,s$^{-1}$ (comparable to the peak
luminosities of SNe~Ia) to $\sim\,5\,\times\,10^{44}$\,erg\,s$^{-1}$
(brighter than some SLSNe). {Additionally, the late-time temperatures of
the events whose SEDs can be constructed at the epochs of $>\,10$\,days no longer
decline and reach a constant. For the events lacking SEDs at epochs of $>\,10$\,days,
the late-time temperature evolution is unknown. Finally,
we find that the early-time radii of the photospheres of the events increased, and reached a ``plateau''
($\sim$ (1--2) $\times 10^{15}$\,cm) in most of events or even receded in some events.}

\section{Modeling the Multiband Light Curves of REOTs}

\label{sec:fit}

{In this section, we use the magnetar and CSI models to fit the multiband
light curves of the events. The ejecta thermalize the input luminosities from
the energy sources and produce the bolometric photosphere luminosities.
The equations reproducing theoretical bolometric light curves are listed by \cite{Wang2015} and
\cite{Dai2016} for the magnetar model, and by \cite{Cha2012} and \cite{Wang2019} for the
CSI models. Another assumption needed to calculate the multiband light curves of SNe or SN-like
optical transients is that their photospheric emission can be described by the blackbody model
or the absorbed blackbody model \citep{Pra2017,Nich2017}.
As shown in Fig. \ref{fig:bbfit}, the SEDs of the events in our sample are consistent with
 blackbody radiation and the blackbody assumption is valid at least at some epochs.
We neglect the ultraviolet absorption effect, since the SEDs of the events we study can be fitted by the
standard blackbody model.
Finally, like \cite{Nich2017}, we assume that the radii of the photospheres of the events were
proportional to time at early epochs, while their temperatures no longer change
at late epoches (see Eqs. (8) and (9) of \citealt{Nich2017}). Fig. \ref{fig:evo} shows that
the assumption of the ``floor temperature'' ($T_{\rm f}$) is valid for the events in our sample.}

{The value of the optical opacity ($\kappa$) of the ejecta depends on the SN type; it can be
0.05--0.20\,cm$^2$\,g$^{-1}$ for the ejecta of H-deficient SNe~Ib/Ic/Ibn (see \citealt{Wang2015},
and the references therein). To reduce the number of the parameters, we assume that $\kappa = 0.1$\,cm$^2$\,g$^{-1}$.
\footnote{For SNe~IIn/II, the value of $\kappa$ can be set to 0.34\,cm$^2$\,g$^{-1}$;
but we still suppose that the mean value of $\kappa$ is 0.1\,cm$^2$\,g$^{-1}$ for our models,
since there are significantly more confirmed luminous rapidly evolving H-deficient SNe~Ibn/Ic
than luminous rapidly evolving SNe~IIn/II in the sample of \cite{Ho2021}; see Table 2 (which provides the spectral classifications
the events of their gold sample), Table 5 (which compiles other rapidly SNe and REOTs), and Table 14
(which presents the peak absolute magnitudes of the events of their sample).}
For the CSI model, we suppose that the densities of the inner and outer components
are respectively $\rho\,\propto\,r^{-1}$ and $\rho\,\propto\,r^{-7}$, and the
density of the CSM is either $\rho\,\propto\,r^{-2}$ (a wind) or $\rho\,=$ constant (a shell).
The definitions, units, and prior ranges of the free parameters of the two models are listed in
Tables \ref{tab:mag-parameters} and \ref{tab:CSI-parameters}.}

{The MCMC method is also used to get the best fits and derive the best-fit parameters of the models.
Figures \ref{fig:LCfits_excellent}, \ref{fig:LCfits_good}, and \ref{fig:LCfits_bad}
present the best fits of the magnetar (left panels), CSI-shell (middle panels),
and CSI-wind (right panels) models.}
The corresponding {best-fit} parameters of the {models} are {presented
in Tables \ref{table:LC_Param_MAG}, \ref{table:LC_Param_CSI-shell}, \ref{table:LC_Param_CSI-wind},
respectively. The reduced $\chi^2$ values ($\chi^2$/dof; dof = degrees of freedom) are also listed in the Tables.
The number of data points for some events is smaller than or equal to the number of parameters of the models, so
the values of dof are $\leq\,0$ and $\chi^2$/dof is infinite or negative. For these cases, $\chi^2$/dof
is meaningless, and we do not list the values.}

{We summarize the validity of the models we use in Table \ref{tab:vali} and determine
the best model for every event by comparing the $\chi^2$/dof values.
We note that the $\chi^2$/dof values can be used to help determine the best model
for an event, but they cannot be used to judge the validity of a model, since they strongly depend
on the data error bars, the number of data points, and the number of
free parameters in the model. The validity of a model
are determined mainly by comparing the data and the best-fit light curves.}

{From the figures and the tables}, we find the following.
\begin{enumerate}
\item {20} luminous REOTs {can be fitted by the magnetar model with excellent quality
(all data points are fitted by the theoretical light curves, or one/two data point(s)
slightly deviate from the theoretical light curves); see the left panels of Fig. \ref{fig:LCfits_excellent}.}

\item {8 REOTs can be fitted by the magnetar model with good quality (two or three data points
slightly deviate from the theoretical light curves, but the major parts of the light curves can be fitted;
or some data points deviate from the theoretical light curves, but can be supposed to be due to the cooling
emission from the shock-heated extended envelopes (DES16E2pv) and/or late-rebrightening (SNLS06D1hc));
see the left panels of Fig. \ref{fig:LCfits_good}.}

\item {25 REOTs can also be fitted by the CSI model; see the middle and right
panels of Figures \ref{fig:LCfits_excellent} and \ref{fig:LCfits_good}.}

\item {For the 25 REOTs that can fitted by both the magnetar and CSI models, the magnetar model get
better fits than the CSI model do for all events}.

\item {The remaining 3 REOTs cannot be explained by both the the magnetar and CSI models (the data around
the peaks cannot be fitted; see Fig. \ref{fig:LCfits_bad}).}
\end{enumerate}

\section{The Nature of REOTs}

\label{sec:dis}

\begin{figure}
\centering
\includegraphics[width=0.48\textwidth,angle=0]{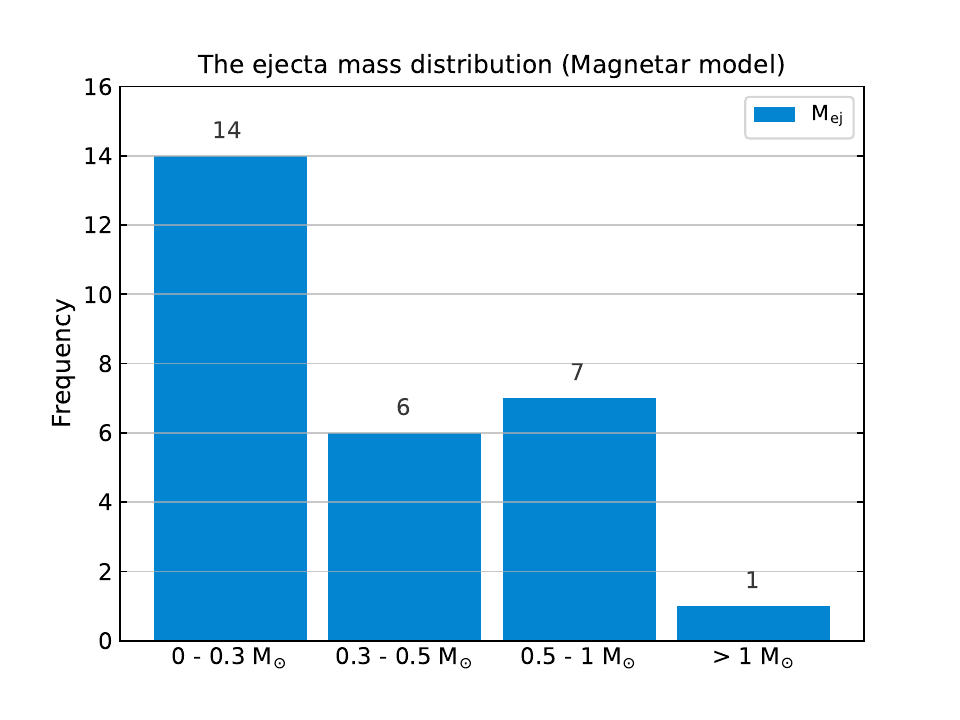}\\
\includegraphics[width=0.48\textwidth,angle=0]{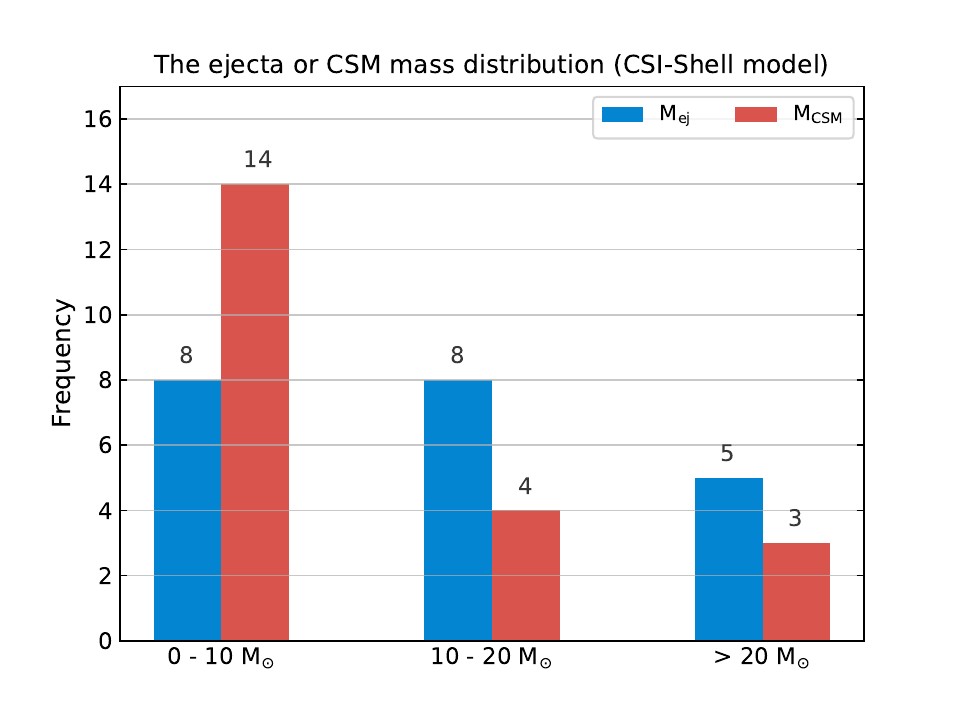}
\includegraphics[width=0.48\textwidth,angle=0]{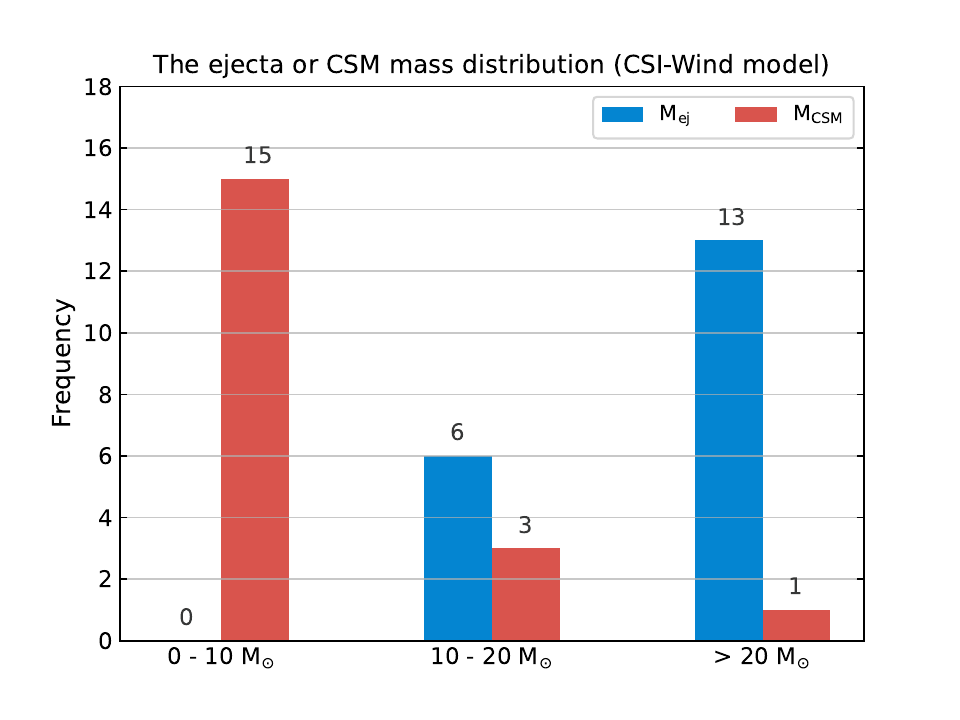}
\caption{{Histograms showing the distribution of the derived masses of the ejecta (and the derived CSM masses)
for the magnetar model (top panel), the CSI-shell model (bottom-left panel), and the CSI-wind model (bottom-right panel).}}
\label{fig:his}
\end{figure}

\begin{figure}
\centering
\includegraphics[width=0.48\textwidth,angle=0]{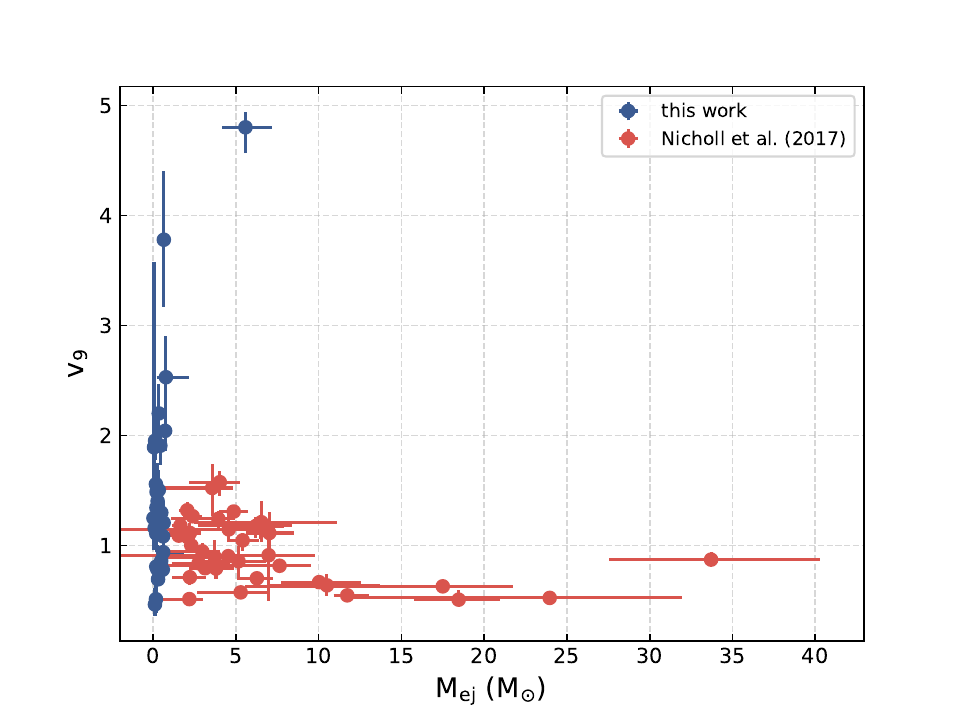}
\includegraphics[width=0.48\textwidth,angle=0]{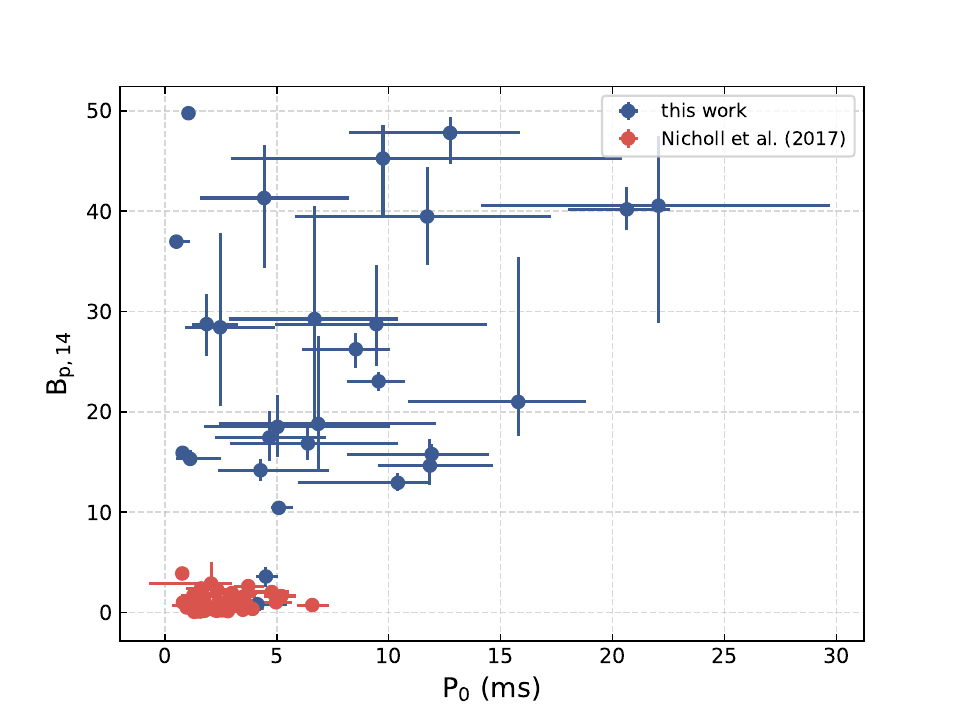}\\
\includegraphics[width=0.48\textwidth,angle=0]{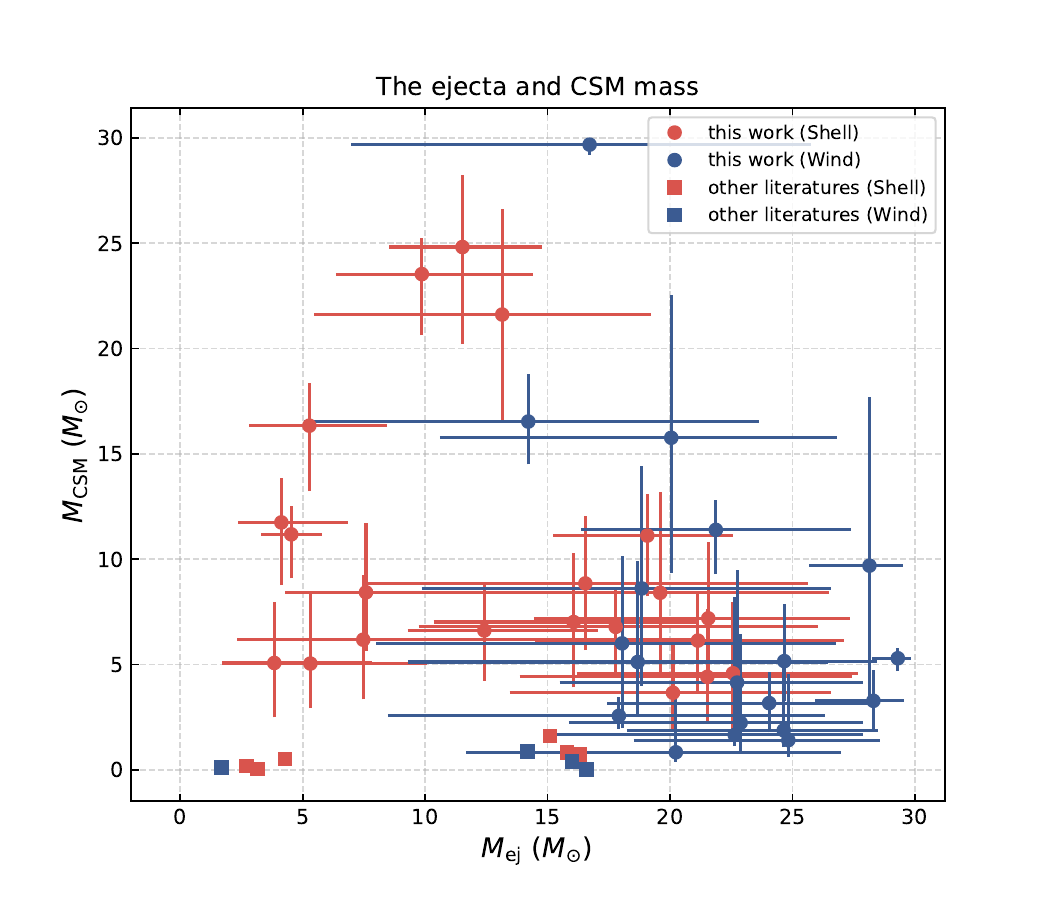}
\caption{{The $v$--$M_{\rm ej}$ space (top-left panel) and
the $P_0$--$B_p$ space (top-right panel) of the magnetar model,
and the $M_{\rm ej}$--$M_{\rm CSM}$ space (bottom panel) of the CSI model.
Parameters are from Tables \ref{table:LC_Param_MAG}, \ref{table:LC_Param_CSI-shell},
and \ref{table:LC_Param_CSI-wind}. For comparison, the same parameters of SLSN~I or
SNe~Ibn in the literature listed in \ref{sec:dis} are also plotted.}}
\label{fig:para}
\end{figure}

{In Section \ref{sec:fit}, we use the magnetar and CSI models to
fit the multiband light curves of 31 luminous REOTs and find that 27
of them can be explained by the magnetar and/or the CSI model.
The best-fit ejecta masses of 14, 6, 7, and 1 REOTs are 0.05--0.3\,M$_\odot$,
0.3--0.5\,M$_\odot$, 0.5--1.0\,M$_\odot$, and $>1.0$\,M$_\odot$, respectively (see Fig. \ref{fig:his}).
To illustrate the parameters, we present the four parameters ($M_{\rm ej}$, $v$, $B_p$, and $P_0$)
of the models in Fig. \ref{fig:para}; for comparison, the same parameters of SLSNe~I that were supposed to
be powered by magnetars \citep{Nich2017} are also plotted. We find that the ejecta masses of luminous REOTs
are significantly smaller than those of the magnetar-powered SLSNe~I, except for HSC17bbaz
($M_{\rm ej}=5.58^{+1.6}_{-1.4}$\,M$_\odot$). On the other hand, the velocity of the ejecta of some luminous
REOTs is larger than that of the magnetar-powered SLSNe~I. Furthermore, the $B_p$--$P_0$ distribution of
the REOTs is significantly larger than that of the magnetar-powered SLSNe~I.}

If mass transfer in a binary system consisting of a helium star and
a compact companion (e.g., a neutron star; NS, hereafter) strips the helium star,
the star would became a stripped core and explode as an {ultrastripped}
CCSN with ejecta mass $\sim 0.1$--0.3\,M$_\odot$
\citep{Tauris2013,Suwa2015,Tauris2015,Moriya2016,Moriya2017}.
By providing a detailed example, \citet{Tauris2013} demonstrated
that the mass-transfer process in a binary consisting of an NS and
a helium star can produce a small, bare core of $\sim 1.5$\,M$_\odot$ that
collapses and produces an SN whose ejecta mass is only $\sim 0.1$\,M$_\odot$.

If the explosion leaves a normal NS, the SN would be rather
faint since the $^{56}$Ni yield must be less than the ejecta mass.
\citet{Drout2013} and \citet{Tauris2013}
used the $^{56}$Ni-powered \citep{Arn1982} ultrastripped SN model
\citep{Tauris2013,Tauris2015,Moriya2016,Moriya2017} to explain the
pseudobolometric light curve of SN~2005ek.
\citet{De2018} reported photometric and spectral observations of iPTF14gqr,
{a rapidly evolving SN~Ic}. By modeling the bolometric light
curve and analyzing the spectra, \citet{De2018} found that the early-time excess
can be explained by the post-shock cooling model involving an extended
He-rich envelope heated by the SN shock, and the second peak of the SN
can be explained by the radioactivity model with $M_{\rm ej} \approx
0.15$--0.30\,M$_\odot$ and $M_{\rm Ni}\approx 0.05$\,M$_\odot$.
{\cite{Yao2020} studied SN~2019dge, a rapidly evolving SN~Ib,
and found that its ejecta mass is also extremely low, $\sim\,0.3$\,M$_\odot$.
It can be expected that the rapidly evolving subluminous SNe~IIb/Ib/Ic
of \cite{Ho2021} might also be $^{56}$Ni-powered ultrastripped CCSNe.}

If the {ultrastripped CCSNe} create rapidly rotating magnetars, the energy
input from the newly-born magnetar would significantly boost the
luminosities of the SNe, producing (super)luminous
REOTs/SNe whose peak luminosities can be significantly larger than
those of {SN~2005ek,} iPTF14gqr{, SN~2019dge,} and
{other} similar events. {To date, there are only two
luminous, rapidly evolving SNe~Ic --- iPTF16asu and SN~2018gep.}
\citet{Whit2017} used four different models ($^{56}$Ni,
magnetar, off-axis afterglow, and post-shock cooling)
to fit the pseudobolometric light curves of the luminous {SN~Ic-BL} iPTF16asu, suggesting that all
of them cannot fit the whole light curve well. \citet{Wang2019} used more complicated
models (magnetar plus interaction model, etc.) to fit the {bolometric} light curve of iPTF16asu.
{\cite{Pritchard2021} used $^{56}$Ni, $^{56}$Ni plus magnetar, and $^{56}$Ni
plus CSI models to fit the multiband light curves of SN~2018gep, demonstrating that
the last model is the best one for explaining the data. All of these works
require additional energy sources to yield the extraordinary luminosities of
the luminous, rapidly evolving SNe~Ic-BL. Our fits suggest that some
luminous REOTs might be luminous, rapidly evolving SNe~Ib/Ic similar to iPTF16asu and SN~2018gep.
Moreover, we demonstrate that some luminous REOTs might be luminous {(ultra)stripped} CCSNe
powered by the nascent magnetars.}

{Our fits also show that 25 luminous REOTs can be explained by the CSI model.
We find that, if they were powered by CSI, the derived masses of the ejecta and the CSM
can be rather large, $\sim 3$--30\,M$_\odot$ and $\sim 2.6$--30\,M$_\odot$ (respectively).
For the CSI-shell model, the respective derived ejecta masses of 8, 8, and 6 events are
$\sim 3$--10\,M$_\odot$, $\sim 10$--20\,M$_\odot$, and $\sim 20$--30\,M$_\odot$;
for the CSI-wind model, the respective derived ejecta masses of 0, 6, and 13 events are
$\sim 3$--10\,M$_\odot$, $\sim 10$--20\,M$_\odot$, and $\sim 20$--30\,M$_\odot$
(see Fig. \ref{fig:his} for the mass distribution).
As shown in Fig. \ref{fig:para}, the masses of the ejecta are consistent
with that of SNe~Ibn ($\sim 2$--20\,M$_\odot$). The mean values of the derived CSM masses are
larger than the values quoted in the literature (see, e.g., \citealt{Karamehmetoglu2017},
\citealt{Vallely2018}, \citealt{Gangopadhyay2020}, \citealt{Wang+Li2020}, \citealt{Kool2020},
\citealt{Xiang2021}), but the CSM masses of some events are lower than $\sim 5$\,M$_\odot$.
Moreover, the sample of SNe~Ibn fitted by the CSI model is too small, and we can expect
that modeling of a larger sample might extend the range of the CSM masses of SNe~Ibn.
We therefore suggest that the derived masses of the ejecta and the CSM favor the
scenario that some events might be luminous SNe~Ibn/IIn which are so-called
``interacting SNe'' exploding in massive, dense CSM.}

{The resemblance between the light curves of REOTs we study here and the roughly dozen rapidly
evolving SNe~Ibn presented by \cite{Ho2021} can provide additional support
for the scenario that at least some luminous REOTs might be interacting SNe.
Although the values of $\chi^2$/dof of the CSI model for some events are
larger than that of the magnetar model for the same events, we cannot conclude
that the magnetar model is the best model for them, since some luminous REOTs
might be luminous SNe~Ibn/IIn for which the CSI model is best.}

\section{Conclusions}

\label{sec:con}

In this paper, we {collect 31} luminous REOTs which cannot
be explained by the {$^{56}$Ni} model that has long
been adopted to account for the light curves of SNe~Ia and normal SNe~Ibc.
We suppose that the REOTs are {either luminous} ultrastripped CCSNe
{or SNe~Ibn/IIn, and we use the magnetar and CSI models to fit
their multiband light curves.}

We find that {28} events {in the sample} can be explained by
the magnetar model; the best-fit ejecta masses of {14, 6, 7, and 1 of them
are (respectively) 0.05--0.3\,M$_\odot$, 0.3--0.5\,M$_\odot$, 0.5--1.0\,M$_\odot$,
and $>1.0$\,M$_\odot$, indicating that most REOTs are ultrastripped SNe
 ($\sim 0.05$--0.20\,M$_\odot$; \citealt{Tauris2013})
or stripped SNe if they were powered by the magnetars. The ejecta masses of the 6 REOTs
with masses $\sim 0.5$--1.0\,M$_\odot$ are comparable to} that of the Type Ic
SN~1994I ($0.6^{+0.3}_{-0.1}$\,M$_\odot$), {while the derived mass of HSC17bbaz
($5.58^{+1.6}_{-1.4}$\,M$_\odot$) is comparable to that of SLSNe~I studied by \cite{Nich2017}.}
Our modeling assumes that the value of the optical opacity is 0.1\,cm$^2$\,g$^{-1}$;
adopting smaller values for the optical opacity (e.g., 0.05\,cm$^2$\,g$^{-1}$),
the inferred masses would be larger, since the derived ejecta mass is inversely
proportional to the optical opacity (see, e.g., \citealt{Wang2015,Lyman2016}).

{In addition, our modeling shows that 25 REOTs can be fitted by the CSI model, all of which
can also be explained by the magnetar model. The magnetar model can provide better fits than
the CSI model for all events that can be explained by both models.
The remaining three events in our sample cannot be explained by either of the models.
Although the fitting quality of the CSI model for most of the luminous REOTs
that can explained by both of the models is lower than that of the magnetar
model, we cannot exclude the CSI model since a fraction of them might be
luminous SNe~Ibn/IIn that favor the CSI model. Moreover, the derived ejecta masses
of the events that can be fitted by the CSI model are consistent with that of some
SNe~Ibn, supporting the scenario that some luminous REOTs might be luminous rapid
evolving SNe~Ibn.}

{The resemblance between the rapidly evolving multiband light curves of the luminous
REOTs and those of some luminous rapidly evolving SNe~Ic (iPTF16asu and SN~2018gep) and
SNe~Ibn also indicates that some of our luminous REOTs might be luminous SNe~Ib/Ic or
SNe~Ibn/IIn. Therefore, we suggest that both the magnetar and the CSI models are
plausible in explaining the luminous REOTs.}

{Unfortunately, the luminous REOTs we study are
spectroscopically unclassified, preventing us from drawing more robust conclusions about
their nature.}
Future high-cadence sky surveys, together with follow-up photometric and spectroscopic observations
of luminous REOTs, might {permit detailed modeling of their multiband light curves,
giving more stringent constraints on their physical properties and helping to distinguish among
different models. The spectra of luminous REOTs would provide key information for determining
their nature and the energy sources. If the spectral classifications show that
luminous REOTs are SNe~Ibn/IIn or SNe~Ib/Ic, then they would favor the CSI model or the magnetar-powered
(ultra)stripped CCSN model, respectively.}

The Large Synoptic Survey Telescope (LSST; \citealt{Ive2008,LSST2009})
can discover $\sim 100,000$ SNe~II per year \citep{LSST2009};
since the rate of SNe~Ibc is $\sim 1/3$ that of SNe~II
(see Table 1 of \citealt{Sma2009}, and references therein),
$\sim 130,000$ CCSNe should be discovered each year by LSST. \citet{Drout2014}
find that the rate of REOTs is 4--7\% of the CCSN rate at redshift $z \approx 0.2$;
based on the 37 DES REOTs with spectroscopic redshifts, \citet{Pur2018}
estimate that the REOT rate is $\gtrsim 1000$\,Gpc$^{-3}$\,yr$^{-1}$, which is
1.5\% of the CCSN rate derived by \citet{Li2011}. As noted by \citet{Pur2018},
this value is only a lower limit; adding 35 DES REOTs that do not have
spectroscopic redshifts, the rate would be doubled to 3\% of the CCSN rate.
We adopt the latter value as a conservative lower limit to estimate the
LSST REOT discovery rate, {finding that} $\gtrsim 4000$ REOTs should be
discovered by LSST per year. According to the PS1 and DES REOT samples, the
ratio of luminous REOTs to all {discovered} REOTs is $\sim70$\% (for the PS1 sample) or
$\sim75$\% (for the DES sample); thus, the number of luminous REOTs that should
be discovered by LSST is $\sim3000$ per year. It can be expected that further
investigation of a significantly larger sample might provide more reliable
values of the ratios of luminous REOTs that can and cannot be explained by
the magnetar {or CSI} models.

\acknowledgments

We thank Miika Pursiainen for providing the DES sample data.
This work is supported by the National Natural Science Foundation of China
(grants 11963001, 12133003, and 11973020 (C0035736)), the Bagui Scholars
Program (LEW), and the Bagui Young Scholars Program (LHJ).
L.J.W. is supported by the National Natural Science Foundation of China (grant U1938201).
Z.G.D. is supported by the National Key Research and Development Program of China
(grant 2017YFA0402600) and the National Natural Science Foundation
of China (grants 11573014 and 11833003).
Z.G.D. is supported by the National Key Research and Development Program of China
(grant 2017YFA0402600), the National SKA Program of China
(grant 2020SKA0120300), and the National Natural
Science Foundation of China (grant 11833003).
A.V.F. is grateful for financial
assistance from the Christopher R. Redlich Fund, the TABASGO Foundation,
and the U.C. Berkeley Miller Institute for Basic Research in Science
(where he is a Senior Miller Fellow).

\clearpage

\begin{table*}[tbp]
\footnotesize
\caption{Basic information for the REOTs and their host galaxies in our sample.$^a$}
\label{tab:info}
\begin{center}
{\small
\begin{tabular}{cccccccccccccccc}
\hline\hline
Name	    & $\alpha$	    & $\delta$           & $z$     &$M_{\rm rest,peak,g}$  & &  SFR               & sSFR       & Offset   &$E_\mathrm{B-V}$	 \\
 		    & (J2000)	    &	(J2000)	         &	       & (mag)                 & &  ($M_{\odot}$\,yr$^{-1}$) &  (Gyr$^{-1}$)  & (kpc)    & (mag) \\
\hline
DES13C3uig  & 03:31:46.55	& $-$27:35:07.96	 & 0.67    & $-19.00\pm0.08$       & & $\sim$11.2         & $\sim$0.26 & 3.80  & 0.007  \\   	
DES13E2lpk  & 00:40:23.80	& $-$43:32:19.74	 & 0.48    & $-19.19\pm0.07$       & & $\sim$6.17         & $\sim$0.42 & 4.77  & 0.0054 \\  	
DES13X1hav  & 02:20:07.80	& $-$05:06:36.53	 & 0.58    & $-19.42\pm0.32$       & & $\sim$0.68         & $\sim$0.48 & 2.51  & 0.0186 \\
DES13X3gms	& 02:23:12.27   & $-$04:29:38.35	 & 0.65	   & $-19.66\pm0.04$       & & $\sim$1.15         & $\sim$0.68 & 6.09  & 0.0257 \\
DES13X3npb  & 02:26:34.11	& $-$04:08:01.96	 & 0.50    & $-19.01\pm0.16$       & & $\sim$14.13        & $\sim$0.31 & 1.06  & 0.0245 \\ 	
DES13X3nyg  & 02:27:58.17	& $-$03:54:48.05	 & 0.71    & $-20.26\pm0.05$       & & $\sim$0.95         & $\sim$0.83 & 2.95  & 0.0224 \\ 	  	
DES15C3lpq	& 03:30:50.89   & $-$28:36:47.08	 & 0.61	   & $-19.84\pm0.08$       & & $\sim$0.81         & $\sim$0.48 & 2.18  & 0.0077 \\
DES15C3lzm  & 03:28:41.86	& $-$28:13:54.96	 & 0.33    & $-18.54\pm0.04$       & & $\sim$2.57         & $\sim$0.36 & 2.43  & 0.0059 \\
DES15C3nat  & 03:31:32.44	& $-$28:43:25.06	 & 0.84    & $-19.37\pm0.14$       & & $\sim$3.24         & $\sim$0.46 & 4.52  & 0.0087 \\
DES15C3opk  & 03:26:38.76   & $-$28:20:50.12	 & 0.57    & $-20.20\pm0.04$       & & $\sim$2.34         & $\sim$0.34 & 3.41  & 0.0125 \\ 	
DES15C3opp  & 03:26:57.53	& $-$28:06:53.61	 & 0.44	   & $-18.11\pm0.15$       & & $\sim$0.38         & $\sim$0.42 & 2.44  & 0.0081 \\
DES15E2nqh	& 00:38:55.59   & $-$43:05:13.14	 & 0.52	   & $-19.74\pm0.06$       & & $\sim$0.44         & $\sim$0.48 & 3.22  & 0.0077 \\
DES15S1fll	& 02:51:09.36   & $-$00:11:48.71	 & 0.23	   & $-18.23\pm0.05$       & & $\sim$0.68         & $\sim$0.36 & 11.01 & 0.0547 \\
DES15X3mxf	& 02:26:57.72   & $-$05:14:22.81	 & 0.44    & $-19.64\pm0.02$       & & $\sim$0.89         & $\sim$0.42 & 9.81  & 0.0235 \\
DES16C1cbd	& 03:39:25.97   & $-$27:40:20.37	 & 0.54	   & $-19.46\pm0.05$       & &      -             &      -     & 6.83  & 0.0100 \\
DES16C3gin	& 03:31:03.06   & $-$28:17:30.98	 & 0.35    & $-19.15\pm0.03$       & & $\sim$1.32         & $\sim$0.36 & 7.93  & 0.0085 \\
DES16E2pv   & 00:36:50.19	& $-$43:31:40.16	 & 0.73    & $-19.84\pm0.08$       & & $\sim$2.51         & $\sim$0.41 & 7.85  & 0.0055 \\
DES16X3cxn	& 02:27:19.32   & $-$04:57:04.27	 & 0.58	   & $-19.62\pm0.05$       & & $\sim$1.07         & $\sim$0.48 & 4.12  & 0.0236 \\
DES16X3ega	& 02:28:23.71   & $-$04:46:36.18	 & 0.26	   & $-19.25\pm0.03$       & & $\sim$3.80         & $\sim$0.36 & 8.93  & 0.0302 \\
HSC17auls	& 09:57:55.13   & +02:25:08.10 	     & 0.339   & $-18.00\pm0.02$       & &     -              & 1.19       & 3.73  & 0.0176 \\
HSC17bbaz	& 09:58:26.32   & +00:53:08.09 	     & 1.480   & $-19.76\pm0.11$($i$)  & &     -              & 8.23       & 25.69 & 0.0233 \\
HSC17bhyl	& 10:01:22.21   & +02:01:53.37       & 0.750   & $-18.69\pm0.04$       & &     -              & 1.69       & 0.43  & 0.0156 \\
HSC17btum	& 09:57:54.02   & +02:39:57.17       & 0.467   & $<-17.73\pm0.05$      & &     -              & 0.51       & 13.01 & 0.0168 \\
HSC17dadp	& 10:00:29.05   & +01:36:27.66       & 0.830   & $-18.74\pm0.03$($i$)  & &     -              & 0.39       & 5.27  & 0.0161 \\
PS1-10bjp   & 23:26:21.402  & $-$01:31:23.11     & 0.113   & $-$18.2               & & $\sim$2.1          & $\sim$1.6  & 2.39  & 0.0483 \\
PS1-11bbq   & 08:42:34.733  & +42:55:49.61       & 0.646   & $-$19.6               & & $>$0.3             & $>$2.4     & 8.81  & 0.026  \\
PS1-11qr    & 09:56:41.767  & +01:53:38.25       & 0.324   & $-$19.5               & & $\sim$4.3          & $\sim$0.3  & 12.58 & 0.0164 \\
PS1-12bv    & 12:25:34.602  & +46:41:26.97       & 0.405   & $-$19.4               & & $>$0.3             & $>$0.03    & 10.29 & 0.0102 \\
PS1-13duy   & 22:21:47.929  & $-$00:14:34.94 	 & 0.270   & $-$18.6               & & $>$0.1             & $>$0.2     & 2.16  & 0.0548 \\
SNLS04D4ec	& 22:16:29.29   & $-$18:11:04.1      & 0.593   & $-19.21\pm0.02$       & & $2.87\pm0.82$(OII) & $\sim$0.44 &   -   & 0.0238 \\
SNLS06D1hc	& 02:24:48.25   & $-$04:56:03.6      & 0.555   & $-19.75\pm0.01$       & & $0.21\pm0.07$(OII) & $\sim$0.36 &   -   & 0.0244 \\
\hline\hline
\end{tabular}%
}
\end{center}
\par
{$^a$ {Except for the values of the Milky Way extinction ($E_\mathrm{B-V}$), which are from \cite{Schlafly2011}}, all information
comes from Tables 3, 4, and 7 of \citet{Pur2018}, {Tables 1, 2, and 3 of \citet{Tampo2020}, Tables 1 and 5 of \citet{Drout2014},
and Tables 1, 6, and 8 of \cite{Arc2016}}. Note that the values of SFR and sSFR of the host galaxies of DES REOTs
are obtained by using the values of log($M_{\star}$/$M_{\odot}$) and log(sSFR) listed in
Table 7 of \citet{Pur2018}, {the values of sSFR of the host galaxies of SNSL04D4ec and SNLS06D1hc
are obtained from the values of log(sSFR) listed in
Table 7 of \citet{Arc2016}, the $g$-band peak magnitudes of HSC REOTs are derived from the $i$-band peak magnitudes listed in
Table 2 of \citet{Tampo2020} and the photometry presented in Table 4 of \citet{Tampo2020}, and
the $g$-band peak magnitudes of SNSL04D4ec and SNLS06D1hc are derived from the $z$- and $i$-band peak magnitudes listed in
Table 6 of \citet{Arc2016} and the photometry presented in Table 2 of \citet{Arc2016}}.}
\end{table*}

\clearpage

\begin{center}

\begin{longtable}{ccccccccccccc}
  \caption{Best-fit parameters of the blackbody model for REOT SEDs.}\label{tab:SED_Param_blackbody} \\

\toprule

\multirow{2}*{Name}  & Phase  & $T_{\rm ph}$ & $R_{\rm ph}$  & $L_{\rm ph}$             &\multirow{2}*{$\chi^{\rm 2}$}    \\
&                                    (days) & ($10^3$ K)   &($10^{15}$\,cm) & ($10^{42}$\,erg\,s$^{-1}$) &               &      \\
\midrule

\multirow{3}*{DES13C3uig} &  0.0  &  $23.11^{+6.3}_{-3.8}$  &  $0.46^{+0.1}_{-0.1}$  &  $42.7^{+25.3}_{-11.2}$  &  0.67 \\
 &  2.3  &  $12.83^{+5.9}_{-2.4}$  &  $0.9^{+0.4}_{-0.4}$  &  $16.83^{+6.1}_{-3.4}$  &  0.34 \\
 &  6.5  &  $9.25^{+3.1}_{-1.7}$  &  $1.34^{+0.7}_{-0.5}$  &  $9.88^{+2.0}_{-1.3}$  &  0.97 \\

\rowskip
\multirow{2}*{DES13E2lpk} &  0.0  &  $12.56^{+1.0}_{-0.8}$  &  $1.28^{+0.1}_{-0.1}$  &  $29.16^{+2.5}_{-2.1}$  &  0.79 \\
 &  2.7  &  $12.99^{+3.4}_{-2.2}$  &  $1.05^{+0.3}_{-0.3}$  &  $22.37^{+8.6}_{-4.5}$  &  0.64 \\
\rowskip
\multirow{5}*{DES13X1hav} &  0.0  &  $13.29^{+6.9}_{-3.4}$  &  $1.32^{+0.9}_{-0.6}$  &  $39.82^{+30.7}_{-9.0}$  &  0.39 \\
 &  5.0  &  $28.68^{+37.5}_{-15.1}$  &  $0.53^{+0.6}_{-0.2}$  &  $135.07^{+1009.9}_{-104.3}$  &  0.23 \\
 &  7.6  &  $11.97^{+1.7}_{-1.3}$  &  $1.4^{+0.3}_{-0.3}$  &  $29.09^{+3.7}_{-2.8}$  &  2.18 \\
 &  10.0  &  $10.03^{+2.7}_{-1.8}$  &  $1.56^{+0.7}_{-0.6}$  &  $18.25^{+3.8}_{-3.3}$  &  0.75 \\
 &  15.1  &  $10.36^{+1.7}_{-1.3}$  &  $1.37^{+0.4}_{-0.3}$  &  $15.69^{+2.0}_{-1.8}$  &  0.12 \\
\rowskip
\multirow{7}*{DES13X3gms} &  0.6  &  $47.83^{+31.3}_{-21.0}$  &  $0.33^{+0.2}_{-0.1}$  &  $409.44^{+1158.2}_{-309.2}$  &  0.16 \\
 &  3.4  &  $14.82^{+1.0}_{-0.9}$  &  $1.16^{+0.1}_{-0.1}$  &  $46.0^{+4.2}_{-3.4}$  &  0.79 \\
 &  6.0  &  $18.2^{+1.3}_{-1.1}$  &  $0.91^{+0.1}_{-0.1}$  &  $64.06^{+7.0}_{-5.5}$  &  2.68 \\
 &  10.6  &  $13.08^{+0.8}_{-0.7}$  &  $1.23^{+0.1}_{-0.1}$  &  $31.4^{+2.0}_{-1.7}$  &  0.24 \\
 &  13.3  &  $8.53^{+1.8}_{-1.3}$  &  $2.15^{+0.8}_{-0.6}$  &  $18.12^{+2.3}_{-1.7}$  &  0.084 \\
 &  20.5  &  $11.88^{+8.2}_{-3.0}$  &  $1.01^{+0.6}_{-0.5}$  &  $14.77^{+16.0}_{-2.8}$  &  1.64 \\
 &  26.7  &  $6.45^{+2.2}_{-1.3}$  &  $2.51^{+2.1}_{-1.2}$  &  $8.32^{+2.8}_{-1.6}$  &  0.29 \\
\rowskip
\multirow{4}*{DES13X3npb} &  4.7  &  $13.3^{+29.1}_{-6.9}$  &  $0.61^{+1.6}_{-0.4}$  &  $9.92^{+89.3}_{-4.2}$  &  0.5 \\
 &  9.4  &  $14.37^{+2.7}_{-1.9}$  &  $0.94^{+0.2}_{-0.2}$  &  $27.1^{+8.4}_{-5.1}$  &  1.07 \\
 &  14.0  &  $12.17^{+3.6}_{-1.9}$  &  $0.86^{+0.3}_{-0.3}$  &  $12.15^{+2.9}_{-1.7}$  &  0.17 \\
 &  15.3  &  $9.3^{+5.1}_{-2.1}$  &  $1.22^{+0.8}_{-0.6}$  &  $8.6^{+3.8}_{-1.1}$  &  0.18 \\
\rowskip
\multirow{4}*{DES13X3nyg} &  0.3  &  $20.52^{+1.9}_{-1.6}$  &  $0.99^{+0.1}_{-0.1}$  &  $124.36^{+19.8}_{-14.6}$  &  0.91 \\
 &  2.3  &  $15.33^{+13.5}_{-3.3}$  &  $1.37^{+0.8}_{-0.8}$  &  $79.85^{+83.8}_{-10.3}$  &  0.65 \\
 &  6.5  &  $11.13^{+0.6}_{-0.6}$  &  $1.84^{+0.2}_{-0.2}$  &  $36.89^{+1.7}_{-1.6}$  &  5.13 \\
 &  9.9  &  $43.96^{+33.5}_{-22.6}$  &  $0.38^{+0.3}_{-0.1}$  &  $392.75^{+1417.4}_{-324.4}$  &  1.25 \\
\rowskip
\multirow{8}*{DES15C3lpq} &  0.0  &  $33.21^{+7.4}_{-4.9}$  &  $0.39^{+0.1}_{-0.1}$  &  $132.55^{+80.6}_{-39.4}$  &  6.27 \\
 &  4.0  &  $29.14^{+9.7}_{-5.6}$  &  $0.52^{+0.1}_{-0.1}$  &  $141.26^{+129.7}_{-50.5}$  &  1.33 \\
 &  9.9  &  $10.61^{+0.8}_{-0.8}$  &  $1.55^{+0.2}_{-0.2}$  &  $21.69^{+1.8}_{-1.6}$  &  3.45 \\
 &  12.4  &  $11.23^{+0.9}_{-0.8}$  &  $1.24^{+0.2}_{-0.2}$  &  $17.62^{+1.5}_{-1.3}$  &  0.08 \\
 &  17.4  &  $9.36^{+1.4}_{-1.1}$  &  $1.38^{+0.4}_{-0.3}$  &  $10.64^{+1.0}_{-0.9}$  &  3.02 \\
 &  21.2  &  $6.97^{+1.4}_{-1.1}$  &  $2.24^{+1.1}_{-0.7}$  &  $8.79^{+1.4}_{-1.1}$  &  3.51 \\
 &  27.9  &  $7.98^{+2.8}_{-1.5}$  &  $1.36^{+0.8}_{-0.6}$  &  $5.77^{+1.1}_{-0.7}$  &  0.51 \\
 &  33.5  &  $6.06^{+0.8}_{-0.7}$  &  $2.53^{+0.9}_{-0.7}$  &  $6.15^{+1.2}_{-0.8}$  &  2.29 \\
\rowskip
\multirow{3}*{DES15C3lzm} &  1.4  &  $15.18^{+1.0}_{-0.9}$  &  $0.68^{+0.1}_{-0.0}$  &  $17.73^{+2.0}_{-1.6}$  &  9.06 \\
 &  4.5  &  $12.36^{+1.0}_{-0.9}$  &  $0.66^{+0.1}_{-0.1}$  &  $7.33^{+0.8}_{-0.7}$  &  14.22 \\
 &  10.6  &  $3.97^{+5.6}_{-1.2}$  &  $3.28^{+8.4}_{-2.8}$  &  $2.64^{+6.8}_{-1.2}$  &  0.39 \\
\rowskip
\multirow{2}*{DES15C3nat} &  0.5  &  $44.16^{+27.1}_{-14.8}$  &  $0.27^{+0.1}_{-0.1}$  &  $201.07^{+475.6}_{-120.2}$  &  2.77 \\
 &  5.4  &  $13.81^{+1.5}_{-1.2}$  &  $0.94^{+0.2}_{-0.1}$  &  $22.87^{+2.5}_{-1.9}$  &  6.16 \\
\rowskip\\\\\\\\\\\\\\

\rowskip
\multirow{6}*{DES15C3opk} &  0.0  &  $15.7^{+1.5}_{-1.3}$  &  $1.31^{+0.2}_{-0.2}$  &  $73.71^{+9.8}_{-7.2}$  &  0.0082 \\
 &  1.2  &  $31.28^{+26.9}_{-10.9}$  &  $0.72^{+0.3}_{-0.3}$  &  $355.31^{+1437.7}_{-224.9}$  &  0.23 \\
 &  2.5  &  $16.8^{+1.3}_{-1.1}$  &  $1.27^{+0.1}_{-0.1}$  &  $92.18^{+10.1}_{-7.7}$  &  0.0052 \\
 &  5.7  &  $14.83^{+1.3}_{-1.1}$  &  $1.19^{+0.1}_{-0.1}$  &  $48.68^{+5.4}_{-4.1}$  &  2.39 \\
 &  10.5  &  $20.13^{+10.2}_{-4.4}$  &  $0.61^{+0.2}_{-0.2}$  &  $43.69^{+61.1}_{-15.4}$  &  0.23 \\
 &  13.3  &  $16.69^{+17.9}_{-5.3}$  &  $0.52^{+0.4}_{-0.3}$  &  $16.0^{+42.2}_{-6.9}$  &  0.44 \\
\rowskip

\multirow{3}*{DES15C3opp} &  0.0  &  $38.72^{+32.5}_{-16.5}$  &  $0.2^{+0.1}_{-0.1}$  &  $61.21^{+233.7}_{-43.2}$  &  0.067 \\
 &  1.4  &  $22.6^{+38.8}_{-12.7}$  &  $0.32^{+0.5}_{-0.2}$  &  $19.78^{+235.2}_{-15.4}$  &  0.26 \\
 &  2.8  &  $42.14^{+32.1}_{-18.5}$  &  $0.16^{+0.1}_{-0.1}$  &  $59.34^{+197.3}_{-43.2}$  &  0.014 \\
\rowskip
\multirow{4}*{DES15E2nqh} &  0.0  &  $23.53^{+5.1}_{-3.3}$  &  $0.66^{+0.1}_{-0.1}$  &  $95.48^{+46.1}_{-22.8}$  &  2.04 \\
 &  4.6  &  $21.76^{+17.6}_{-6.4}$  &  $0.68^{+0.4}_{-0.3}$  &  $73.97^{+185.8}_{-31.9}$  &  0.28 \\
 &  12.5  &  $8.63^{+1.5}_{-1.3}$  &  $1.85^{+0.7}_{-0.5}$  &  $13.98^{+1.7}_{-1.6}$  &  3.87 \\
 &  15.1  &  $9.4^{+2.8}_{-1.6}$  &  $1.47^{+0.8}_{-0.6}$  &  $12.79^{+2.4}_{-2.1}$  &  0.23 \\
\rowskip
\multirow{4}*{DES15S1fll} &  0.0  &  $38.45^{+24.3}_{-11.4}$  &  $0.26^{+0.1}_{-0.1}$  &  $102.53^{+263.4}_{-58.0}$  &  1.13 \\
 &  5.6  &  $12.72^{+0.8}_{-0.7}$  &  $0.84^{+0.1}_{-0.1}$  &  $13.21^{+1.3}_{-1.1}$  &  3.72 \\
 &  8.9  &  $10.96^{+1.1}_{-0.9}$  &  $0.9^{+0.1}_{-0.1}$  &  $8.35^{+1.0}_{-0.8}$  &  1.04 \\
 &  23.5  &  $18.75^{+39.0}_{-12.3}$  &  $0.29^{+0.8}_{-0.2}$  &  $7.76^{+132.5}_{-6.2}$  &  0.19 \\
\rowskip
\multirow{4}*{DES15X3mxf} &  6.6  &  $15.54^{+0.4}_{-0.4}$  &  $1.16^{+0.0}_{-0.0}$  &  $55.5^{+2.2}_{-2.0}$  &  65.25 \\
 &  13.6  &  $10.62^{+0.6}_{-0.5}$  &  $1.04^{+0.1}_{-0.1}$  &  $9.81^{+0.5}_{-0.5}$  &  6.17 \\
 &  18.1  &  $24.69^{+34.2}_{-11.8}$  &  $0.27^{+0.3}_{-0.1}$  &  $19.27^{+154.1}_{-13.8}$  &  1.12 \\
 &  28.1  &  $33.68^{+36.4}_{-17.8}$  &  $0.11^{+0.1}_{-0.0}$  &  $11.97^{+63.8}_{-9.2}$  &  0.12 \\
\rowskip
\multirow{4}*{DES16C1cbd} &  0.0  &  $18.29^{+19.8}_{-5.9}$  &  $0.5^{+0.4}_{-0.3}$  &  $19.84^{+63.7}_{-7.9}$  &  1.19 \\
 &  4.5  &  $14.43^{+1.1}_{-0.9}$  &  $1.14^{+0.1}_{-0.1}$  &  $40.46^{+3.6}_{-2.8}$  &  0.62 \\
 &  9.0  &  $12.29^{+1.2}_{-1.0}$  &  $1.31^{+0.2}_{-0.2}$  &  $27.88^{+2.5}_{-2.0}$  &  0.48 \\
 &  12.9  &  $8.25^{+11.5}_{-2.8}$  &  $2.32^{+3.0}_{-1.6}$  &  $20.45^{+46.7}_{-3.7}$  &  0.41 \\
\rowskip
\multirow{7}*{DES16C3gin} &  0.0  &  $27.98^{+10.5}_{-5.1}$  &  $0.31^{+0.1}_{-0.1}$  &  $42.24^{+44.1}_{-13.9}$  &  0.17 \\
 &  0.8  &  $39.9^{+25.5}_{-12.3}$  &  $0.29^{+0.1}_{-0.1}$  &  $152.86^{+418.4}_{-91.5}$  &  0.14 \\
 &  6.3  &  $14.44^{+0.5}_{-0.5}$  &  $0.93^{+0.0}_{-0.0}$  &  $27.04^{+1.5}_{-1.4}$  &  3.39 \\
 &  8.9  &  $11.89^{+1.1}_{-0.9}$  &  $1.12^{+0.1}_{-0.1}$  &  $17.83^{+2.3}_{-1.6}$  &  0.019 \\
 &  10.4  &  $10.01^{+0.3}_{-0.3}$  &  $1.35^{+0.1}_{-0.1}$  &  $13.09^{+0.4}_{-0.4}$  &  0.0025 \\
 &  14.1  &  $8.29^{+1.8}_{-1.3}$  &  $1.45^{+0.5}_{-0.4}$  &  $7.2^{+1.7}_{-1.0}$  &  0.035 \\
 &  22.2  &  $6.74^{+2.4}_{-1.6}$  &  $1.43^{+1.7}_{-0.7}$  &  $3.24^{+1.8}_{-0.7}$  &  0.11 \\
\rowskip
\multirow{3}*{DES16E2pv} &  3.5  &  $19.35^{+3.4}_{-2.4}$  &  $0.86^{+0.2}_{-0.2}$  &  $74.25^{+20.7}_{-11.9}$  &  1.77 \\
 &  5.8  &  $15.18^{+2.7}_{-1.9}$  &  $1.15^{+0.3}_{-0.3}$  &  $50.44^{+9.8}_{-6.2}$  &  0.84 \\
 &  8.1  &  $30.62^{+29.8}_{-11.6}$  &  $0.4^{+0.3}_{-0.2}$  &  $101.96^{+377.1}_{-55.6}$  &  0.78 \\
\rowskip
\multirow{6}*{DES16X3cxn} &  0.0  &  $18.94^{+1.8}_{-1.5}$  &  $0.82^{+0.1}_{-0.1}$  &  $61.62^{+10.4}_{-7.7}$  &  1.47 \\
 &  4.5  &  $11.73^{+1.0}_{-0.9}$  &  $1.51^{+0.2}_{-0.2}$  &  $31.0^{+3.2}_{-2.5}$  &  0.83 \\
 &  9.5  &  $8.97^{+0.6}_{-0.6}$  &  $1.94^{+0.3}_{-0.2}$  &  $17.38^{+0.9}_{-0.9}$  &  2.37 \\
 &  13.0  &  $7.97^{+0.8}_{-0.7}$  &  $2.1^{+0.5}_{-0.4}$  &  $12.77^{+1.1}_{-1.1}$  &  7.32 \\
 &  15.2  &  $7.41^{+0.8}_{-0.8}$  &  $2.26^{+0.7}_{-0.5}$  &  $11.08^{+1.4}_{-1.3}$  &  0.02 \\
 &  16.8  &  $7.01^{+2.0}_{-1.1}$  &  $2.41^{+1.3}_{-1.0}$  &  $10.61^{+2.2}_{-1.7}$  &  0.14 \\
\rowskip\\\\\\\\\\\\\\
\rowskip

\multirow{10}*{DES16X3ega} &  3.9  &  $13.59^{+0.7}_{-0.6}$  &  $0.88^{+0.1}_{-0.1}$  &  $18.67^{+1.5}_{-1.3}$  &  0.007 \\
 &  5.6  &  $12.74^{+0.9}_{-0.8}$  &  $1.05^{+0.1}_{-0.1}$  &  $20.81^{+2.5}_{-1.9}$  &  0.01 \\
 &  9.5  &  $14.43^{+0.6}_{-0.5}$  &  $1.06^{+0.0}_{-0.0}$  &  $34.63^{+2.4}_{-2.1}$  &  27.17 \\
 &  17.1  &  $10.32^{+0.3}_{-0.3}$  &  $1.31^{+0.1}_{-0.1}$  &  $13.82^{+0.6}_{-0.5}$  &  26.19 \\
 &  20.6  &  $9.65^{+0.2}_{-0.2}$  &  $1.32^{+0.0}_{-0.0}$  &  $10.74^{+0.3}_{-0.3}$  &  0.0034 \\
 &  23.0  &  $8.59^{+0.5}_{-0.5}$  &  $1.48^{+0.1}_{-0.1}$  &  $8.53^{+0.4}_{-0.3}$  &  0.0052 \\
 &  25.4  &  $8.05^{+0.3}_{-0.3}$  &  $1.44^{+0.1}_{-0.1}$  &  $6.21^{+0.3}_{-0.2}$  &  0.0035 \\
 &  27.0  &  $6.46^{+0.5}_{-0.4}$  &  $2.09^{+0.3}_{-0.3}$  &  $5.47^{+0.2}_{-0.2}$  &  0.0069 \\
 &  30.5  &  $6.84^{+0.4}_{-0.4}$  &  $1.66^{+0.2}_{-0.2}$  &  $4.32^{+0.3}_{-0.3}$  &  0.013 \\
 &  43.2  &  $5.68^{+17.4}_{-2.4}$  &  $0.76^{+3.3}_{-0.7}$  &  $0.81^{+3.6}_{-0.4}$  &  0.86 \\
\rowskip
\multirow{1}*{HSC17auls} &  0.0  &  $19.42^{+0.7}_{-0.6}$  &  $0.44^{+0.0}_{-0.0}$  &  $19.82^{+1.2}_{-1.1}$  &  1.08 \\

\rowskip
\multirow{3}*{HSC17bbaz} &  8.4  &  $7.3^{+0.5}_{-0.4}$  &  $12.67^{+3.0}_{-2.3}$  &  $324.61^{+65.3}_{-49.3}$  &  0.0025 \\
 &  11.3  &  $8.39^{+1.2}_{-1.0}$  &  $7.43^{+2.9}_{-2.0}$  &  $196.5^{+34.4}_{-18.2}$  &  0.024 \\
 &  12.5  &  $6.97^{+0.5}_{-0.5}$  &  $13.19^{+4.1}_{-2.9}$  &  $292.47^{+79.7}_{-53.4}$  &  0.0044 \\
\rowskip
\multirow{3}*{HSC17bhyl} &  1.1  &  $20.79^{+3.3}_{-2.4}$  &  $0.63^{+0.1}_{-0.1}$  &  $53.35^{+17.9}_{-10.6}$  &  0.025 \\
 &  5.1  &  $28.0^{+32.5}_{-12.0}$  &  $0.45^{+0.3}_{-0.2}$  &  $87.59^{+549.5}_{-59.6}$  &  0.26 \\
 &  6.8  &  $8.99^{+0.6}_{-0.6}$  &  $1.54^{+0.3}_{-0.2}$  &  $11.1^{+0.6}_{-0.5}$  &  3.32 \\
\rowskip
\multirow{3}*{HSC17btum} &  0.7  &  $72.7^{+17.9}_{-18.4}$  &  $0.12^{+0.0}_{-0.0}$  &  $300.66^{+246.1}_{-159.8}$  &  0.35 \\
 &  2.1  &  $51.03^{+27.6}_{-17.6}$  &  $0.19^{+0.1}_{-0.0}$  &  $178.51^{+401.4}_{-117.7}$  &  0.013 \\
 &  16.3  &  $8.97^{+1.2}_{-0.9}$  &  $0.96^{+0.3}_{-0.2}$  &  $4.36^{+0.4}_{-0.3}$  &  0.031 \\
\rowskip
\multirow{1}*{HSC17dadp} &  3.8  &  $17.15^{+6.6}_{-3.5}$  &  $0.88^{+0.3}_{-0.2}$  &  $47.83^{+42.4}_{-14.5}$  &  0.12 \\
\rowskip
\multirow{7}*{PS1-10bjp} &  0.0  &  $41.91^{+33.1}_{-19.1}$  &  $0.095^{+0.1}_{-0.0}$  &  $19.7^{+75.1}_{-15.2}$  &  0.0081 \\
 &  2.8  &  $19.49^{+1.8}_{-1.5}$  &  $0.51^{+0.0}_{-0.0}$  &  $26.78^{+5.5}_{-4.0}$  &  0.012 \\
 &  5.4  &  $13.7^{+1.2}_{-1.0}$  &  $0.76^{+0.1}_{-0.1}$  &  $14.45^{+2.2}_{-1.6}$  &  0.017 \\
 &  8.0  &  $12.89^{+0.7}_{-0.6}$  &  $0.69^{+0.0}_{-0.0}$  &  $9.44^{+0.8}_{-0.6}$  &  0.0041 \\
 &  12.8  &  $8.8^{+1.2}_{-1.0}$  &  $0.95^{+0.2}_{-0.1}$  &  $3.88^{+0.8}_{-0.6}$  &  0.004 \\
 &  21.5  &  $13.83^{+21.7}_{-5.9}$  &  $0.29^{+0.4}_{-0.2}$  &  $2.26^{+15.5}_{-1.0}$  &  0.28 \\
 &  36.2  &  $1.94^{+0.4}_{-0.3}$  &  $23.31^{+28.8}_{-13.8}$  &  $5.48^{+9.6}_{-3.5}$  &  0.051 \\
\rowskip
\multirow{3}*{PS1-11bbq} &  0.9  &  $27.51^{+37.2}_{-15.3}$  &  $0.73^{+1.0}_{-0.3}$  &  $216.92^{+1749.3}_{-171.6}$  &  0.092 \\
 &  1.9  &  $29.18^{+12.1}_{-6.2}$  &  $0.72^{+0.2}_{-0.2}$  &  $266.82^{+315.7}_{-100.8}$  &  0.53 \\
 &  3.5  &  $16.58^{+17.3}_{-4.5}$  &  $1.48^{+0.9}_{-0.8}$  &  $118.48^{+359.8}_{-35.5}$  &  0.49 \\
\rowskip
\multirow{5}*{PS1-11qr} &  0.8  &  $15.61^{+2.5}_{-1.8}$  &  $1.12^{+0.2}_{-0.2}$  &  $53.09^{+13.2}_{-7.6}$  &  0.047 \\
 &  1.9  &  $18.27^{+34.7}_{-10.0}$  &  $0.98^{+1.6}_{-0.5}$  &  $77.3^{+1056.8}_{-55.3}$  &  0.5 \\
 &  3.0  &  $17.15^{+1.3}_{-1.1}$  &  $1.2^{+0.1}_{-0.1}$  &  $89.17^{+11.6}_{-8.9}$  &  2.52 \\
 &  9.7  &  $11.2^{+1.7}_{-1.3}$  &  $1.59^{+0.4}_{-0.3}$  &  $28.5^{+3.9}_{-2.4}$  &  0.027 \\
 &  18.8  &  $11.82^{+5.8}_{-2.6}$  &  $0.78^{+0.4}_{-0.3}$  &  $8.78^{+5.4}_{-1.4}$  &  0.15 \\
\rowskip
\multirow{5}*{PS1-12bv} &  1.4  &  $18.43^{+2.6}_{-2.0}$  &  $1.01^{+0.1}_{-0.1}$  &  $83.92^{+22.3}_{-13.9}$  &  4.69 \\
 &  2.8  &  $16.58^{+2.6}_{-1.8}$  &  $1.2^{+0.2}_{-0.2}$  &  $77.53^{+18.3}_{-10.8}$  &  0.066 \\
 &  5.0  &  $19.99^{+3.7}_{-2.6}$  &  $0.89^{+0.2}_{-0.2}$  &  $90.94^{+33.8}_{-18.4}$  &  0.046 \\
 &  7.5  &  $17.0^{+2.2}_{-1.7}$  &  $0.99^{+0.1}_{-0.1}$  &  $58.38^{+14.4}_{-9.5}$  &  0.024 \\
 &  10.4  &  $7.88^{+21.6}_{-3.2}$  &  $2.12^{+4.5}_{-1.6}$  &  $16.86^{+109.1}_{-4.5}$  &  0.86 \\
\rowskip
\multirow{3}*{PS1-13duy} &  0.8  &  $35.73^{+29.3}_{-12.9}$  &  $0.41^{+0.2}_{-0.1}$  &  $197.92^{+743.5}_{-127.1}$  &  0.16 \\
 &  1.6  &  $22.31^{+15.7}_{-6.2}$  &  $0.73^{+0.3}_{-0.2}$  &  $94.17^{+250.9}_{-46.7}$  &  0.23 \\
 &  3.6  &  $37.58^{+35.4}_{-19.5}$  &  $0.6^{+0.5}_{-0.2}$  &  $513.75^{+2681.5}_{-428.4}$  &  0.0063 \\
\rowskip\\\\\\\\

\rowskip
\multirow{7}*{SNLS04D4ec} &  0.0  &  $15.05^{+1.0}_{-0.8}$  &  $1.29^{+0.1}_{-0.1}$  &  $61.05^{+6.5}_{-5.0}$  &  6.9 \\
 &  1.9  &  $12.09^{+0.5}_{-0.5}$  &  $1.7^{+0.1}_{-0.1}$  &  $44.21^{+2.1}_{-1.8}$  &  3.74 \\
 &  5.0  &  $10.98^{+0.2}_{-0.2}$  &  $1.82^{+0.1}_{-0.1}$  &  $34.52^{+0.4}_{-0.4}$  &  0.44 \\
 &  8.8  &  $9.5^{+0.2}_{-0.2}$  &  $1.97^{+0.1}_{-0.1}$  &  $22.52^{+0.4}_{-0.4}$  &  16.58 \\
 &  20.1  &  $19.87^{+34.3}_{-9.7}$  &  $0.38^{+0.6}_{-0.2}$  &  $15.81^{+147.6}_{-8.9}$  &  0.45 \\
 &  22.6  &  $8.93^{+14.1}_{-3.1}$  &  $1.08^{+1.9}_{-0.8}$  &  $7.13^{+13.0}_{-1.9}$  &  0.52 \\
 &  24.5  &  $4.28^{+1.2}_{-1.0}$  &  $6.86^{+12.1}_{-3.7}$  &  $11.4^{+20.2}_{-5.1}$  &  0.052 \\
\rowskip
\multirow{14}*{SNLS06D1hc} &  9.1  &  $15.76^{+0.3}_{-0.3}$  &  $1.19^{+0.0}_{-0.0}$  &  $61.8^{+1.6}_{-1.5}$  &  4.52 \\
 &  11.0  &  $14.28^{+0.3}_{-0.3}$  &  $1.38^{+0.0}_{-0.0}$  &  $56.51^{+1.3}_{-1.2}$  &  6.18 \\
 &  13.6  &  $13.71^{+0.2}_{-0.2}$  &  $1.43^{+0.0}_{-0.0}$  &  $51.75^{+0.9}_{-0.8}$  &  24.02 \\
 &  15.5  &  $12.72^{+0.2}_{-0.2}$  &  $1.49^{+0.0}_{-0.0}$  &  $41.26^{+0.7}_{-0.7}$  &  1.96 \\
 &  17.4  &  $12.41^{+0.3}_{-0.3}$  &  $1.43^{+0.1}_{-0.1}$  &  $34.4^{+0.7}_{-0.6}$  &  4.95 \\
 &  27.0  &  $8.87^{+0.3}_{-0.3}$  &  $1.79^{+0.1}_{-0.1}$  &  $14.13^{+0.3}_{-0.3}$  &  1.0 \\
 &  30.2  &  $8.54^{+0.6}_{-0.6}$  &  $1.86^{+0.3}_{-0.3}$  &  $13.24^{+0.8}_{-0.8}$  &  9.04 \\
 &  32.1  &  $8.03^{+0.4}_{-0.4}$  &  $1.83^{+0.2}_{-0.2}$  &  $9.89^{+0.5}_{-0.5}$  &  0.32 \\
 &  34.1  &  $7.45^{+0.6}_{-0.5}$  &  $2.09^{+0.4}_{-0.4}$  &  $9.63^{+1.1}_{-0.9}$  &  5.07 \\
 &  36.0  &  $7.11^{+0.5}_{-0.5}$  &  $2.21^{+0.4}_{-0.3}$  &  $8.92^{+0.6}_{-0.5}$  &  4.1 \\
 &  47.6  &  $7.41^{+1.3}_{-1.0}$  &  $1.71^{+0.6}_{-0.5}$  &  $6.48^{+0.6}_{-0.5}$  &  7.06 \\
 &  50.1  &  $4.22^{+0.8}_{-0.6}$  &  $7.95^{+7.2}_{-3.4}$  &  $14.36^{+12.6}_{-5.3}$  &  0.016 \\
 &  52.1  &  $5.25^{+0.8}_{-0.6}$  &  $3.56^{+1.7}_{-1.2}$  &  $6.9^{+2.1}_{-1.4}$  &  2.92 \\
 &  54.0  &  $7.26^{+2.2}_{-1.4}$  &  $1.66^{+1.1}_{-0.7}$  &  $5.81^{+1.1}_{-0.7}$  &  0.042 \\

\bottomrule
\end{longtable}
\end{center}

\clearpage

\begin{table*}
\tabletypesize{\footnotesize}
\caption{\label{tab:mag-parameters}Definitions, units, and prior ranges of parameters of the magnetar model.}
\begin{tabular}{c c c c c c c}
\hline
\hline
	\colhead{Parameter}             & \colhead{Definition}                                 &  \colhead{Unit}           &     \colhead{Posterior} \\
     \hline
	$M_{\rm ej}$                    & the ejecta mass                                      &   M$_\odot$               &    $[0.001, 50]$          \\
	$P_0$                           & the initial period of the magnetar                   &   ms                      &    $[0.2, 50]$          \\
	$B_p$                           & the magnetic field strength of the magnetar          &   $10^{14}$ G             &    $[0.5, 50]$         \\
	$v$                             & the ejecta velocity                                  &   $10^9$ cm s$^{-1}$      &    $[0.1, 5.0]$         \\
	$\kappa_{\rm \gamma, mag}$      & gamma-ray opacity of magnetar photons                &   cm$^2$g$^{-1}$          &    $[-1.5686, 10,000] $   \\
	$T_{\rm f}$                     & the temperature floor of the photosphere             &   K                       &    $[1000, 10,000] $      \\
	$t_{\rm shift}$                 & the explosion time relative to the first data        &   days                    &    $[-20, 0]$           \\
	$A_{{\rm host}}$                  & Extinction in the host galaxy                      &   mag                     &    $[0, 0.5]$             \\
\hline
\hline
\noalign{\smallskip}
\end{tabular}
\end{table*}

\clearpage

\begin{table*}
\tabletypesize{\footnotesize}
\caption{\label{tab:CSI-parameters}Definitions, units, and prior ranges of parameters of the CSI model.}
\begin{tabular}{c c c c c c c}
\hline
\hline
	\colhead{Parameter}             & \colhead{Definition}                                 &  \colhead{Unit}           &     \colhead{Posterior} \\\hline
	$M_{\rm ej}$                    & the ejecta mass                                      &   M$_\odot$               &    $[0.1, 30]$          \\
	$v$                             & the ejecta velocity                                  &   $10^9$\,cm\,s$^{-1}$      &    $[0.1, 4.0]$         \\
	$M_{\rm CSM}$                   & the CSM mass                                         &   M$_\odot$               &    $[0.1, 30]$          \\
	$\rho_{\rm CSM,in}$             & the mass density of the innermost CSM                &  $10^{-12}$\,g\,cm$^{-3}$   &    $[0.001, 100]$       \\
	$R_{\rm in}$                    & the radius of the innermost CSM                      &  $10^{14}$\,cm             &    $[0.01, 100]$        \\
    $\epsilon$                      & the conversion efficiency of the kinetic energy to radiation   &  -              &    $[0.01, 0.5]$        \\
	$x_{\rm 0}$                     & the ratio of the radius of the inner ejecta to the radius of the ejecta    &  -  &    $[0.01, 0.5]$        \\
	$T_{\rm f}$                     & the temperature floor of the photosphere             &   K                       &    $[1000, 10,000] $      \\
	$t_{\rm shift}$                 & the explosion time relative to the first data        &   days                    &    $[-20, 0]$           \\
	$A_{{\rm host}}$                & Extinction in the host galaxy                        &   mag                     &    $[0, 0.5]$           \\
\hline
\hline
\noalign{\smallskip}
\end{tabular}
\end{table*}

\clearpage

\begin{table*}
\footnotesize
\setlength{\tabcolsep}{4pt}
\renewcommand{\arraystretch}{1.1}

\centering\caption{\label{table:LC_Param_MAG}Best-fit parameters of the luminous REOT light curves with the magnetar model.\footnote{The units of $M_{\rm ej}$, $P_0$, $v$, $\kappa_{\gamma}$, $T_{\rm f}$, and $t_{\rm shift}$ are respectively $M_{\odot}$, ms, 10$^{14}$\,G, 10$^9$\,cm\,s$^{-1}$, cm$^{2}$\,g$^{-1}$, $10^3$\,K, and days; $B_p=B_{p,14}\times10^{14}$\,G.}}
\begin{tabular}{ccccccccccccc}
\hline\hline
\colhead{Name}&\colhead{$M_{\rm ej}$}&\colhead{$P_{\rm 0}$}&\colhead{$B_{p,14}$}&\colhead{$v$}&\colhead{log($\kappa_{\gamma}$)}&\colhead{$T_{\rm f}$}&\colhead{$t_{\rm shift}$}&\colhead{$A_{\rm host}$}  & \colhead{$\chi^{\rm 2}$/dof} \\
\hline

DES13C3uig & $0.17^{+0.1}_{-0.1}$ &  $5.04^{+5.0}_{-3.3}$ &  $18.51^{+3.2}_{-3.0}$ &  $0.51^{+0.2}_{-0.1}$ &  $1.42^{+1.8}_{-1.8}$ &  $5.08^{+2.8}_{-2.8}$ &  $-12.556^{+3.5}_{-4.1}$ &  $0.23^{+0.2}_{-0.2}$ &  3.98 \\
DES13E2lpk & $0.61^{+1.2}_{-0.4}$ &  $6.85^{+5.2}_{-4.4}$ &  $18.8^{+8.8}_{-4.5}$ &  $0.94^{+0.3}_{-0.2}$ &  $1.33^{+1.8}_{-1.8}$ &  $7.52^{+1.9}_{-4.4}$ &  $-15.013^{+4.1}_{-3.3}$ &  $0.2^{+0.2}_{-0.1}$ &  15.96 \\
{DES13X1hav} & $0.24^{+0.4}_{-0.2}$ &  $16.41^{+0.7}_{-14.4}$ &  $11.37^{+5.0}_{-1.9}$ &  $3.9^{+0.7}_{-2.5}$ &  $1.83^{+1.5}_{-1.5}$ &  $9.88^{+0.1}_{-0.4}$ &  $-3.7^{+0.4}_{-0.7}$ &  $0.032^{+0.4}_{-0.0}$ &  1.58 \\
DES13X3gms & $0.46^{+0.1}_{-0.1}$ &  $4.28^{+3.1}_{-1.9}$ &  $14.17^{+1.2}_{-1.0}$ &  $0.84^{+0.1}_{-0.1}$ &  $1.28^{+1.9}_{-1.8}$ &  $5.21^{+2.9}_{-2.9}$ &  $-6.55^{+1.0}_{-1.3}$ &  $0.33^{+0.1}_{-0.2}$ &  1.33 \\
DES13X3npb & $0.49^{+0.3}_{-0.1}$ &  $9.45^{+5.0}_{-4.5}$ &  $28.74^{+5.9}_{-4.2}$ &  $0.86^{+0.1}_{-0.1}$ &  $1.12^{+2.0}_{-2.0}$ &  $5.11^{+2.8}_{-2.8}$ &  $-1.87^{+0.4}_{-0.5}$ &  $0.26^{+0.2}_{-0.2}$ &  7.61 \\
DES13X3nyg & $0.27^{+0.3}_{-0.2}$ &  $4.67^{+2.5}_{-2.4}$ &  $17.44^{+2.7}_{-2.4}$ &  $1.4^{+0.3}_{-0.2}$ &  $1.21^{+1.9}_{-1.9}$ &  $5.59^{+3.1}_{-3.1}$ &  $-7.19^{+2.0}_{-2.0}$ &  $0.12^{+0.1}_{-0.1}$ &  6.42 \\
DES15C3lpq & $0.19^{+0.0}_{-0.0}$ &  $1.13^{+1.4}_{-0.6}$ &  $15.31^{+0.9}_{-0.7}$ &  $0.81^{+0.1}_{-0.1}$ &  $1.75^{+1.5}_{-1.5}$ &  $7.53^{+0.7}_{-0.9}$ &  $-4.76^{+0.7}_{-0.7}$ &  $0.44^{+0.0}_{-0.1}$ &  2.42 \\
{DES15C3lzm} & $0.02^{+0.0}_{-0.0}$ &  $8.2^{+3.9}_{-2.6}$ &  $1.9^{+2.3}_{-1.0}$ &  $2.35^{+0.4}_{-0.2}$ &  $-1.46^{+0.3}_{-0.1}$ &  $5.67^{+4.1}_{-0.9}$ &  $-1.48^{+0.2}_{-0.3}$ &  $0.091^{+0.1}_{-0.1}$ &  49.21 \\
{DES15C3nat} & $0.12^{+0.2}_{-0.1}$ &  $6.12^{+4.4}_{-3.7}$ &  $14.67^{+3.1}_{-2.9}$ &  $0.55^{+0.2}_{-0.2}$ &  $1.48^{+1.7}_{-1.7}$ &  $5.44^{+2.9}_{-3.0}$ &  $-9.49^{+3.3}_{-5.0}$ &  $0.18^{+0.2}_{-0.1}$ &  5.56 \\
DES15C3opk & $0.44^{+0.1}_{-0.1}$ &  $9.55^{+1.2}_{-1.4}$ &  $23.04^{+0.9}_{-1.0}$ &  $1.91^{+0.2}_{-0.2}$ &  $1.39^{+1.8}_{-1.8}$ &  $9.94^{+0.0}_{-0.1}$ &  $-5.2^{+0.5}_{-0.5}$ &  $0.016^{+0.0}_{-0.0}$ &  8.24 \\

DES15C3opp & $0.11^{+0.1}_{-0.1}$ &  $9.75^{+10.7}_{-6.8}$ &  $45.26^{+3.4}_{-5.7}$ &  $0.46^{+0.1}_{-0.1}$ &  $1.06^{+2.0}_{-1.9}$ &  $5.51^{+3.1}_{-3.1}$ &  $-8.44^{+2.3}_{-2.0}$ &  $0.24^{+0.2}_{-0.2}$ &  -- \\  
DES15E2nqh & $0.19^{+0.1}_{-0.1}$ &  $6.39^{+4.0}_{-3.5}$ &  $16.84^{+1.8}_{-1.6}$ &  $1.11^{+0.2}_{-0.2}$ &  $1.59^{+1.6}_{-1.7}$ &  $6.9^{+2.3}_{-4.0}$ &  $-6.72^{+1.5}_{-1.1}$ &  $0.24^{+0.2}_{-0.2}$ &  1.64 \\
DES15S1fll & $0.26^{+0.2}_{-0.2}$ &  $11.72^{+5.5}_{-5.9}$ &  $39.48^{+4.9}_{-4.8}$ &  $0.78^{+0.2}_{-0.1}$ &  $1.34^{+1.8}_{-1.8}$ &  $6.33^{+2.4}_{-3.5}$ &  $-5.72^{+2.4}_{-1.6}$ &  $0.15^{+0.1}_{-0.1}$ &  2.28 \\
DES15X3mxf & $0.17^{+0.0}_{-0.0}$ &  $0.52^{+0.6}_{-0.2}$ &  $36.98^{+0.7}_{-0.8}$ &  $1.56^{+0.1}_{-0.1}$ &  $1.73^{+1.5}_{-1.5}$ &  $9.87^{+0.1}_{-0.2}$ &  $-0.88^{+0.1}_{-0.1}$ &  $0.015^{+0.0}_{-0.0}$ &  4.5 \\
DES16C1cbd & $0.34^{+0.2}_{-0.2}$ &  $11.84^{+2.8}_{-2.3}$ &  $14.64^{+2.7}_{-2.0}$ &  $1.5^{+0.2}_{-0.2}$ &  $1.49^{+1.7}_{-1.7}$ &  $6.12^{+2.7}_{-3.5}$ &  $-3.36^{+0.6}_{-0.8}$ &  $0.31^{+0.1}_{-0.2}$ &  1.54 \\
DES16C3gin & $0.022^{+0.0}_{-0.0}$ &  $4.14^{+1.3}_{-0.8}$ &  $0.81^{+0.5}_{-0.2}$ &  $1.25^{+0.0}_{-0.0}$ &  $-1.5^{+0.2}_{-0.1}$ &  $6.51^{+0.5}_{-0.4}$ &  $-2.48^{+0.3}_{-0.3}$ &  $0.085^{+0.1}_{-0.1}$ &  3.14 \\
DES16E2pv & $0.63^{+0.4}_{-0.3}$ &  $2.48^{+2.5}_{-1.6}$ &  $28.43^{+9.4}_{-7.9}$ &  $1.21^{+0.2}_{-0.2}$ &  $1.21^{+1.9}_{-1.9}$ &  $5.52^{+3.0}_{-3.1}$ &  $-3.87^{+0.5}_{-0.6}$ &  $0.2^{+0.2}_{-0.1}$ &  4.05 \\
DES16X3cxn & $0.2^{+0.2}_{-0.1}$ &  $11.92^{+2.6}_{-3.8}$ &  $15.79^{+1.0}_{-1.0}$ &  $1.34^{+0.2}_{-0.2}$ &  $1.93^{+1.4}_{-1.5}$ &  $7.04^{+0.7}_{-0.8}$ &  $-7.37^{+1.3}_{-1.7}$ &  $0.16^{+0.2}_{-0.1}$ &  1.39 \\
DES16X3ega & $0.3^{+0.0}_{-0.0}$ &  $0.8^{+0.2}_{-0.2}$ &  $15.91^{+0.5}_{-0.5}$ &  $0.69^{+0.0}_{-0.0}$ &  $-1.61^{+0.1}_{-0.1}$ &  $3.07^{+1.2}_{-1.4}$ &  $-1.3^{+0.1}_{-0.1}$ &  $0.5^{+0.0}_{-0.0}$ &  11.05 \\
HSC17auls & $0.057^{+0.1}_{-0.0}$ &  $22.06^{+7.7}_{-7.9}$ &  $40.56^{+7.0}_{-11.7}$ &  $1.89^{+1.7}_{-0.9}$ &  $-1.22^{+0.6}_{-0.3}$ &  $5.27^{+1.7}_{-1.3}$ &  $-2.41^{+1.0}_{-2.2}$ &  $0.27^{+0.2}_{-0.2}$ &  -- \\ 

HSC17bbaz & $5.58^{+1.6}_{-1.4}$ &  $4.52^{+0.6}_{-0.4}$ &  $3.58^{+0.9}_{-1.1}$ &  $4.8^{+0.1}_{-0.2}$ &  $0.22^{+2.6}_{-1.6}$ &  $4.33^{+2.2}_{-2.3}$ &  $-4.13^{+0.6}_{-0.7}$ &  $0.31^{+0.1}_{-0.2}$ &  24.15 \\
HSC17bhyl & $0.12^{+0.0}_{-0.0}$ &  $20.64^{+1.9}_{-2.6}$ &  $40.19^{+2.2}_{-2.1}$ &  $1.95^{+0.2}_{-0.2}$ &  $2.11^{+1.3}_{-1.3}$ &  $8.25^{+0.7}_{-0.7}$ &  $-2.77^{+0.3}_{-0.3}$ &  $0.044^{+0.1}_{-0.0}$ &  8.61 \\
HSC17btum & $0.58^{+0.1}_{-0.1}$ &  $12.75^{+3.1}_{-4.5}$ &  $47.82^{+1.6}_{-3.1}$ &  $0.78^{+0.1}_{-0.1}$ &  $1.11^{+2.0}_{-2.0}$ &  $9.77^{+0.2}_{-0.3}$ &  $-3.47^{+0.4}_{-0.5}$ &  $0.074^{+0.1}_{-0.1}$ &  9.61 \\
HSC17dadp & $0.77^{+1.4}_{-0.6}$ &  $15.79^{+3.0}_{-4.9}$ &  $20.99^{+14.5}_{-3.4}$ &  $2.53^{+0.4}_{-0.3}$ &  $1.5^{+1.7}_{-1.8}$ &  $6.43^{+1.6}_{-3.0}$ &  $-1.93^{+0.4}_{-0.4}$ &  $0.18^{+0.2}_{-0.1}$ &  -- \\  
PS1-10bjp & $0.094^{+0.0}_{-0.0}$ &  $1.06^{+0.1}_{-0.0}$ &  $49.78^{+0.2}_{-0.3}$ &  $1.16^{+0.0}_{-0.0}$ &  $1.95^{+1.4}_{-1.4}$ &  $6.48^{+0.3}_{-0.4}$ &  $-0.68^{+0.0}_{-0.0}$ &  $0.49^{+0.0}_{-0.0}$ &  13.83 \\
PS1-11bbq & $0.65^{+0.3}_{-0.2}$ &  $6.69^{+3.7}_{-3.8}$ &  $29.26^{+11.3}_{-10.3}$ &  $3.78^{+0.6}_{-0.6}$ &  $1.18^{+1.9}_{-2.0}$ &  $5.48^{+3.1}_{-3.1}$ &  $-1.84^{+0.3}_{-0.4}$ &  $0.23^{+0.2}_{-0.2}$ &  16.91 \\
PS1-11qr & $0.72^{+0.2}_{-0.1}$ &  $8.54^{+1.5}_{-2.4}$ &  $26.24^{+1.6}_{-1.9}$ &  $2.04^{+0.2}_{-0.2}$ &  $1.37^{+1.8}_{-1.8}$ &  $9.77^{+0.2}_{-0.4}$ &  $-3.43^{+0.3}_{-0.3}$ &  $0.065^{+0.1}_{-0.0}$ &  2.29 \\
PS1-12bv & $0.49^{+0.1}_{-0.1}$ &  $1.87^{+1.4}_{-0.7}$ &  $28.75^{+3.0}_{-3.1}$ &  $1.3^{+0.1}_{-0.1}$ &  $1.11^{+2.0}_{-2.0}$ &  $5.68^{+2.6}_{-3.1}$ &  $-5.03^{+0.5}_{-0.6}$ &  $0.16^{+0.1}_{-0.1}$ &  3.03 \\
PS1-13duy & $0.34^{+0.2}_{-0.1}$ &  $4.44^{+3.8}_{-2.9}$ &  $41.33^{+5.3}_{-7.0}$ &  $2.2^{+0.3}_{-0.2}$ &  $1.12^{+2.0}_{-1.9}$ &  $5.85^{+2.9}_{-3.3}$ &  $-1.46^{+0.3}_{-0.3}$ &  $0.13^{+0.1}_{-0.1}$ &  3.48 \\
SNLS04D4ec & $0.21^{+0.1}_{-0.1}$ &  $10.41^{+1.5}_{-4.5}$ &  $12.92^{+1.0}_{-0.9}$ &  $1.49^{+0.1}_{-0.1}$ &  $-0.7^{+0.5}_{-0.3}$ &  $9.82^{+0.1}_{-2.1}$ &  $-9.5^{+0.7}_{-0.5}$ &  $0.12^{+0.3}_{-0.1}$ &  3.0 \\
SNLS06D1hc & $0.6^{+0.0}_{-0.0}$ &  $5.09^{+0.6}_{-0.3}$ &  $10.43^{+0.2}_{-0.2}$ &  $1.08^{+0.0}_{-0.0}$ &  $1.83^{+1.5}_{-1.5}$ &  $9.27^{+0.2}_{-0.2}$ &  $-1.28^{+0.1}_{-0.0}$ &  $0.47^{+0.0}_{-0.0}$ &  5.99 \\

\noalign{\smallskip}\hline\noalign{\smallskip}
\end{tabular}
\smallskip
\end{table*}

\clearpage

\begin{table*}
\footnotesize
\setlength{\tabcolsep}{3.7pt}
\renewcommand{\arraystretch}{1.1}

\centering\caption{\label{table:LC_Param_CSI-shell}Best-fit parameters of the luminous REOT light curves with the CSI-shell model. \footnote{The units of $M_{\rm ej}$, $v$, $M_{\rm CSM}$, $\rho_{\rm CSM}$, $R_{\rm in}$, $T_{\rm f}$, and $t_{\rm shift}$ are respectively $M_{\odot}$, $10^9$\,cm\,s$^{-1}$, $M_{\odot}$, $10^{-12}$\,g\,cm$^{-3}$, $10^{14}$\,cm, $10^{3}$\,K, and days.}}
\begin{tabular}{ccccccccccccc}
\hline\hline
\colhead{Name}&\colhead{$M_{\rm ej}$}&\colhead{$v$}&\colhead{$M_{\rm CSM}$}&\colhead{$\rho_{\rm CSM}$}&\colhead{$R_{\rm in}$}&\colhead{$\epsilon$}&\colhead{$x_0$}&\colhead{$T_{\rm f}$} &  \colhead{$t_{\rm shift}$} & \colhead{$A_{\rm host}$}  & \colhead{$\chi^{\rm 2}$/dof}\\
\hline
DES13C3uig & $22.57^{+5.1}_{-6.4}$ &  $0.88^{+0.5}_{-0.3}$ &  $4.59^{+3.4}_{-2.7}$ &  $48.22^{+33.1}_{-27.9}$ &  $58.06^{+27.5}_{-28.9}$ &  $0.28^{+0.1}_{-0.1}$ &  $0.29^{+0.1}_{-0.1}$ &  $5.04^{+3.1}_{-2.8}$ &  $-8.29^{+3.4}_{-3.9}$ &  $0.12^{+0.1}_{-0.1}$ &  24.89 \\
DES13E2lpk & $19.6^{+6.9}_{-7.5}$ &  $1.0^{+0.4}_{-0.2}$ &  $8.41^{+4.8}_{-3.8}$ &  $47.87^{+35.0}_{-31.0}$ &  $67.36^{+22.9}_{-29.3}$ &  $0.28^{+0.1}_{-0.1}$ &  $0.3^{+0.1}_{-0.1}$ &  $6.16^{+2.8}_{-3.5}$ &  $-13.997^{+4.4}_{-3.9}$ &  $0.24^{+0.2}_{-0.2}$ &  --\\  
{DES13X1hav} & $20.85^{+6.3}_{-8.1}$ &  $1.85^{+1.4}_{-0.3}$ &  $5.34^{+5.1}_{-2.5}$ &  $56.19^{+30.2}_{-33.1}$ &  $40.8^{+36.0}_{-30.9}$ &  $0.17^{+0.2}_{-0.1}$ &  $0.2^{+0.1}_{-0.1}$ &  $9.66^{+0.3}_{-0.7}$ &  $-4.31^{+0.4}_{-0.6}$ &  $0.078^{+0.1}_{-0.1}$ &  7.49 \\
DES13X3gms & $19.07^{+3.5}_{-3.8}$ &  $0.8^{+0.1}_{-0.0}$ &  $11.12^{+2.0}_{-2.9}$ &  $36.94^{+30.5}_{-18.5}$ &  $83.97^{+11.6}_{-20.0}$ &  $0.37^{+0.1}_{-0.1}$ &  $0.35^{+0.1}_{-0.1}$ &  $5.48^{+2.8}_{-3.1}$ &  $-8.77^{+0.9}_{-0.6}$ &  $0.055^{+0.1}_{-0.0}$ &  3.03 \\
DES13X3npb & $17.78^{+8.2}_{-8.0}$ &  $0.8^{+0.1}_{-0.1}$ &  $6.79^{+1.9}_{-2.3}$ &  $22.59^{+36.2}_{-15.4}$ &  $80.04^{+14.4}_{-24.0}$ &  $0.36^{+0.1}_{-0.1}$ &  $0.33^{+0.1}_{-0.1}$ &  $5.26^{+2.9}_{-2.9}$ &  $-2.23^{+0.5}_{-0.6}$ &  $0.24^{+0.1}_{-0.1}$ &  15.43 \\
DES13X3nyg & $7.58^{+5.3}_{-3.3}$ &  $1.3^{+0.3}_{-0.2}$ &  $8.43^{+3.3}_{-2.8}$ &  $61.79^{+27.8}_{-41.6}$ &  $77.75^{+15.9}_{-24.7}$ &  $0.34^{+0.1}_{-0.1}$ &  $0.42^{+0.1}_{-0.1}$ &  $5.55^{+3.1}_{-3.1}$ &  $-8.59^{+1.9}_{-1.5}$ &  $0.058^{+0.1}_{-0.0}$ &  30.35 \\
{DES15C3lpq} & $25.85^{+2.9}_{-4.8}$ &  $0.99^{+0.0}_{-0.0}$ &  $7.14^{+2.1}_{-2.3}$ &  $32.32^{+31.4}_{-20.0}$ &  $68.14^{+19.3}_{-20.8}$ &  $0.35^{+0.1}_{-0.1}$ &  $0.24^{+0.1}_{-0.0}$ &  $3.64^{+1.7}_{-1.8}$ &  $-7.1^{+0.2}_{-0.1}$ &  $0.008^{+0.0}_{-0.0}$ &  5.38 \\
{DES15C3lzm} & $20.24^{+6.5}_{-7.9}$ &  $2.36^{+0.4}_{-0.3}$ &  $1.85^{+2.0}_{-1.0}$ &  $64.68^{+24.7}_{-31.8}$ &  $32.96^{+28.0}_{-18.9}$ &  $0.031^{+0.1}_{-0.0}$ &  $0.15^{+0.0}_{-0.0}$ &  $9.91^{+0.1}_{-0.1}$ &  $-1.93^{+0.4}_{-0.5}$ &  $0.023^{+0.0}_{-0.0}$ &  122.73 \\
{DES15C3nat} & $24.42^{+3.9}_{-5.7}$ &  $0.72^{+0.3}_{-0.2}$ &  $6.07^{+3.2}_{-2.6}$ &  $48.62^{+31.9}_{-27.8}$ &  $61.65^{+24.7}_{-25.1}$ &  $0.36^{+0.1}_{-0.1}$ &  $0.37^{+0.1}_{-0.1}$ &  $5.51^{+3.0}_{-3.1}$ &  $-7.9^{+2.8}_{-3.5}$ &  $0.073^{+0.1}_{-0.1}$ &  34.06 \\
DES15C3opk & $4.13^{+2.7}_{-1.8}$ &  $1.75^{+0.1}_{-0.1}$ &  $11.75^{+2.1}_{-3.0}$ &  $75.23^{+17.6}_{-23.0}$ &  $81.98^{+12.9}_{-20.2}$ &  $0.11^{+0.1}_{-0.0}$ &  $0.42^{+0.1}_{-0.1}$ &  $9.93^{+0.1}_{-0.1}$ &  $-6.02^{+0.3}_{-0.1}$ &  $0.011^{+0.0}_{-0.0}$ &  12.61 \\

{DES15C3opp}  & $16.63^{+9.2}_{-9.8}$ &  $0.49^{+0.2}_{-0.1}$ &  $4.27^{+2.3}_{-2.0}$ &  $64.58^{+25.0}_{-32.4}$ &  $62.31^{+25.3}_{-27.8}$ &  $0.28^{+0.1}_{-0.1}$ &  $0.4^{+0.1}_{-0.1}$ &  $5.56^{+3.0}_{-3.1}$ &  $-8.81^{+2.9}_{-2.4}$ &  $0.1^{+0.1}_{-0.1}$ &  --\\ 
DES15E2nqh & $21.14^{+6.0}_{-6.7}$ &  $1.25^{+0.3}_{-0.2}$ &  $6.14^{+2.3}_{-2.4}$ &  $31.78^{+38.6}_{-22.4}$ &  $71.77^{+19.8}_{-26.8}$ &  $0.32^{+0.1}_{-0.1}$ &  $0.24^{+0.1}_{-0.1}$ &  $5.68^{+2.9}_{-3.2}$ &  $-7.21^{+1.4}_{-1.0}$ &  $0.071^{+0.1}_{-0.1}$ &  3.16 \\
DES15S1fll & $12.42^{+4.6}_{-3.1}$ &  $0.73^{+0.1}_{-0.1}$ &  $6.62^{+2.3}_{-2.4}$ &  $60.74^{+26.5}_{-26.8}$ &  $76.2^{+17.0}_{-24.5}$ &  $0.24^{+0.2}_{-0.1}$ &  $0.37^{+0.1}_{-0.1}$ &  $6.32^{+2.6}_{-3.5}$ &  $-6.81^{+1.5}_{-1.9}$ &  $0.089^{+0.1}_{-0.1}$ &  5.15 \\
{DES15X3mxf}  & $3.79^{+1.5}_{-1.7}$ &  $1.55^{+0.1}_{-0.1}$ &  $13.24^{+0.4}_{-0.7}$ &  $72.96^{+19.0}_{-24.5}$ &  $96.97^{+2.2}_{-4.6}$ &  $0.091^{+0.0}_{-0.0}$ &  $0.4^{+0.1}_{-0.1}$ &  $9.96^{+0.0}_{-0.1}$ &  $-2.99^{+0.4}_{-0.3}$ &  $0.004^{+0.0}_{-0.0}$ &  59.78 \\
DES16C1cbd & $21.56^{+5.8}_{-7.1}$ &  $1.36^{+0.2}_{-0.2}$ &  $7.19^{+3.6}_{-3.0}$ &  $50.1^{+33.2}_{-30.6}$ &  $58.84^{+26.6}_{-27.5}$ &  $0.25^{+0.2}_{-0.1}$ &  $0.26^{+0.1}_{-0.1}$ &  $5.63^{+3.0}_{-3.2}$ &  $-4.02^{+0.8}_{-1.1}$ &  $0.37^{+0.1}_{-0.2}$ &  4.4 \\
DES16C3gin & $16.06^{+5.1}_{-5.7}$ &  $1.03^{+0.0}_{-0.0}$ &  $7.02^{+3.3}_{-3.1}$ &  $59.68^{+27.6}_{-30.0}$ &  $65.62^{+24.4}_{-27.9}$ &  $0.15^{+0.2}_{-0.1}$ &  $0.27^{+0.1}_{-0.0}$ &  $4.44^{+1.6}_{-2.3}$ &  $-4.8^{+0.2}_{-0.3}$ &  $0.015^{+0.0}_{-0.0}$ &  6.15 \\
DES16E2pv & $16.54^{+9.1}_{-9.0}$ &  $1.05^{+0.2}_{-0.2}$ &  $8.85^{+3.2}_{-3.2}$ &  $50.8^{+32.7}_{-29.5}$ &  $74.2^{+18.3}_{-26.4}$ &  $0.33^{+0.1}_{-0.1}$ &  $0.38^{+0.1}_{-0.1}$ &  $5.5^{+3.1}_{-3.1}$ &  $-5.82^{+1.4}_{-1.5}$ &  $0.16^{+0.1}_{-0.1}$ &   12.58\\
DES16X3cxn & $11.52^{+3.2}_{-3.0}$ &  $1.14^{+0.1}_{-0.1}$ &  $24.81^{+3.4}_{-4.6}$ &  $89.81^{+7.4}_{-12.7}$ &  $79.21^{+12.0}_{-16.5}$ &  $0.056^{+0.0}_{-0.0}$ &  $0.48^{+0.0}_{-0.0}$ &  $6.87^{+0.5}_{-0.4}$ &  $-11.762^{+1.0}_{-0.6}$ &  $0.02^{+0.0}_{-0.0}$ &  3.71 \\
{DES16X3ega} & $24.69^{+3.5}_{-5.6}$ &  $1.5^{+0.0}_{-0.0}$ &  $25.94^{+2.7}_{-3.1}$ &  $93.39^{+4.8}_{-8.3}$ &  $68.17^{+8.9}_{-10.4}$ &  $0.013^{+0.0}_{-0.0}$ &  $0.49^{+0.0}_{-0.0}$ &  $9.94^{+0.0}_{-0.1}$ &  $-1.63^{+0.1}_{-0.1}$ &  $0.01^{+0.0}_{-0.0}$ &  19.72 \\
HSC17auls & $3.84^{+4.0}_{-2.1}$ &  $1.17^{+0.5}_{-0.3}$ &  $5.07^{+2.9}_{-2.6}$ &  $73.71^{+18.6}_{-28.2}$ &  $59.68^{+27.9}_{-32.1}$ &  $0.15^{+0.2}_{-0.1}$ &  $0.41^{+0.1}_{-0.1}$ &  $8.54^{+1.0}_{-1.6}$ &  $-4.01^{+1.2}_{-1.6}$ &  $0.21^{+0.2}_{-0.1}$ &  --\\  

HSC17bbaz & $13.15^{+6.1}_{-7.7}$ &  $3.84^{+0.1}_{-0.2}$ &  $21.61^{+5.0}_{-5.1}$ &  $10.35^{+12.3}_{-6.4}$ &  $8.27^{+8.2}_{-5.7}$ &  $0.36^{+0.1}_{-0.1}$ &  $0.19^{+0.1}_{-0.0}$ &  $4.53^{+2.4}_{-2.4}$ &  $-7.23^{+1.1}_{-1.2}$ &  $0.34^{+0.1}_{-0.2}$ &  -- \\  
HSC17bhyl & $20.12^{+6.4}_{-6.7}$ &  $1.73^{+0.2}_{-0.2}$ &  $3.66^{+2.4}_{-1.8}$ &  $55.09^{+30.4}_{-32.0}$ &  $56.87^{+27.9}_{-26.1}$ &  $0.19^{+0.2}_{-0.1}$ &  $0.16^{+0.1}_{-0.0}$ &  $3.9^{+3.3}_{-2.0}$ &  $-3.69^{+0.4}_{-0.4}$ &  $0.048^{+0.1}_{-0.0}$ &  53.02 \\
HSC17btum & $5.31^{+4.8}_{-2.4}$ &  $0.73^{+0.1}_{-0.1}$ &  $5.05^{+3.4}_{-2.1}$ &  $83.27^{+12.2}_{-20.8}$ &  $42.96^{+31.4}_{-22.2}$ &  $0.2^{+0.2}_{-0.1}$ &  $0.46^{+0.0}_{-0.1}$ &  $9.55^{+0.3}_{-0.7}$ &  $-4.03^{+0.5}_{-0.6}$ &  $0.037^{+0.1}_{-0.0}$ &  -- \\ 
HSC17dadp & $21.52^{+5.9}_{-7.7}$ &  $2.23^{+0.6}_{-0.3}$ &  $4.42^{+3.2}_{-2.1}$ &  $56.69^{+29.8}_{-33.0}$ &  $43.19^{+32.4}_{-24.2}$ &  $0.12^{+0.2}_{-0.1}$ &  $0.16^{+0.1}_{-0.0}$ &  $5.09^{+2.3}_{-2.8}$ &  $-1.69^{+0.2}_{-0.1}$ &  $0.22^{+0.2}_{-0.1}$ &  -- \\  
{PS1-10bjp} & $3.36^{+1.2}_{-0.9}$ &  $2.81^{+0.1}_{-0.1}$ &  $10.9^{+0.2}_{-0.2}$ &  $98.5^{+1.1}_{-2.3}$ &  $8.9^{+1.0}_{-0.9}$ &  $0.01^{+0.0}_{-0.0}$ &  $0.44^{+0.0}_{-0.0}$ &  $9.99^{+0.0}_{-0.0}$ &  $-1.31^{+0.1}_{-0.1}$ &  $0.0015^{+0.0}_{-0.0}$ &  39.1 \\
PS1-11bbq & $7.47^{+13.5}_{-5.1}$ &  $3.08^{+0.5}_{-0.5}$ &  $6.18^{+3.1}_{-2.8}$ &  $52.44^{+31.9}_{-32.9}$ &  $64.95^{+24.7}_{-33.2}$ &  $0.22^{+0.2}_{-0.1}$ &  $0.25^{+0.1}_{-0.1}$ &  $5.51^{+3.1}_{-3.0}$ &  $-2.53^{+0.4}_{-0.4}$ &  $0.29^{+0.1}_{-0.2}$ &  -- \\ 
PS1-11qr & $5.27^{+3.2}_{-2.4}$ &  $2.14^{+0.2}_{-0.1}$ &  $16.34^{+2.0}_{-3.1}$ &  $74.43^{+18.1}_{-24.0}$ &  $85.95^{+10.2}_{-17.4}$ &  $0.046^{+0.0}_{-0.0}$ &  $0.41^{+0.1}_{-0.1}$ &  $9.79^{+0.2}_{-0.3}$ &  $-4.31^{+0.5}_{-0.3}$ &  $0.018^{+0.0}_{-0.0}$ &  6.58 \\
PS1-12bv & $4.53^{+1.3}_{-1.2}$ &  $1.01^{+0.1}_{-0.1}$ &  $11.18^{+1.4}_{-2.1}$ &  $71.5^{+18.3}_{-19.0}$ &  $87.51^{+8.9}_{-14.9}$ &  $0.4^{+0.1}_{-0.1}$ &  $0.45^{+0.0}_{-0.0}$ &  $6.33^{+1.5}_{-3.3}$ &  $-7.87^{+0.7}_{-0.6}$ &  $0.05^{+0.1}_{-0.0}$ &  6.18 \\
{PS1-13duy} & $8.91^{+6.9}_{-4.9}$ &  $1.99^{+0.3}_{-0.3}$ &  $5.66^{+2.4}_{-2.4}$ &  $65.54^{+24.4}_{-31.7}$ &  $69.4^{+21.6}_{-30.1}$ &  $0.19^{+0.2}_{-0.1}$ &  $0.37^{+0.1}_{-0.1}$ &  $5.42^{+3.1}_{-3.0}$ &  $-1.86^{+0.4}_{-0.4}$ &  $0.17^{+0.1}_{-0.1}$ &  -- \\ 
SNLS04D4ec & $9.86^{+4.5}_{-3.5}$ &  $1.58^{+0.0}_{-0.0}$ &  $23.53^{+1.7}_{-2.9}$ &  $82.36^{+12.6}_{-19.0}$ &  $90.7^{+6.8}_{-12.1}$ &  $0.044^{+0.0}_{-0.0}$ &  $0.45^{+0.0}_{-0.0}$ &  $9.52^{+0.2}_{-0.2}$ &  $-10.019^{+0.3}_{-0.1}$ &  $0.021^{+0.0}_{-0.0}$ &  6.17 \\

{SNLS06D1hc}  & $13.12^{+0.8}_{-1.0}$ &  $1.26^{+0.0}_{-0.0}$ &  $29.85^{+0.1}_{-0.2}$ &  $95.1^{+3.5}_{-5.3}$ &  $78.1^{+1.2}_{-1.3}$ &  $0.056^{+0.0}_{-0.0}$ &  $0.49^{+0.0}_{-0.0}$ &  $7.3^{+0.1}_{-0.1}$ &  $-1.31^{+0.0}_{-0.0}$ &  $0.0012^{+0.0}_{-0.0}$ &  25.29 \\
\noalign{\smallskip}\hline\noalign{\smallskip}
\end{tabular}
\smallskip
\end{table*}

\clearpage

\begin{table*}
\footnotesize
\setlength{\tabcolsep}{3.7pt}
\renewcommand{\arraystretch}{1.1}

\centering\caption{\label{table:LC_Param_CSI-wind}Best-fit parameters of the luminous REOT light curves with the CSI-wind model. \footnote{The units of $M_{\rm ej}$, $v$, $M_{\rm CSM}$, $\rho_{\rm CSM}$, $R_{\rm in}$, $T_{\rm f}$, and $t_{\rm shift}$ are respectively $M_{\odot}$, $10^9$\,cm\,s$^{-1}$, $M_{\odot}$, $10^{-12}$\,g\,cm$^{-3}$, $10^{14}$\,cm, $10^{3}$\,K, and days.}}
\begin{tabular}{cccccccccccccc}

\hline\hline\colhead{Name}&\colhead{$M_{\rm ej}$}&\colhead{$v$}&\colhead{$M_{\rm CSM}$}&\colhead{$\rho_{\rm CSM}$}&\colhead{$R_{\rm in}$}&\colhead{$\epsilon$}&\colhead{$x_0$}&\colhead{$T_{\rm f}$} &  \colhead{$t_{\rm shift}$} & \colhead{$A_{\rm host}$}  & \colhead{$\chi^{\rm 2}$/dof}\\
\hline

{DES13C3uig} & $21.6^{+5.9}_{-8.3}$ &  $1.95^{+0.6}_{-0.4}$ &  $0.49^{+1.8}_{-0.3}$ &  $0.71^{+24.1}_{-0.6}$ &  $9.32^{+33.7}_{-8.3}$ &  $0.37^{+0.1}_{-0.1}$ &  $0.4^{+0.1}_{-0.1}$ &  $5.73^{+2.2}_{-3.0}$ &  $-3.21^{+0.9}_{-1.2}$ &  $0.13^{+0.2}_{-0.1}$ &  43.56 \\
{DES13E2lpk} & $20.04^{+7.0}_{-9.5}$ &  $1.8^{+1.1}_{-0.6}$ &  $10.84^{+10.5}_{-7.8}$ &  $8.32^{+37.8}_{-7.5}$ &  $25.2^{+42.8}_{-21.5}$ &  $0.38^{+0.1}_{-0.1}$ &  $0.41^{+0.1}_{-0.1}$ &  $7.94^{+1.6}_{-4.1}$ &  $-8.22^{+2.9}_{-4.0}$ &  $0.092^{+0.1}_{-0.1}$ &  -- \\  
{DES13X1hav} & $17.34^{+8.9}_{-10.5}$ &  $3.71^{+0.2}_{-0.5}$ &  $20.54^{+6.4}_{-6.9}$ &  $32.62^{+40.2}_{-25.8}$ &  $32.8^{+23.2}_{-23.0}$ &  $0.33^{+0.1}_{-0.2}$ &  $0.35^{+0.1}_{-0.2}$ &  $9.85^{+0.1}_{-0.2}$ &  $-3.24^{+0.3}_{-0.4}$ &  $0.038^{+0.1}_{-0.0}$ &  4.62 \\
DES13X3gms & $28.14^{+1.4}_{-2.5}$ &  $0.85^{+0.3}_{-0.1}$ &  $9.69^{+8.0}_{-6.3}$ &  $0.11^{+0.4}_{-0.0}$ &  $53.07^{+29.6}_{-33.9}$ &  $0.48^{+0.0}_{-0.0}$ &  $0.49^{+0.0}_{-0.0}$ &  $7.99^{+0.8}_{-2.6}$ &  $-7.92^{+3.6}_{-1.5}$ &  $0.02^{+0.0}_{-0.0}$ &  12.67 \\
DES13X3npb & $18.84^{+7.7}_{-9.0}$ &  $0.92^{+0.1}_{-0.1}$ &  $8.61^{+5.8}_{-4.6}$ &  $0.11^{+0.2}_{-0.1}$ &  $51.48^{+32.6}_{-31.8}$ &  $0.36^{+0.1}_{-0.1}$ &  $0.38^{+0.1}_{-0.1}$ &  $5.06^{+2.7}_{-2.7}$ &  $-2.49^{+0.5}_{-0.6}$ &  $0.17^{+0.2}_{-0.1}$ &  25.04 \\
DES13X3nyg & $22.74^{+5.2}_{-7.2}$ &  $1.64^{+0.3}_{-0.2}$ &  $4.15^{+5.3}_{-2.7}$ &  $0.11^{+1.8}_{-0.1}$ &  $34.23^{+42.6}_{-29.2}$ &  $0.42^{+0.1}_{-0.1}$ &  $0.45^{+0.0}_{-0.1}$ &  $5.78^{+3.2}_{-3.2}$ &  $-6.5^{+1.3}_{-1.6}$ &  $0.041^{+0.1}_{-0.0}$ &  16.5 \\
DES15C3lpq & $28.3^{+1.2}_{-2.4}$ &  $1.2^{+0.1}_{-0.1}$ &  $3.28^{+1.5}_{-1.4}$ &  $0.25^{+0.4}_{-0.1}$ &  $27.53^{+13.2}_{-13.5}$ &  $0.48^{+0.0}_{-0.0}$ &  $0.49^{+0.0}_{-0.0}$ &  $3.71^{+1.8}_{-1.8}$ &  $-3.89^{+0.3}_{-0.4}$ &  $0.019^{+0.0}_{-0.0}$ &  3.11 \\
{DES15C3lzm}  & $18.28^{+8.0}_{-8.5}$ &  $1.43^{+0.2}_{-0.2}$ &  $3.05^{+2.9}_{-2.0}$ &  $0.028^{+0.0}_{-0.0}$ &  $53.07^{+31.5}_{-30.4}$ &  $0.35^{+0.1}_{-0.1}$ &  $0.37^{+0.1}_{-0.1}$ &  $4.78^{+2.8}_{-2.6}$ &  $-2.56^{+0.5}_{-0.6}$ &  $0.43^{+0.0}_{-0.1}$ &  123.96 \\
{DES15C3nat} & $24.19^{+4.2}_{-6.6}$ &  $1.66^{+0.3}_{-0.2}$ &  $0.56^{+1.4}_{-0.3}$ &  $2.35^{+31.8}_{-2.1}$ &  $5.45^{+21.8}_{-4.4}$ &  $0.43^{+0.1}_{-0.1}$ &  $0.46^{+0.0}_{-0.1}$ &  $5.32^{+2.6}_{-2.9}$ &  $-2.47^{+0.7}_{-1.0}$ &  $0.056^{+0.1}_{-0.0}$ &  43.78 \\
DES15C3opk & $21.87^{+5.5}_{-5.5}$ &  $1.63^{+0.0}_{-0.1}$ &  $11.39^{+1.4}_{-2.1}$ &  $0.034^{+0.0}_{-0.0}$ &  $87.18^{+9.2}_{-14.5}$ &  $0.33^{+0.1}_{-0.1}$ &  $0.35^{+0.1}_{-0.1}$ &  $9.97^{+0.0}_{-0.0}$ &  $-6.09^{+0.1}_{-0.1}$ &  $0.0063^{+0.0}_{-0.0}$ &  9.94 \\
{DES15C3opp} & $23.26^{+4.7}_{-7.2}$ &  $0.79^{+0.2}_{-0.2}$ &  $2.92^{+4.3}_{-2.3}$ &  $0.16^{+0.9}_{-0.1}$ &  $38.01^{+39.3}_{-31.3}$ &  $0.43^{+0.1}_{-0.1}$ &  $0.45^{+0.0}_{-0.1}$ &  $6.15^{+2.7}_{-3.6}$ &  $-4.51^{+1.5}_{-2.2}$ &  $0.057^{+0.1}_{-0.0}$ &  -- \\  
DES15E2nqh & $24.83^{+3.7}_{-6.3}$ &  $1.61^{+0.3}_{-0.2}$ &  $1.41^{+3.1}_{-0.8}$ &  $1.2^{+27.5}_{-1.0}$ &  $9.13^{+29.9}_{-7.8}$ &  $0.44^{+0.0}_{-0.1}$ &  $0.46^{+0.0}_{-0.1}$ &  $8.8^{+0.5}_{-0.6}$ &  $-5.01^{+1.1}_{-1.2}$ &  $0.089^{+0.1}_{-0.1}$ &  5.23 \\
DES15S1fll & $22.88^{+5.0}_{-7.0}$ &  $0.85^{+0.1}_{-0.1}$ &  $2.24^{+4.2}_{-1.3}$ &  $0.43^{+7.4}_{-0.3}$ &  $17.15^{+37.5}_{-14.6}$ &  $0.41^{+0.1}_{-0.1}$ &  $0.44^{+0.0}_{-0.1}$ &  $7.85^{+1.5}_{-2.2}$ &  $-5.16^{+0.6}_{-0.6}$ &  $0.04^{+0.1}_{-0.0}$ &  5.44 \\
{DES15X3mxf} & $19.23^{+7.3}_{-6.6}$ &  $1.43^{+0.1}_{-0.1}$ &  $6.63^{+3.0}_{-3.1}$ &  $0.031^{+0.0}_{-0.0}$ &  $65.85^{+23.7}_{-29.7}$ &  $0.35^{+0.1}_{-0.1}$ &  $0.38^{+0.1}_{-0.1}$ &  $9.87^{+0.1}_{-0.2}$ &  $-2.4^{+0.3}_{-0.4}$ &  $0.013^{+0.0}_{-0.0}$ &  17.03 \\
{DES16C1cbd} & $22.36^{+5.5}_{-9.3}$ &  $1.62^{+0.2}_{-0.3}$ &  $4.24^{+12.6}_{-3.3}$ &  $0.49^{+32.8}_{-0.4}$ &  $22.09^{+51.9}_{-20.4}$ &  $0.39^{+0.1}_{-0.1}$ &  $0.42^{+0.1}_{-0.1}$ &  $7.72^{+1.5}_{-4.4}$ &  $-2.7^{+0.7}_{-1.8}$ &  $0.26^{+0.1}_{-0.2}$ &  28.38 \\
DES16C3gin & $24.67^{+3.8}_{-5.7}$ &  $1.18^{+0.0}_{-0.0}$ &  $5.16^{+2.7}_{-1.9}$ &  $0.13^{+0.1}_{-0.0}$ &  $45.31^{+21.1}_{-16.5}$ &  $0.39^{+0.1}_{-0.1}$ &  $0.42^{+0.1}_{-0.1}$ &  $3.78^{+1.9}_{-1.9}$ &  $-2.7^{+0.2}_{-0.2}$ &  $0.081^{+0.1}_{-0.0}$ &  6.77 \\
DES16E2pv & $22.64^{+5.2}_{-7.7}$ &  $1.38^{+0.1}_{-0.1}$ &  $1.66^{+6.5}_{-0.5}$ &  $2.53^{+36.7}_{-2.4}$ &  $4.98^{+50.8}_{-3.9}$ &  $0.43^{+0.1}_{-0.1}$ &  $0.46^{+0.0}_{-0.1}$ &  $5.5^{+3.1}_{-3.1}$ &  $-3.56^{+0.5}_{-0.5}$ &  $0.067^{+0.1}_{-0.0}$ &  15.87 \\
DES16X3cxn & $24.05^{+4.4}_{-6.6}$ &  $1.41^{+0.1}_{-0.1}$ &  $3.17^{+1.5}_{-1.4}$ &  $0.57^{+1.8}_{-0.3}$ &  $20.78^{+12.3}_{-12.3}$ &  $0.42^{+0.1}_{-0.1}$ &  $0.45^{+0.0}_{-0.0}$ &  $3.64^{+1.4}_{-1.8}$ &  $-6.78^{+0.8}_{-0.8}$ &  $0.054^{+0.1}_{-0.0}$ &  4.1 \\
{DES16X3ega} & $14.66^{+9.3}_{-9.7}$ &  $1.48^{+0.0}_{-0.0}$ &  $28.54^{+1.1}_{-2.0}$ &  $0.35^{+0.1}_{-0.1}$ &  $76.58^{+3.9}_{-6.5}$ &  $0.33^{+0.1}_{-0.1}$ &  $0.23^{+0.1}_{-0.1}$ &  $9.94^{+0.0}_{-0.1}$ &  $-1.65^{+0.1}_{-0.1}$ &  $0.0084^{+0.0}_{-0.0}$ &  30.6 \\
HSC17auls & $18.68^{+7.8}_{-9.4}$ &  $2.74^{+0.7}_{-0.7}$ &  $5.12^{+4.8}_{-2.5}$ &  $17.56^{+41.9}_{-14.5}$ &  $38.39^{+38.2}_{-24.4}$ &  $0.37^{+0.1}_{-0.2}$ &  $0.4^{+0.1}_{-0.1}$ &  $6.87^{+1.5}_{-1.1}$ &  $-1.78^{+0.4}_{-0.6}$ &  $0.15^{+0.2}_{-0.1}$ & --\\ 
HSC17bbaz & $14.21^{+9.4}_{-9.1}$ &  $3.92^{+0.1}_{-0.1}$ &  $16.53^{+2.2}_{-2.0}$ &  $39.31^{+39.1}_{-28.3}$ &  $2.21^{+2.0}_{-0.7}$ &  $0.25^{+0.2}_{-0.1}$ &  $0.27^{+0.1}_{-0.1}$ &  $4.53^{+2.4}_{-2.4}$ &  $-6.25^{+0.5}_{-0.5}$ &  $0.46^{+0.0}_{-0.1}$ & --\\ 
HSC17bhyl & $20.24^{+6.8}_{-8.5}$ &  $1.89^{+0.1}_{-0.1}$ &  $0.84^{+2.7}_{-0.5}$ &  $0.56^{+23.9}_{-0.5}$ &  $10.36^{+39.3}_{-9.3}$ &  $0.33^{+0.1}_{-0.1}$ &  $0.36^{+0.1}_{-0.1}$ &  $5.51^{+0.9}_{-0.6}$ &  $-2.49^{+0.2}_{-0.2}$ &  $0.069^{+0.1}_{-0.1}$ &  109.44 \\
HSC17btum & $17.91^{+8.4}_{-9.4}$ &  $0.97^{+0.1}_{-0.1}$ &  $2.57^{+0.9}_{-0.6}$ &  $8.55^{+20.0}_{-6.5}$ &  $8.9^{+8.4}_{-4.6}$ &  $0.43^{+0.0}_{-0.1}$ &  $0.45^{+0.0}_{-0.1}$ &  $9.83^{+0.1}_{-0.2}$ &  $-2.74^{+0.2}_{-0.5}$ &  $0.015^{+0.0}_{-0.0}$ &  --\\ 
HSC17dadp & $20.05^{+6.8}_{-9.5}$ &  $2.82^{+0.1}_{-0.1}$ &  $15.76^{+6.8}_{-6.4}$ &  $9.21^{+24.0}_{-6.4}$ &  $57.1^{+29.8}_{-32.2}$ &  $0.39^{+0.1}_{-0.1}$ &  $0.4^{+0.1}_{-0.1}$ &  $6.28^{+0.4}_{-0.4}$ &  $-0.98^{+0.1}_{-0.1}$ &  $0.053^{+0.1}_{-0.0}$ &  --\\ 
PS1-10bjp & $16.72^{+9.0}_{-9.7}$ &  $3.49^{+0.1}_{-0.2}$ &  $29.67^{+0.2}_{-0.5}$ &  $53.67^{+31.7}_{-32.3}$ &  $76.27^{+3.8}_{-2.7}$ &  $0.25^{+0.1}_{-0.1}$ &  $0.31^{+0.1}_{-0.1}$ &  $9.99^{+0.0}_{-0.0}$ &  $-0.37^{+0.0}_{-0.0}$ &  $0.0012^{+0.0}_{-0.0}$ &  38.6 \\
{PS1-11bbq} & $15.84^{+10.1}_{-10.9}$ &  $2.98^{+0.3}_{-0.3}$ &  $1.04^{+0.3}_{-0.2}$ &  $28.59^{+44.6}_{-24.6}$ &  $0.94^{+2.2}_{-0.4}$ &  $0.29^{+0.1}_{-0.1}$ &  $0.3^{+0.1}_{-0.1}$ &  $5.46^{+3.1}_{-3.1}$ &  $-2.96^{+0.3}_{-0.3}$ &  $0.24^{+0.2}_{-0.2}$ &  -- \\
{PS1-11qr} & $29.93^{+0.1}_{-0.1}$ &  $4.0^{+0.0}_{-0.0}$ &  $17.54^{+1.4}_{-0.4}$ &  $97.78^{+1.6}_{-3.4}$ &  $66.49^{+3.5}_{-1.1}$ &  $0.5^{+0.0}_{-0.0}$ &  $0.5^{+0.0}_{-0.0}$ &  $9.41^{+0.3}_{-0.4}$ &  $-2.08^{+0.1}_{-0.1}$ &  $0.00086^{+0.0}_{-0.0}$ &  23.27 \\
PS1-12bv & $24.64^{+3.9}_{-6.4}$ &  $1.26^{+0.1}_{-0.1}$ &  $1.88^{+3.0}_{-0.8}$ &  $0.16^{+0.8}_{-0.1}$ &  $14.38^{+31.5}_{-9.9}$ &  $0.45^{+0.0}_{-0.1}$ &  $0.47^{+0.0}_{-0.0}$ &  $5.53^{+2.4}_{-3.1}$ &  $-5.76^{+0.8}_{-1.0}$ &  $0.096^{+0.1}_{-0.1}$ &  6.9 \\
PS1-13duy & $18.04^{+8.7}_{-10.0}$ &  $2.48^{+0.2}_{-0.5}$ &  $6.01^{+4.2}_{-4.0}$ &  $0.32^{+0.9}_{-0.2}$ &  $56.35^{+30.0}_{-39.3}$ &  $0.39^{+0.1}_{-0.1}$ &  $0.41^{+0.1}_{-0.1}$ &  $5.66^{+2.9}_{-3.2}$ &  $-1.31^{+0.2}_{-0.5}$ &  $0.05^{+0.1}_{-0.0}$ & --\\
{SNLS04D4ec} & $29.65^{+0.3}_{-0.5}$ &  $3.99^{+0.0}_{-0.0}$ &  $28.6^{+1.0}_{-1.4}$ &  $89.75^{+6.5}_{-6.8}$ &  $85.83^{+4.3}_{-4.6}$ &  $0.5^{+0.0}_{-0.0}$ &  $0.5^{+0.0}_{-0.0}$ &  $9.97^{+0.0}_{-0.0}$ &  $-4.33^{+0.1}_{-0.1}$ &  $0.0022^{+0.0}_{-0.0}$ & 11.61 \\
SNLS06D1hc & $29.3^{+0.5}_{-1.0}$ &  $1.17^{+0.0}_{-0.0}$ &  $5.3^{+0.5}_{-0.6}$ &  $0.42^{+0.2}_{-0.1}$ &  $22.07^{+3.1}_{-3.9}$ &  $0.49^{+0.0}_{-0.0}$ &  $0.49^{+0.0}_{-0.0}$ &  $8.93^{+0.2}_{-0.2}$ &  $-0.97^{+0.0}_{-0.0}$ &  $0.17^{+0.0}_{-0.0}$ &  7.33 \\

\noalign{\smallskip}\hline\noalign{\smallskip}

\end{tabular}
\smallskip
\end{table*}

\clearpage

\begin{table*}[tbp]
\caption{Validity of our models. Also shown are the values of $\chi^2$/dof of the models which are obtained from Tables
\ref{table:LC_Param_MAG}, \ref{table:LC_Param_CSI-shell}, and \ref{table:LC_Param_CSI-wind}.}\label{tab:vali}

\begin{center}
\begin{tabular}{cccccccccccccccc}
\hline\hline
Name	   & Mag model ($\chi^2$/dof) & CSI-shell model ($\chi^2$/dof) & CSI-wind model ($\chi^2$/dof) & The best model	 \\
\hline
DES13C3uig     & $\surd$ (3.98)   	& $\surd$ (24.89)	  & X (43.56)           &  Mag                \\   	
DES13E2lpk     & $\surd$ (15.96)   & $\surd$ (-)	       & X (-)               &  Mag                \\  	
{DES13X1hav}  & X (1.58)    & X (7.49)	       & X (4.62)            &  {cannot be fitted} \\
DES13X3gms	& $\surd$ (1.33)    & $\surd$ (3.03)	  & $\surd$ (12.67)	    &  Mag                \\
DES13X3npb     & $\surd$ (7.61)	& $\surd$ (15.43)	  & $\surd$ (25.04)     &  Mag                \\ 	
DES13X3nyg     & $\surd$ (6.42)	& $\surd$ (30.35)	  & $\surd$ (16.5)      &  Mag                \\ 	  	
DES15C3lpq	& $\surd$ (2.42)    & X	(5.38)           & $\surd$ (3.11)	    &  Mag                \\
{DES15C3lzm} & X (49.21)    & X (122.73)	       & X (123.96)          &  {cannot be fitted}  \\
{DES15C3nat} & X (5.56)	& X (34.06)	       & X (43.78)           &  {cannot be fitted}  \\
DES15C3opk     & $\surd$ (8.24)    & $\surd$ (12.61)	  & $\surd$ (9.94)      &  Mag                \\ 	
DES15C3opp     & $\surd$ (-)	     & X (-)	            & X (-)	              &  Mag                \\
DES15E2nqh	& $\surd$ (1.64)    & $\surd$ (3.16)	  & $\surd$ (5.23)	    &  Mag                \\
DES15S1fll	& $\surd$ (2.28)    & $\surd$ (5.15)	  & $\surd$ (5.44)	    &  Mag                \\
DES15X3mxf	& $\surd$ (4.5)     & X (59.78)	       & X (17.03)           &  Mag                \\ 
DES16C1cbd	& $\surd$ (1.54)    & $\surd$ (4.4)	  & X (28.38)	         &  Mag                \\
DES16C3gin	& $\surd$ (3.14)    & $\surd$ (6.15)	  & $\surd$ (6.77)      &  Mag                \\
DES16E2pv      & $\surd$ (4.05)	& $\surd$ (12.58)	  & $\surd$ (15.87)     &  Mag                \\
DES16X3cxn	& $\surd$ (1.39)    & $\surd$ (3.71)	  & $\surd$ (4.1)	    &  Mag                \\
DES16X3ega	& $\surd$ (11.05)   & X (19.72)	       & X (30.6)	         &  Mag                \\
HSC17auls  	& $\surd$ (-)       & $\surd$ (-) 	       & $\surd$ (-)         &  -                  \\
HSC17bbaz	     & $\surd$ (24.15)   & $\surd$ (-) 	       & $\surd$ (-)         &  -                  \\
HSC17bhyl  	& $\surd$ (8.61)    & $\surd$ (53.02)     & $\surd$ (109.44)    &  Mag                \\
HSC17btum	     & $\surd$ (9.61)    & $\surd$ (-)         & $\surd$ (-)         &  -                  \\
HSC17dadp  	& $\surd$ (-)       & $\surd$ (-)         & $\surd$ (-)         &  -                  \\
PS1-10bjp      & $\surd$ (13.83)   & X (39.1)            & $\surd$ (38.6)      &  Mag                \\ 
PS1-11bbq      & $\surd$ (16.91)         & $\surd$ (-)         & $\surd$ (-)               &  Mag          \\ 
PS1-11qr       & $\surd$ (2.29)    & $\surd$ (6.58)      & X (23.27)           &  Mag          \\
PS1-12bv       & $\surd$ (3.03)    & $\surd$ (6.18)      & $\surd$ (6.9)       &  Mag                \\
PS1-13duy      & $\surd$ (3.48)    & X (-) 	            & $\surd$ (-)         &  Mag    \\
SNLS04D4ec	& $\surd$ (3.0)     & $\surd$ (5.58)      & X (11.61)           &  Mag          \\
SNLS06D1hc	& $\surd$ (5.99)    & X (25.29)           & $\surd$ (7.33)      &  Mag                \\
\hline\hline
\end{tabular}
\end{center}
\end{table*}

\clearpage

\clearpage

\clearpage

\begin{figure}
\centering
\includegraphics[width=0.32\textwidth,angle=0]{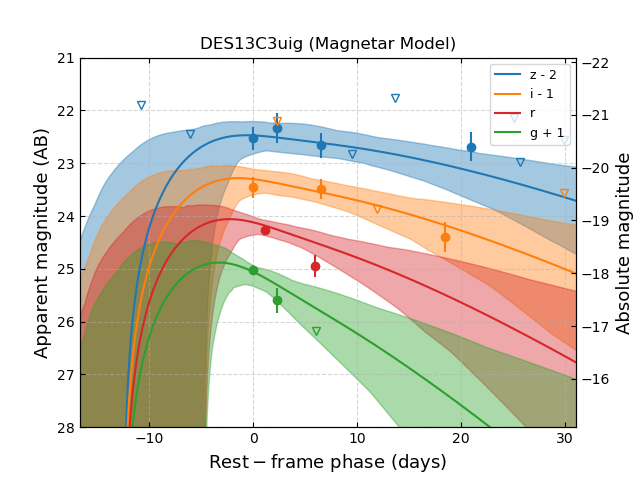}
\includegraphics[width=0.32\textwidth,angle=0]{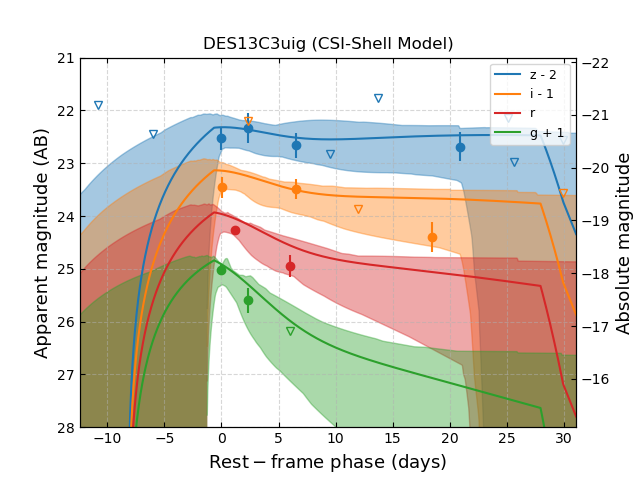}
\includegraphics[width=0.32\textwidth,angle=0]{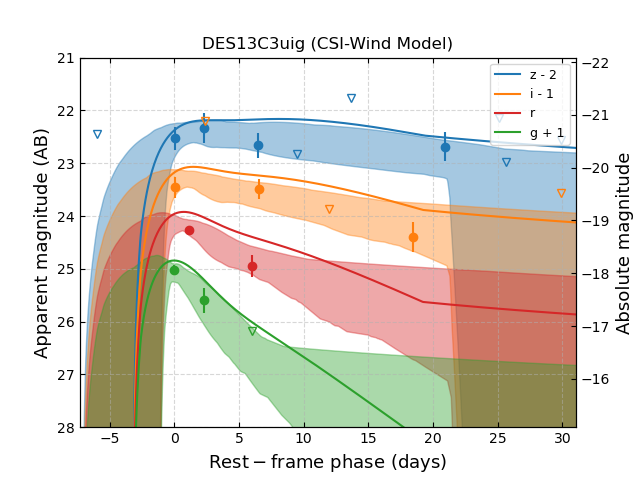}
\includegraphics[width=0.32\textwidth,angle=0]{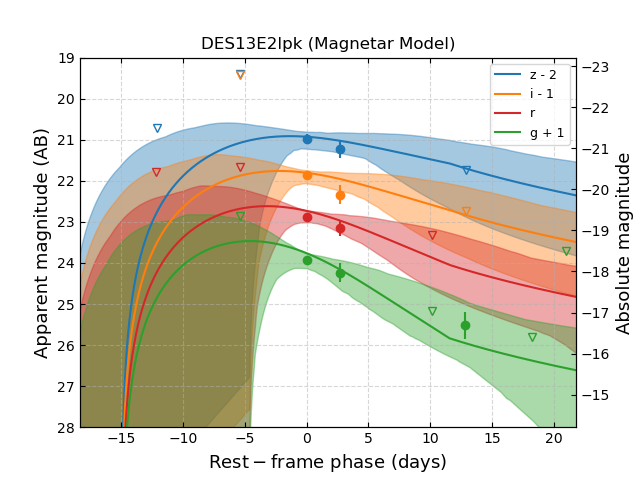}
\includegraphics[width=0.32\textwidth,angle=0]{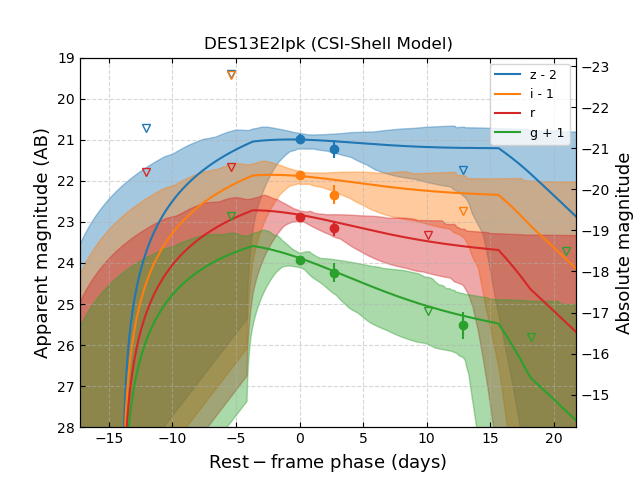}
\includegraphics[width=0.32\textwidth,angle=0]{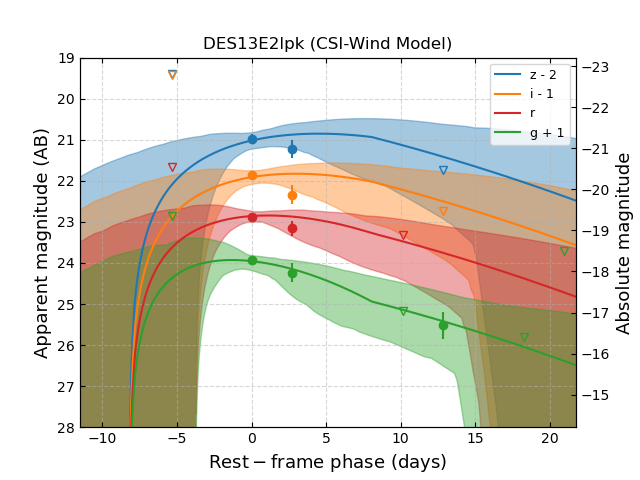}
\includegraphics[width=0.32\textwidth,angle=0]{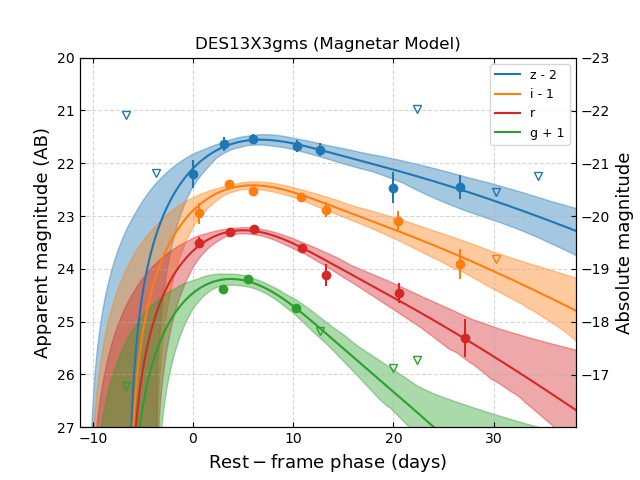}
\includegraphics[width=0.32\textwidth,angle=0]{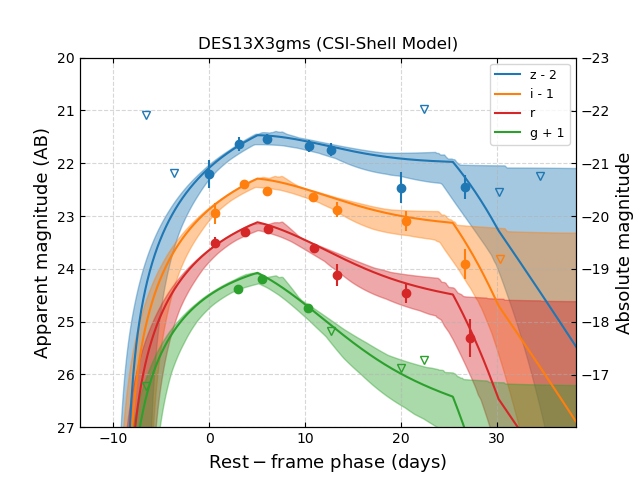}
\includegraphics[width=0.32\textwidth,angle=0]{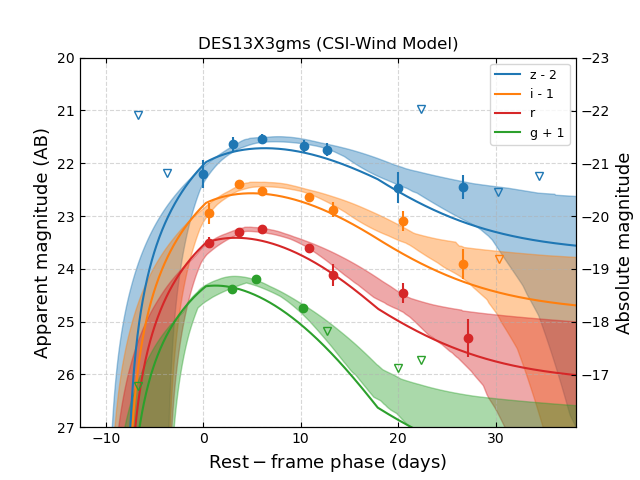}
\includegraphics[width=0.32\textwidth,angle=0]{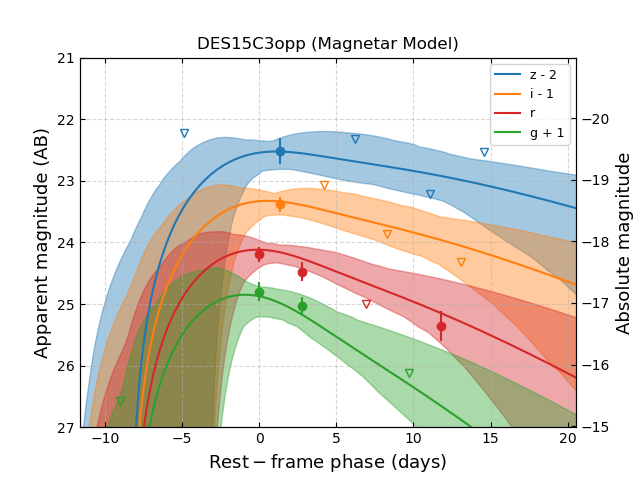}
\includegraphics[width=0.32\textwidth,angle=0]{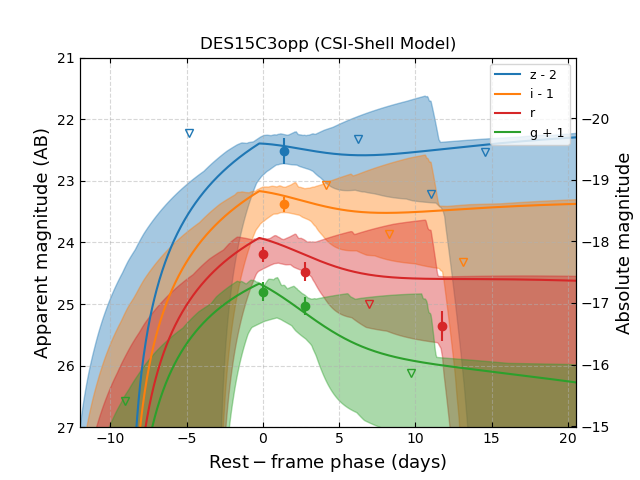}
\includegraphics[width=0.32\textwidth,angle=0]{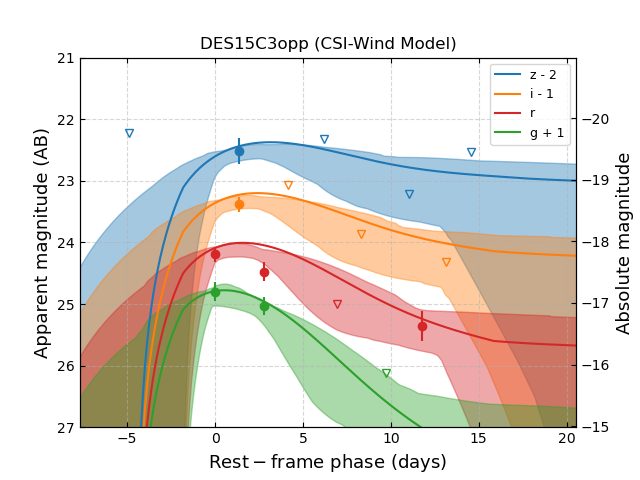}
\includegraphics[width=0.32\textwidth,angle=0]{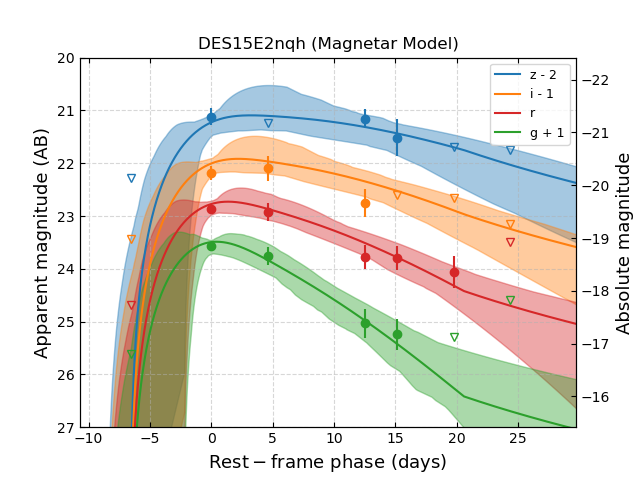}
\includegraphics[width=0.32\textwidth,angle=0]{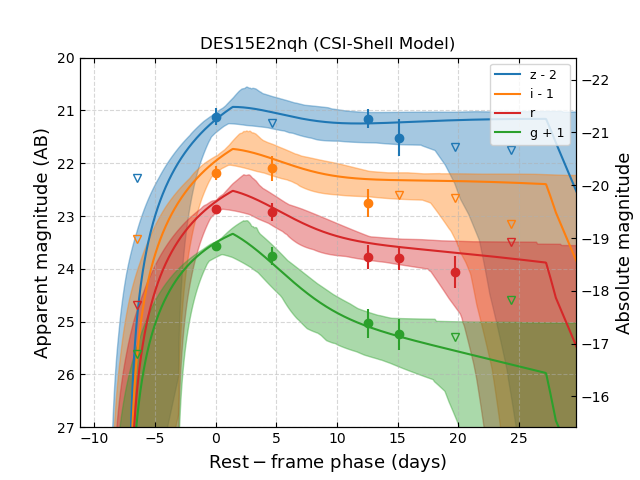}
\includegraphics[width=0.32\textwidth,angle=0]{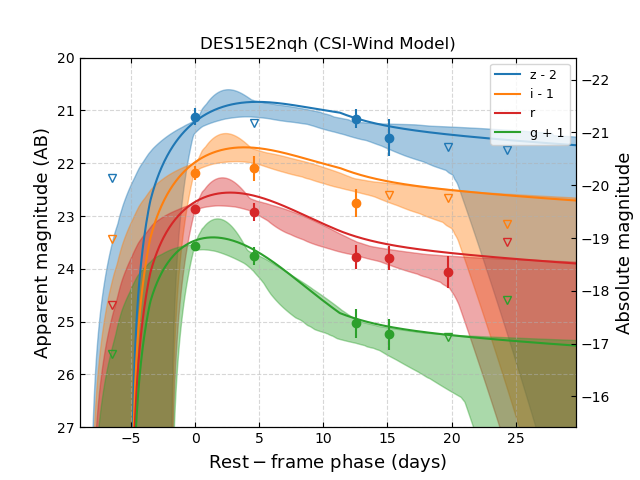}
\caption{{The fits for the multiband light curves of 20 REOTs using the magnetar model
(left panels) and the CSI model (middle and right panels, which respectively correspond to
the shell case and the wind case). The 20 REOTs can be fitted with excellent quality.
The solid lines and the shaded regions present the best-fit
light curves and $3\sigma$ uncertainties, respectively.}
The abscissa represents time since the explosion in the rest frame.}  
\label{fig:LCfits_excellent}
\end{figure}

\clearpage

\begin{figure}
\ContinuedFloat
\centering
\includegraphics[width=0.32\textwidth,angle=0]{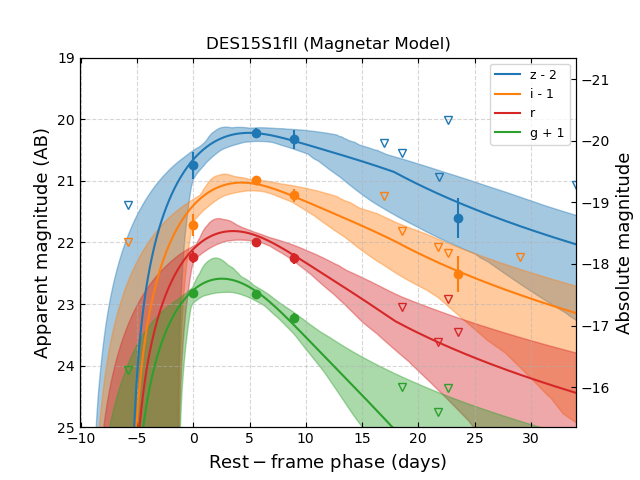}
\includegraphics[width=0.32\textwidth,angle=0]{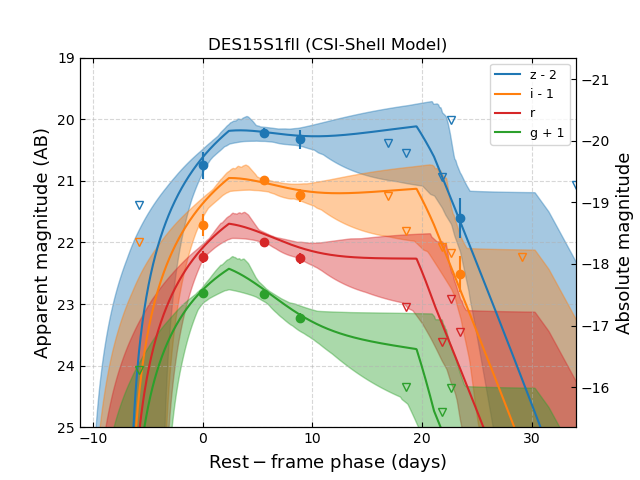}
\includegraphics[width=0.32\textwidth,angle=0]{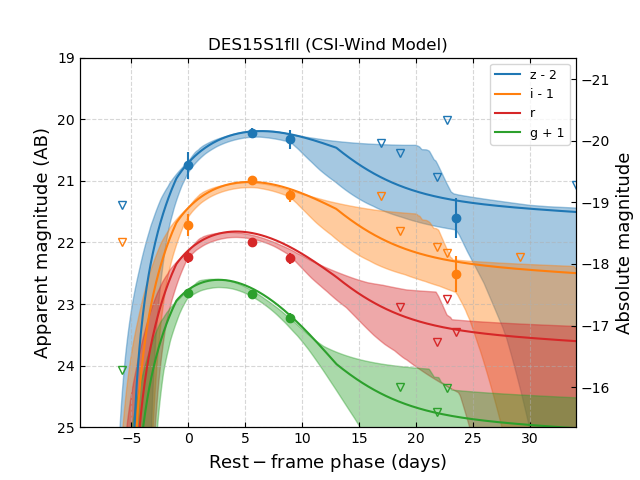}
\includegraphics[width=0.32\textwidth,angle=0]{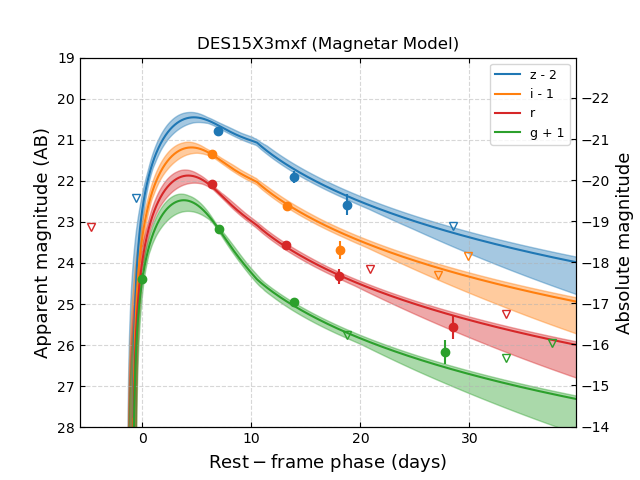}
\includegraphics[width=0.32\textwidth,angle=0]{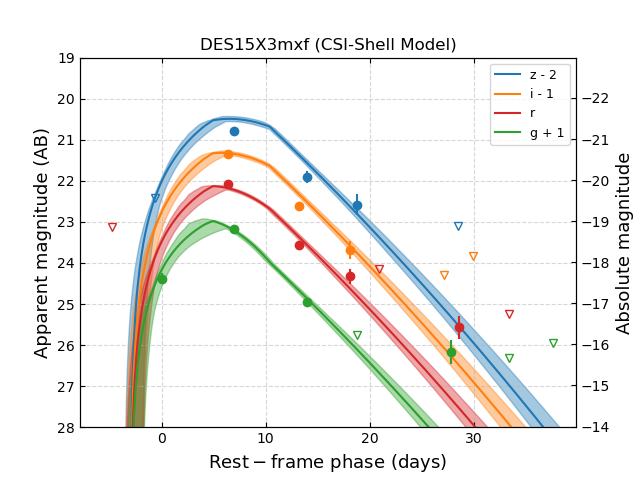}
\includegraphics[width=0.32\textwidth,angle=0]{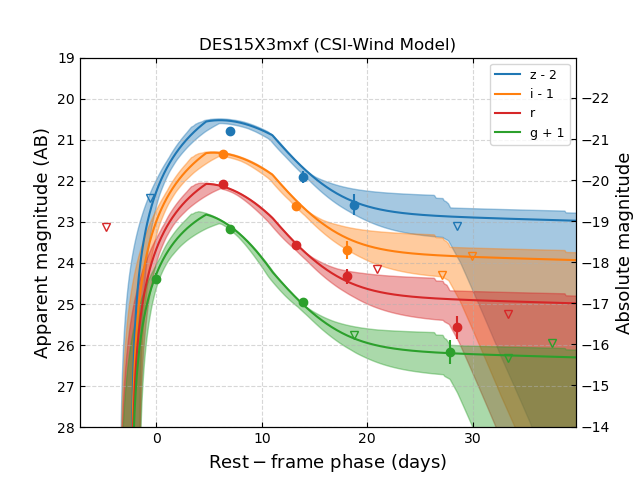}
\includegraphics[width=0.32\textwidth,angle=0]{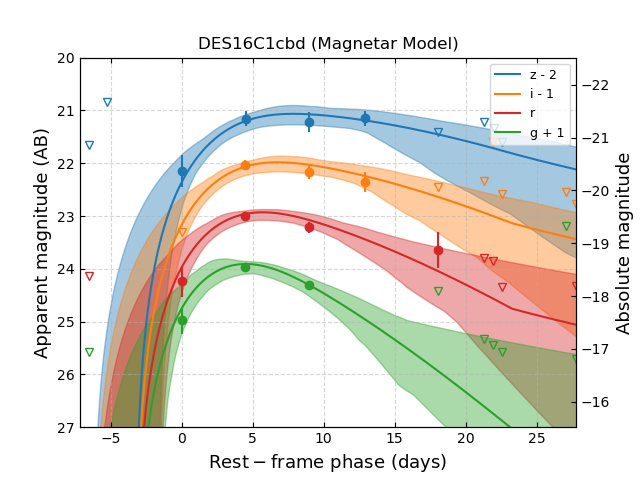}
\includegraphics[width=0.32\textwidth,angle=0]{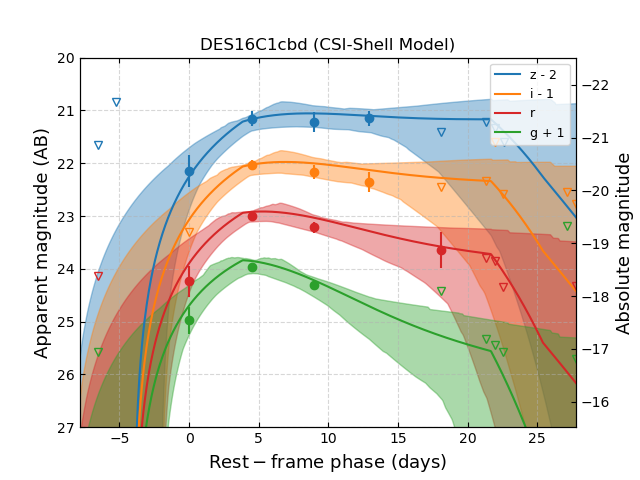}
\includegraphics[width=0.32\textwidth,angle=0]{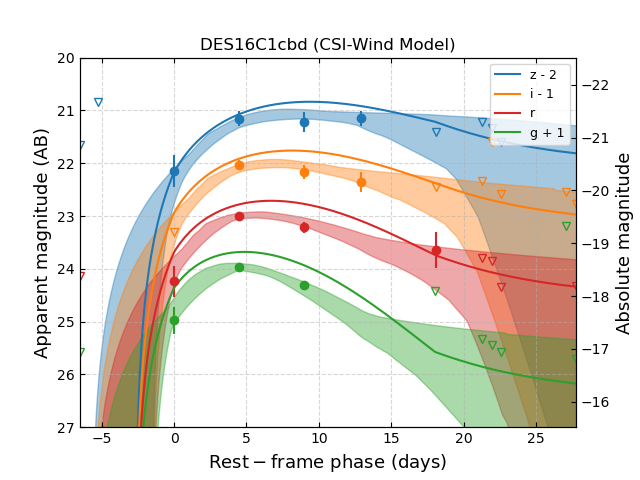}
\includegraphics[width=0.32\textwidth,angle=0]{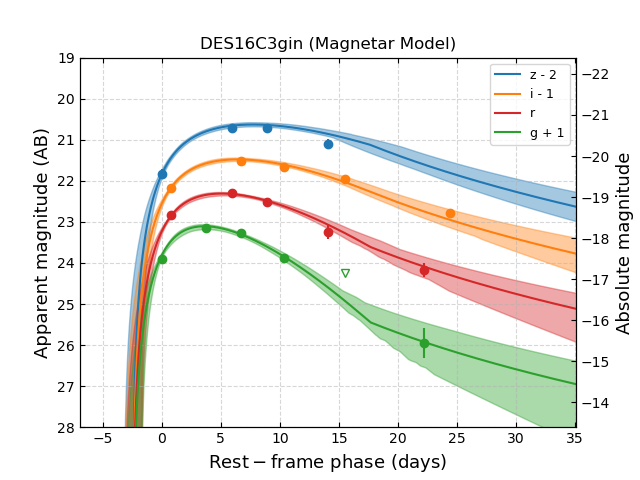}
\includegraphics[width=0.32\textwidth,angle=0]{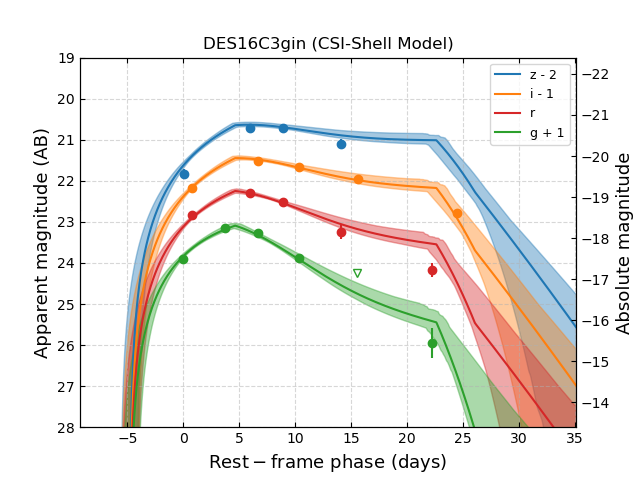}
\includegraphics[width=0.32\textwidth,angle=0]{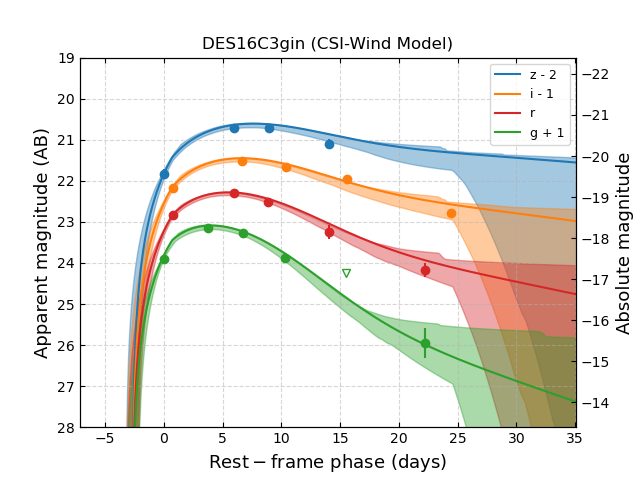}
\includegraphics[width=0.32\textwidth,angle=0]{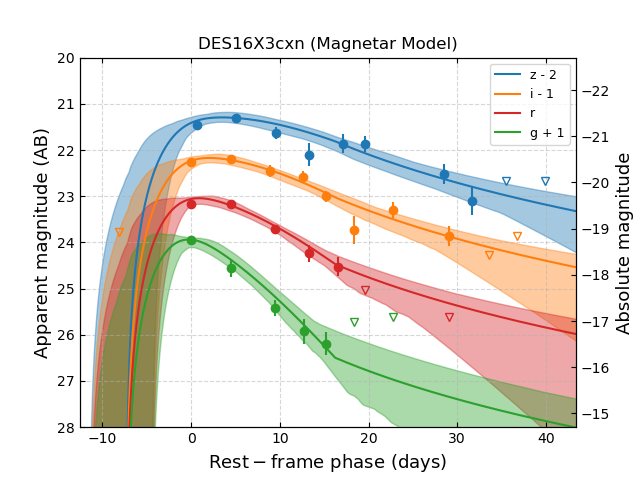}
\includegraphics[width=0.32\textwidth,angle=0]{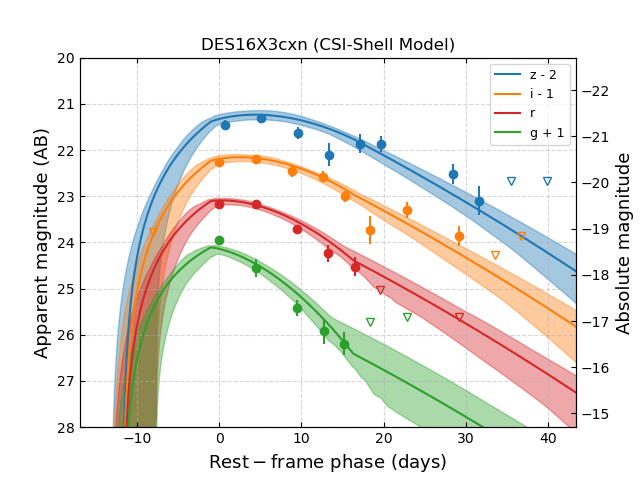}
\includegraphics[width=0.32\textwidth,angle=0]{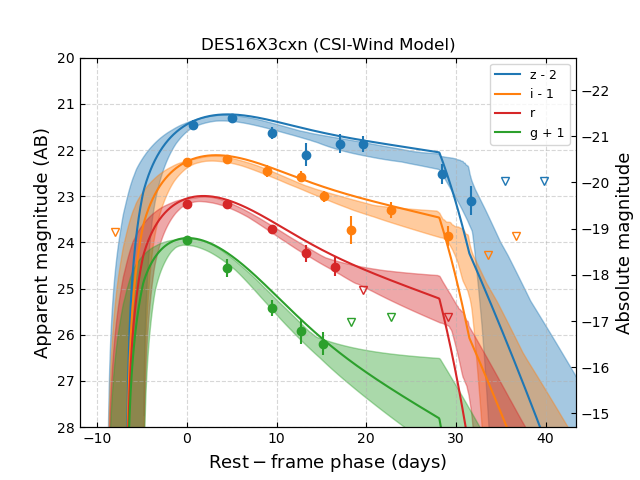}
\caption{(Continued).}
\end{figure}

\clearpage

\begin{figure}
\ContinuedFloat
\centering
\includegraphics[width=0.32\textwidth,angle=0]{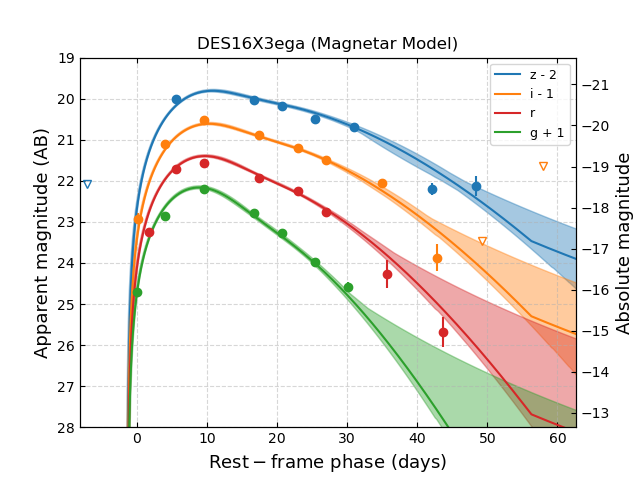}
\includegraphics[width=0.32\textwidth,angle=0]{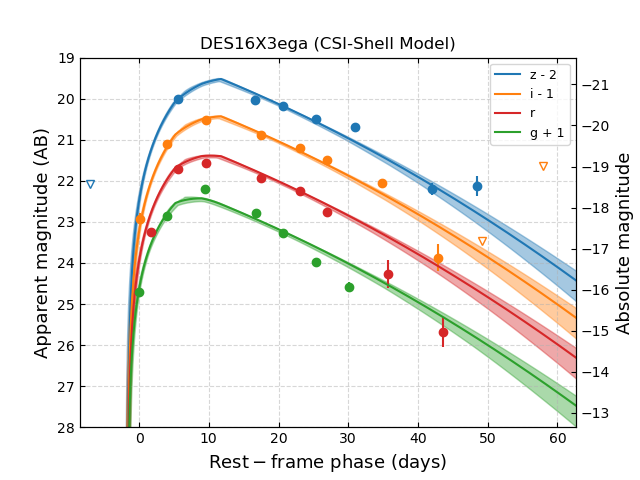}
\includegraphics[width=0.32\textwidth,angle=0]{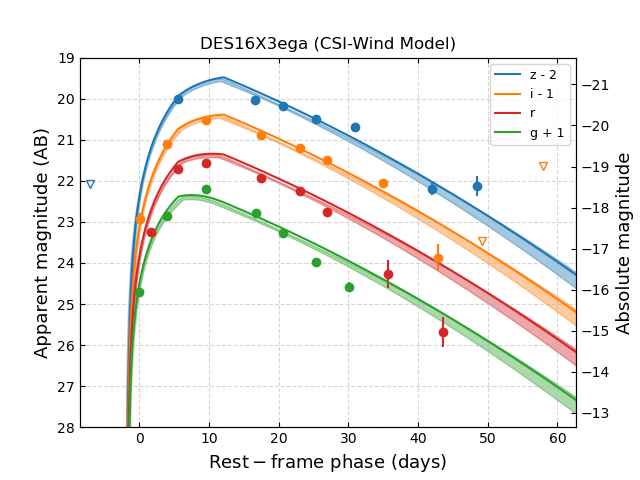}
\includegraphics[width=0.32\textwidth,angle=0]{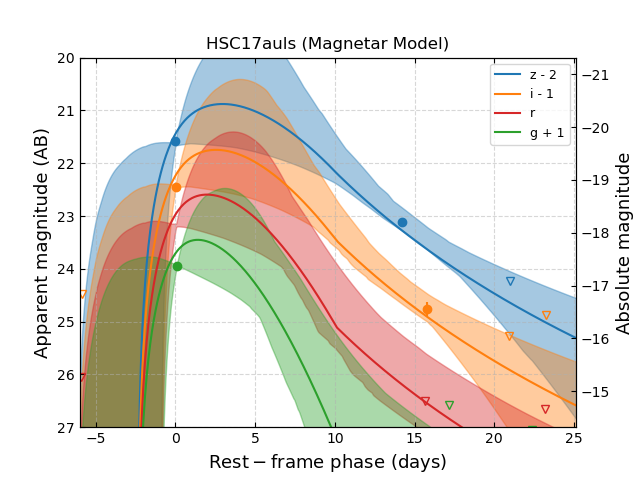}
\includegraphics[width=0.32\textwidth,angle=0]{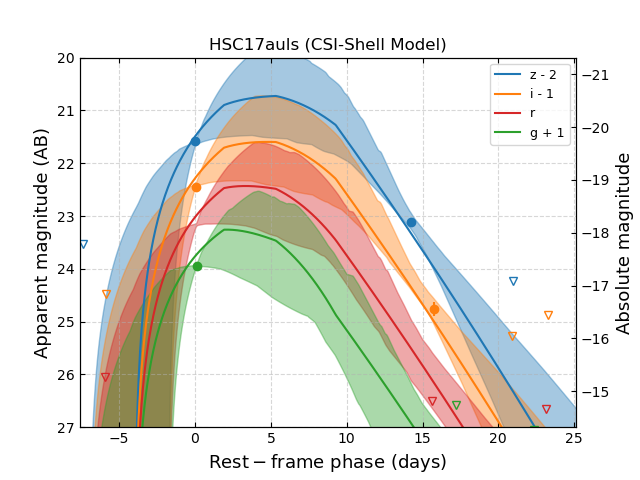}
\includegraphics[width=0.32\textwidth,angle=0]{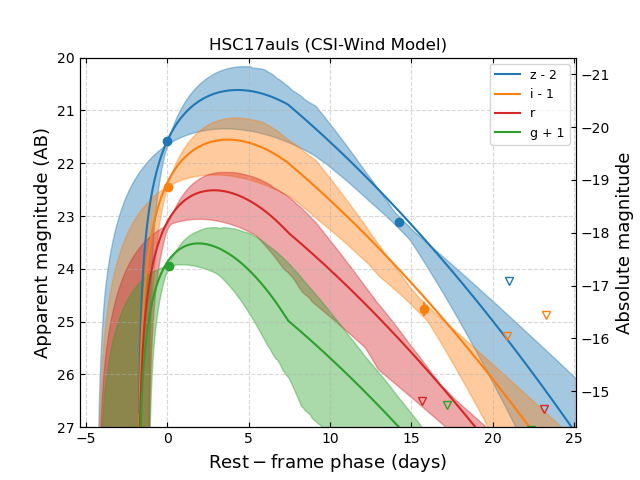}
\includegraphics[width=0.32\textwidth,angle=0]{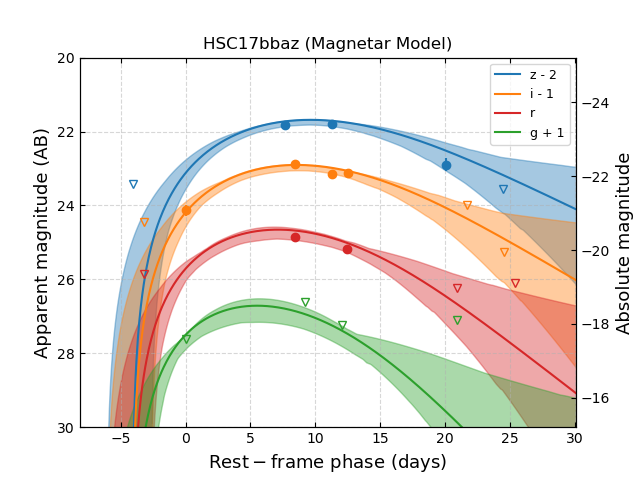}
\includegraphics[width=0.32\textwidth,angle=0]{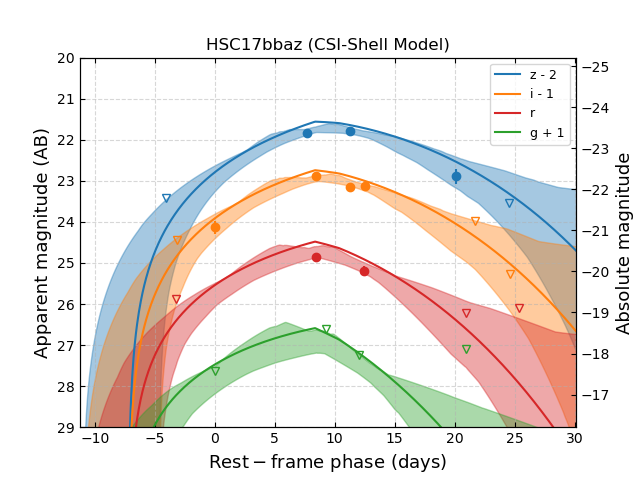}
\includegraphics[width=0.32\textwidth,angle=0]{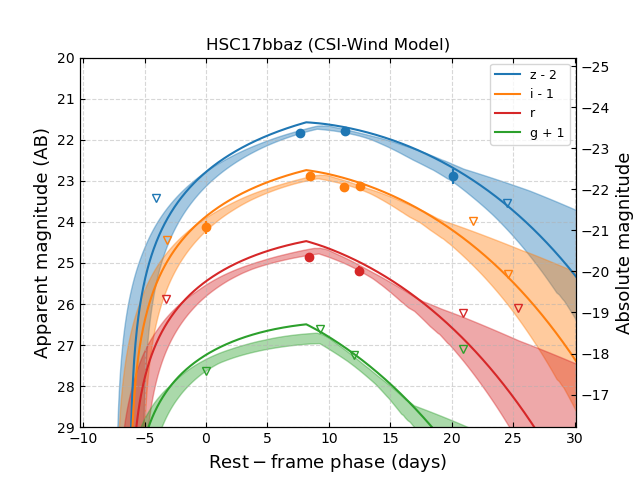}
\includegraphics[width=0.32\textwidth,angle=0]{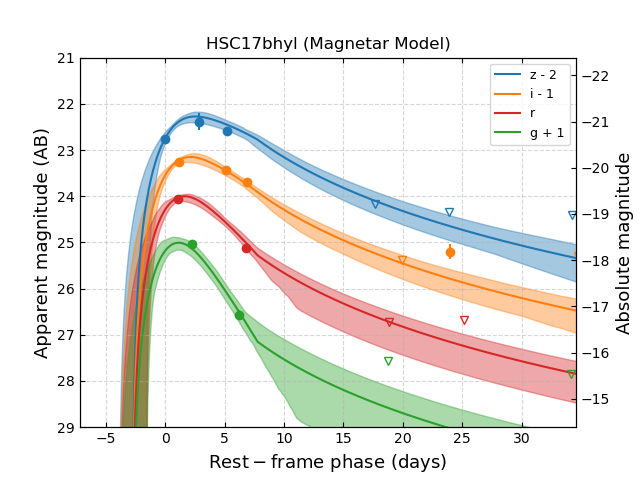}
\includegraphics[width=0.32\textwidth,angle=0]{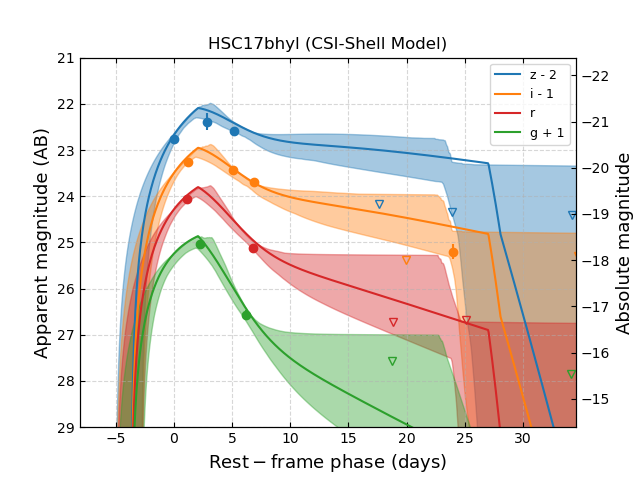}
\includegraphics[width=0.32\textwidth,angle=0]{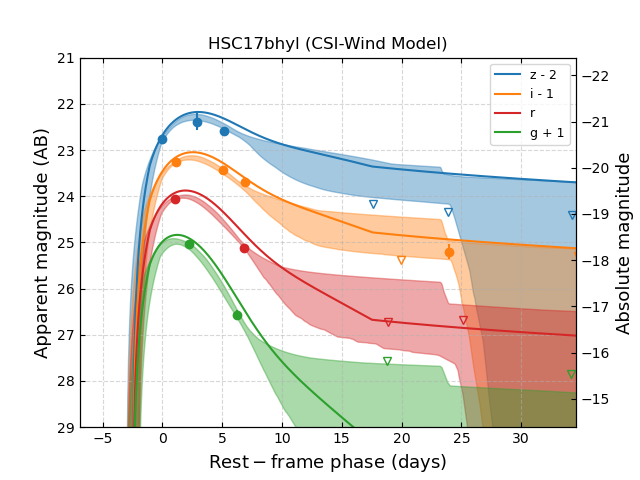}
\includegraphics[width=0.32\textwidth,angle=0]{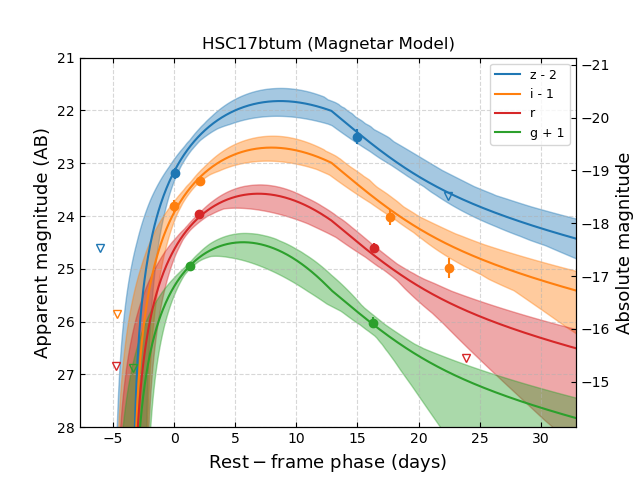}
\includegraphics[width=0.32\textwidth,angle=0]{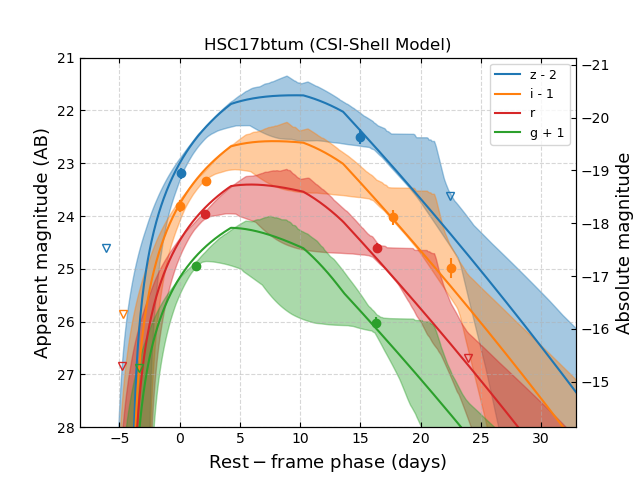}
\includegraphics[width=0.32\textwidth,angle=0]{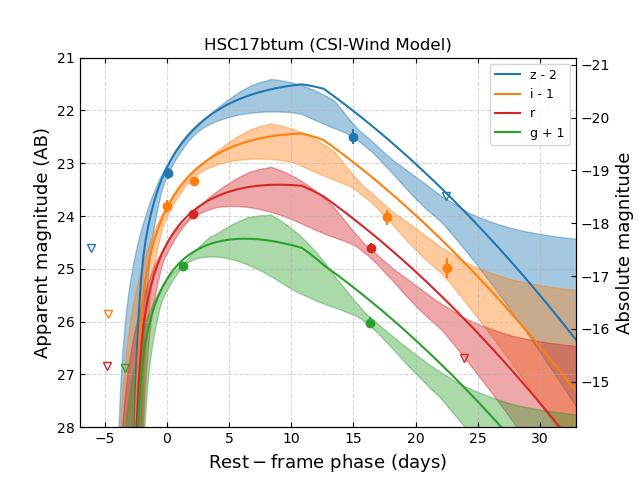}
\caption{(Continued).}
\end{figure}

\clearpage

\begin{figure}
\ContinuedFloat
\centering
\includegraphics[width=0.32\textwidth,angle=0]{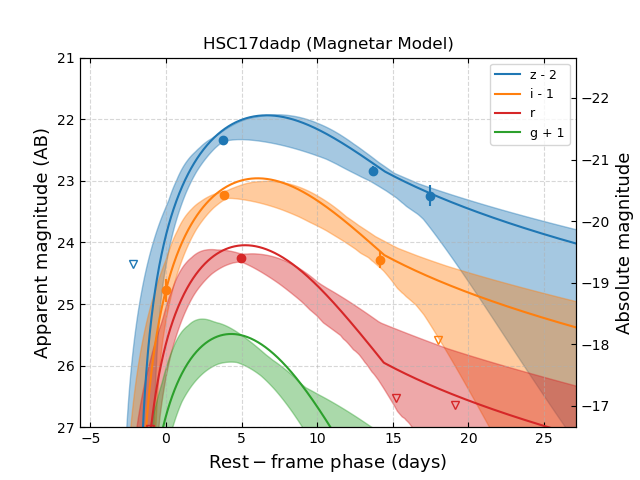}
\includegraphics[width=0.32\textwidth,angle=0]{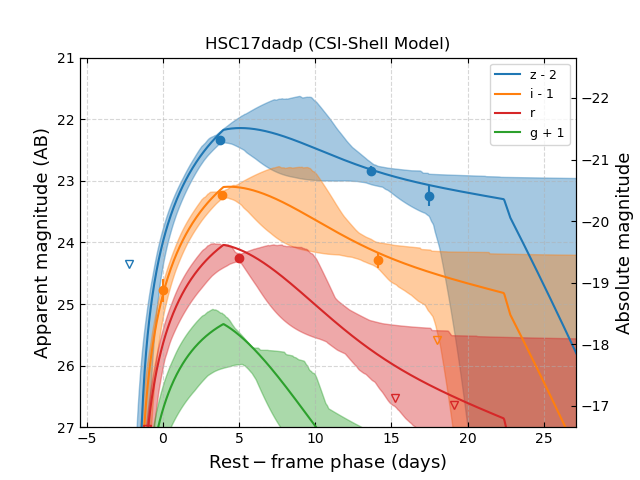}
\includegraphics[width=0.32\textwidth,angle=0]{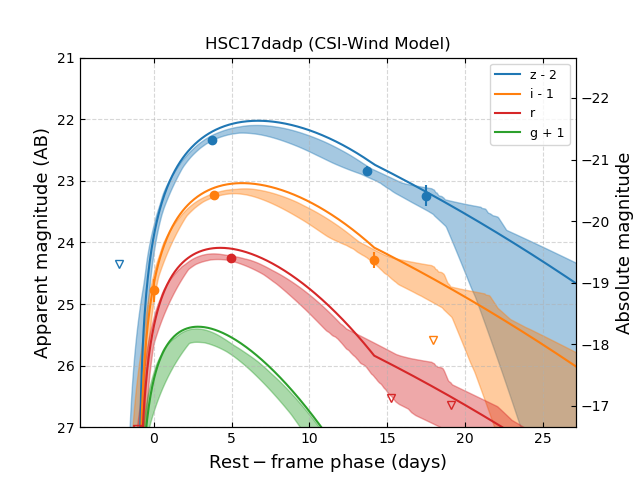}
\includegraphics[width=0.32\textwidth,angle=0]{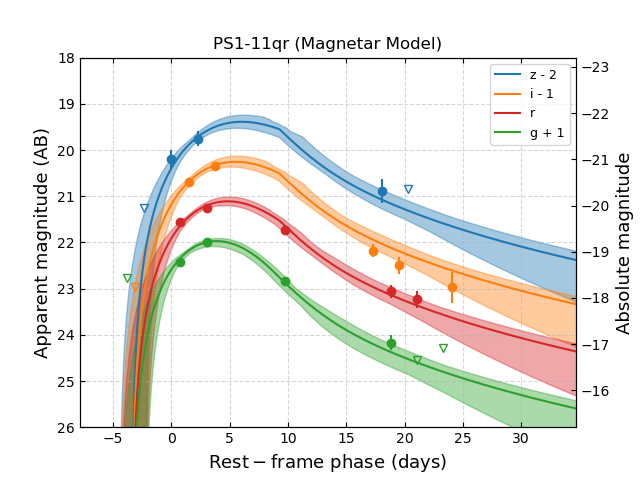}
\includegraphics[width=0.32\textwidth,angle=0]{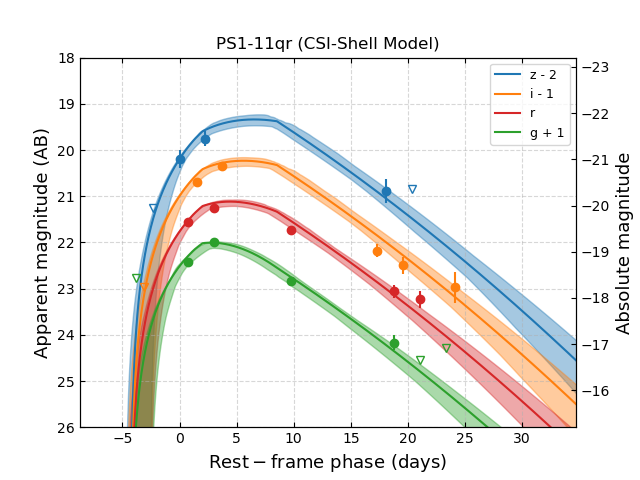}
\includegraphics[width=0.32\textwidth,angle=0]{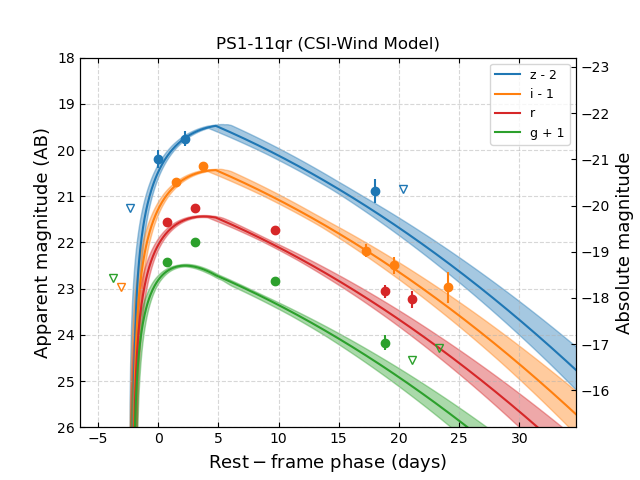}
\includegraphics[width=0.32\textwidth,angle=0]{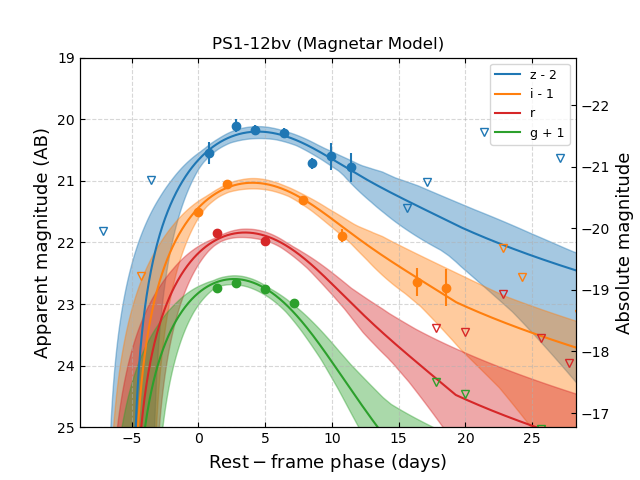}
\includegraphics[width=0.32\textwidth,angle=0]{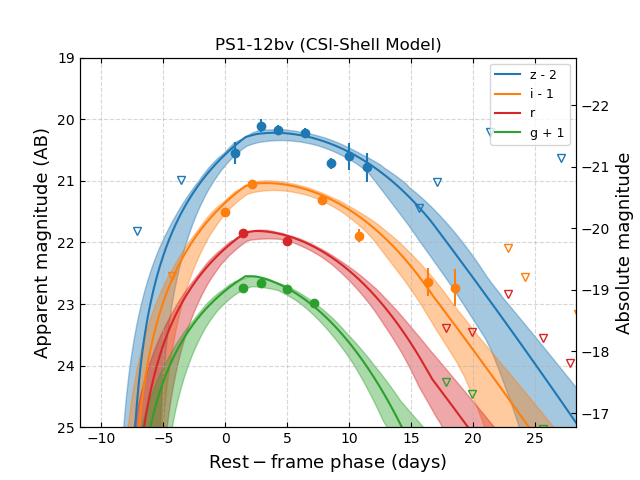}
\includegraphics[width=0.32\textwidth,angle=0]{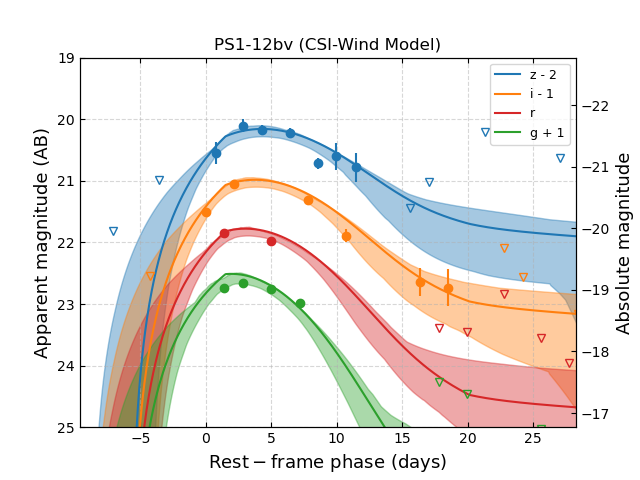}
\includegraphics[width=0.32\textwidth,angle=0]{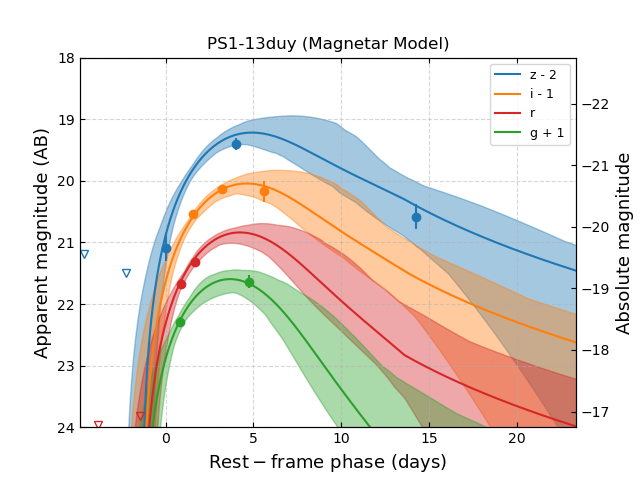}
\includegraphics[width=0.32\textwidth,angle=0]{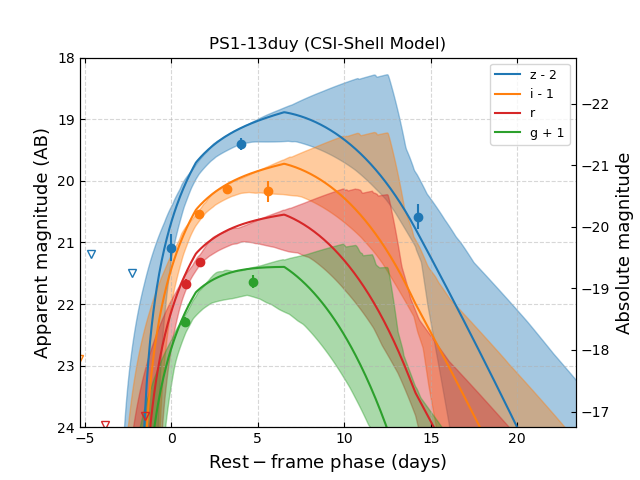}
\includegraphics[width=0.32\textwidth,angle=0]{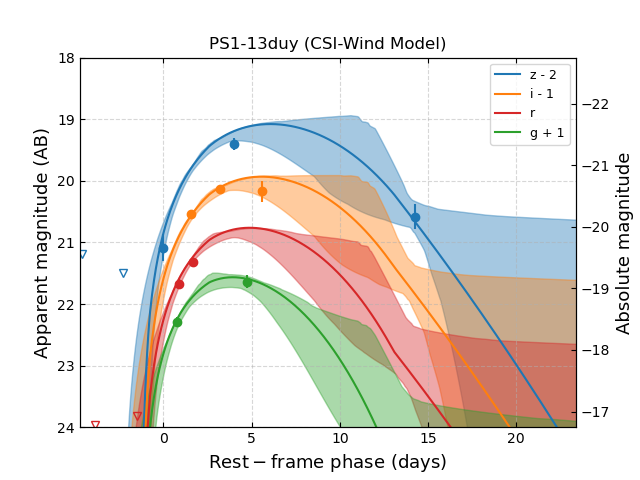}
\includegraphics[width=0.32\textwidth,angle=0]{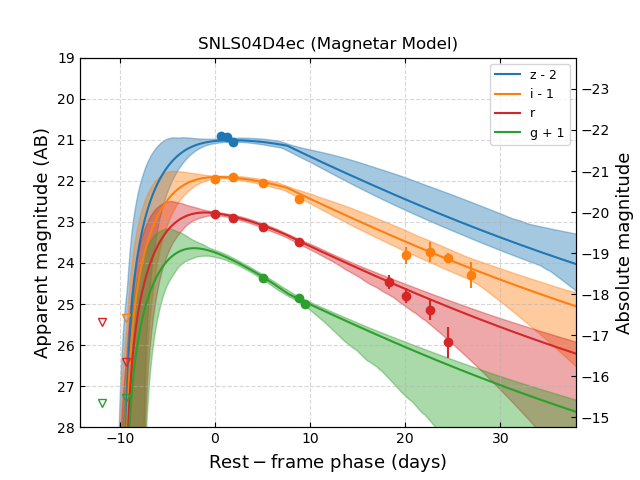}
\includegraphics[width=0.32\textwidth,angle=0]{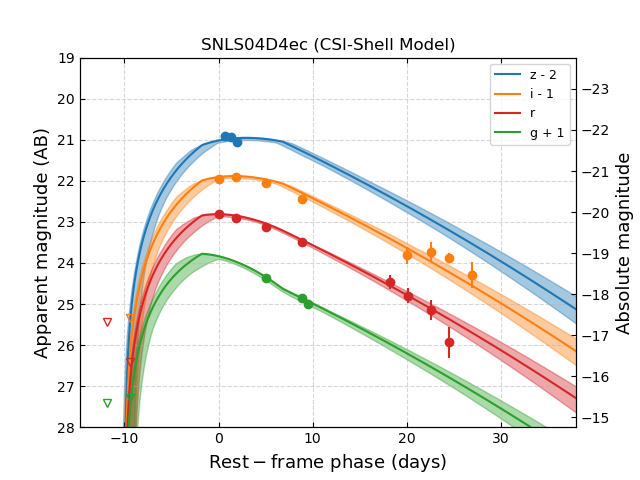}
\includegraphics[width=0.32\textwidth,angle=0]{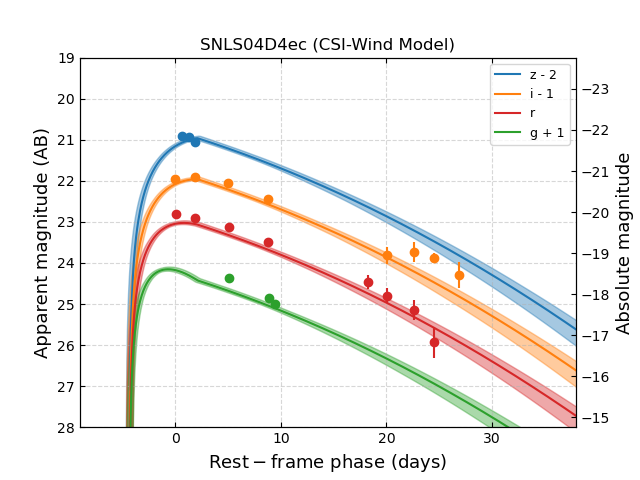}
\caption{(Continued).}
\end{figure}

\clearpage

\begin{figure}
\centering
\includegraphics[width=0.32\textwidth,angle=0]{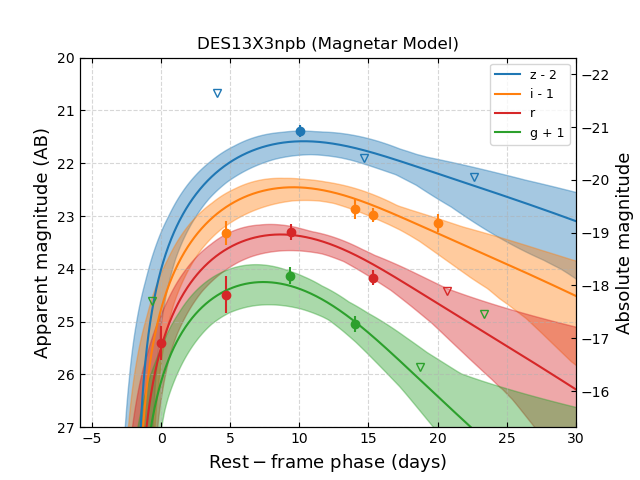}
\includegraphics[width=0.32\textwidth,angle=0]{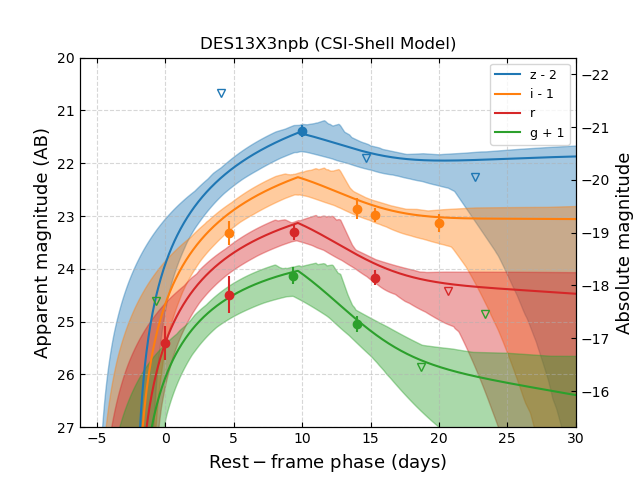}
\includegraphics[width=0.32\textwidth,angle=0]{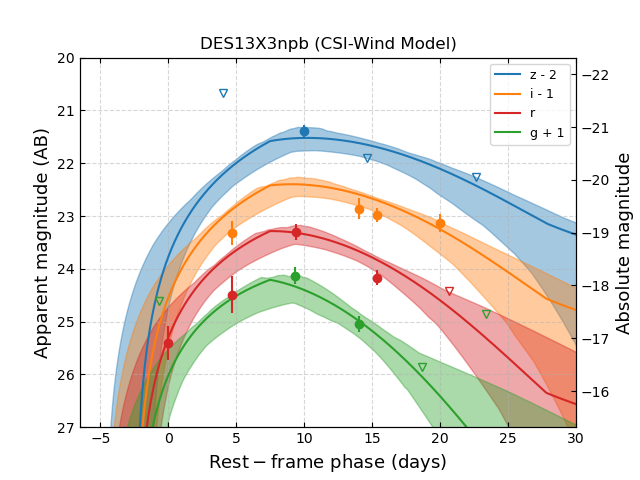}
\includegraphics[width=0.32\textwidth,angle=0]{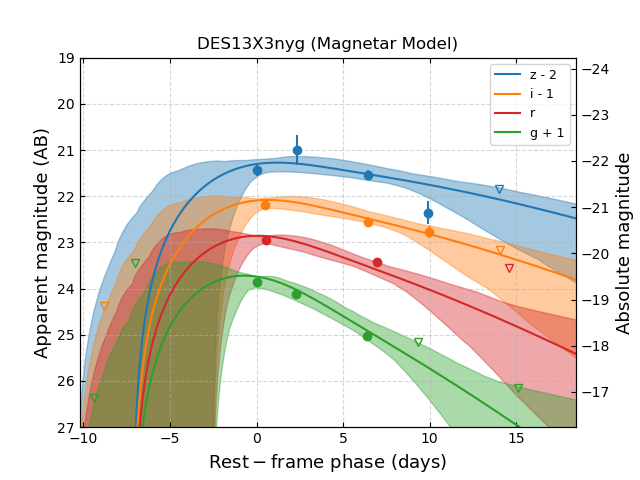}
\includegraphics[width=0.32\textwidth,angle=0]{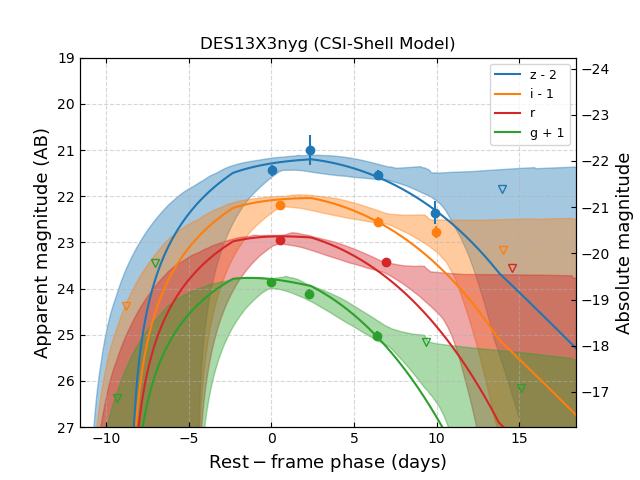}
\includegraphics[width=0.32\textwidth,angle=0]{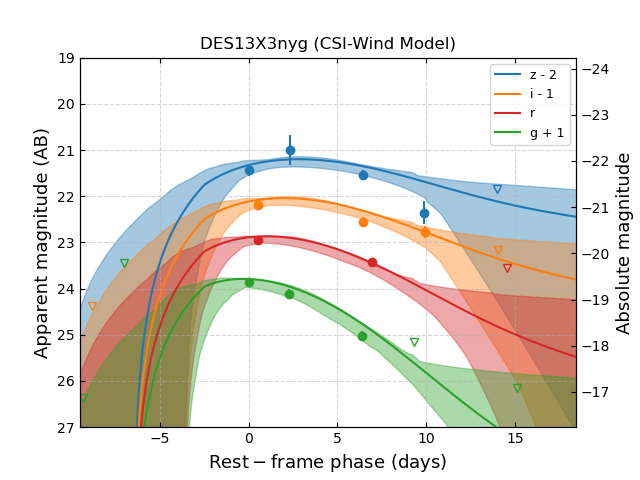}
\includegraphics[width=0.32\textwidth,angle=0]{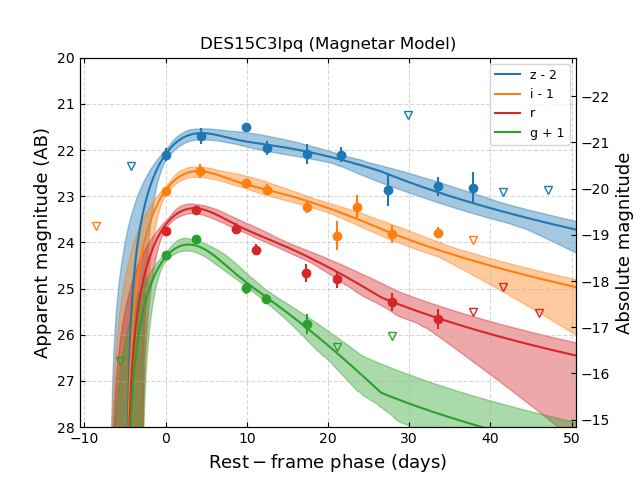}
\includegraphics[width=0.32\textwidth,angle=0]{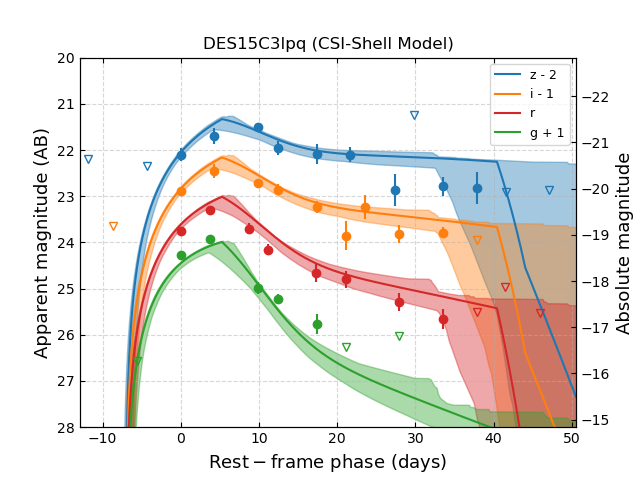}
\includegraphics[width=0.32\textwidth,angle=0]{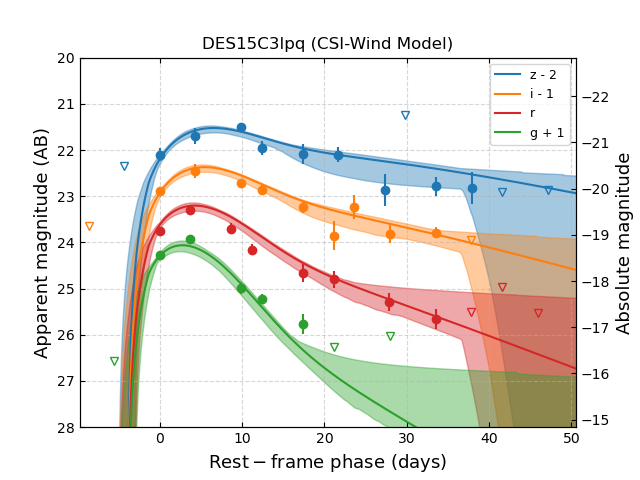}
\includegraphics[width=0.32\textwidth,angle=0]{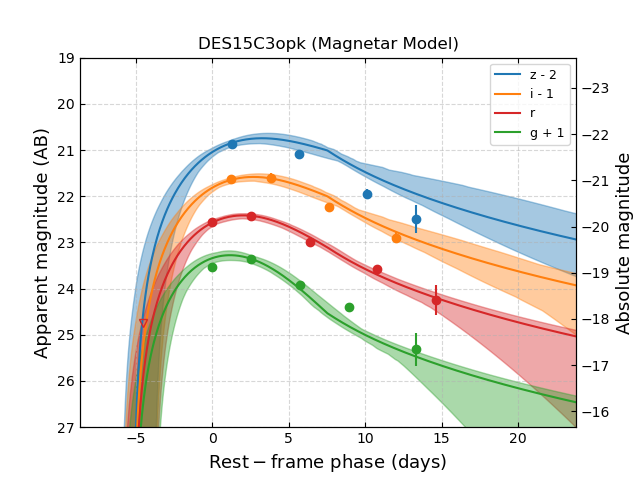}
\includegraphics[width=0.32\textwidth,angle=0]{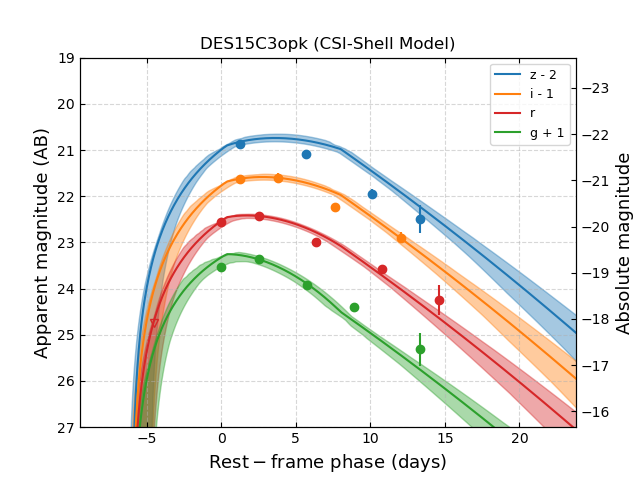}
\includegraphics[width=0.32\textwidth,angle=0]{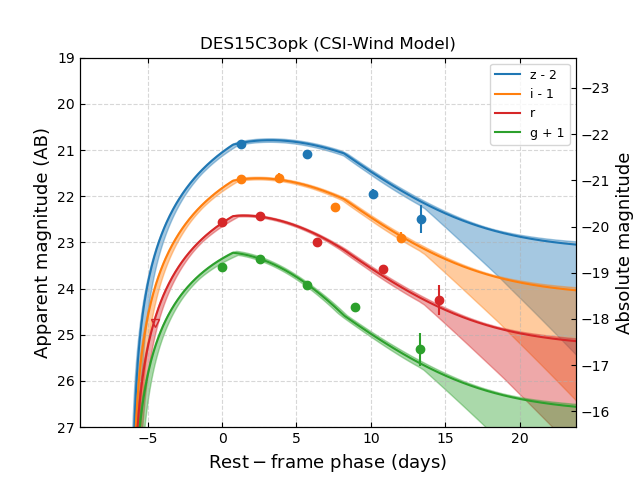}
\includegraphics[width=0.32\textwidth,angle=0]{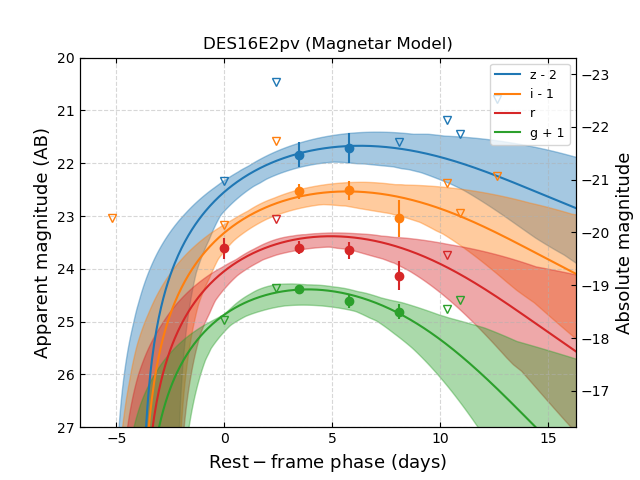}
\includegraphics[width=0.32\textwidth,angle=0]{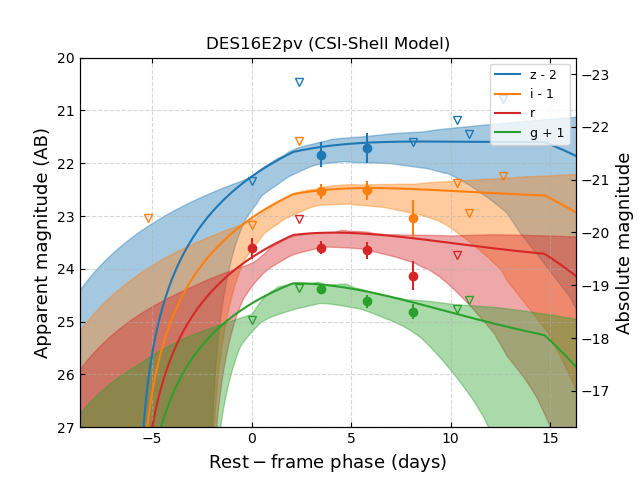}
\includegraphics[width=0.32\textwidth,angle=0]{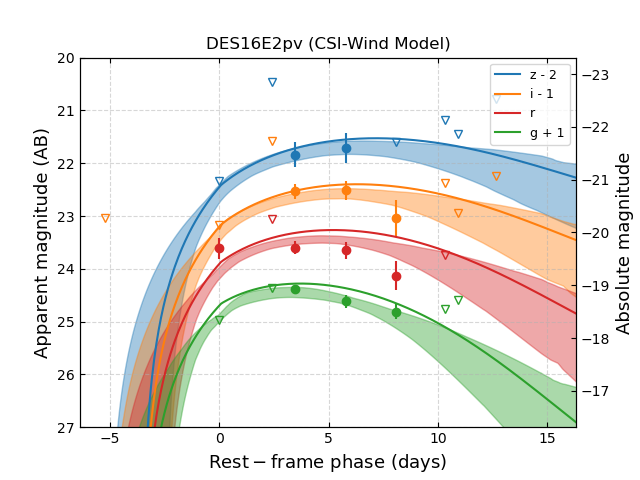}
\caption{{Same as Fig. \ref{fig:LCfits_excellent}, but for the 8 REOTs fitted with good quality.}}
\label{fig:LCfits_good}
\end{figure}

\clearpage

\begin{figure}
\centering
\ContinuedFloat
\includegraphics[width=0.32\textwidth,angle=0]{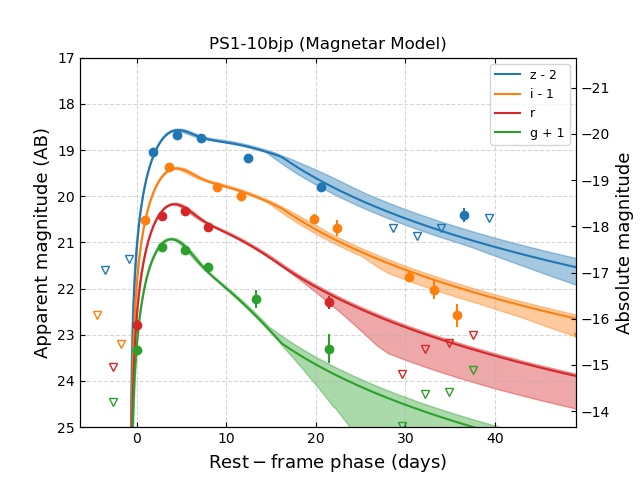}
\includegraphics[width=0.32\textwidth,angle=0]{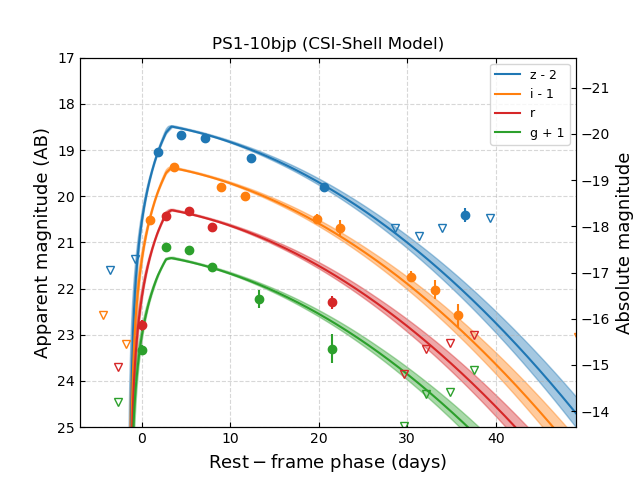}
\includegraphics[width=0.32\textwidth,angle=0]{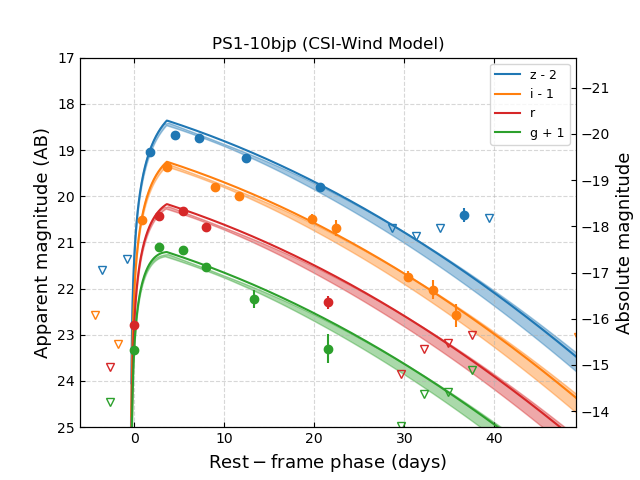}
\includegraphics[width=0.32\textwidth,angle=0]{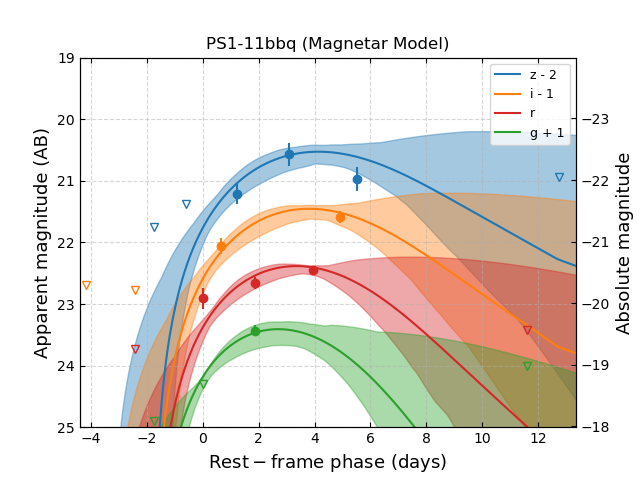}
\includegraphics[width=0.32\textwidth,angle=0]{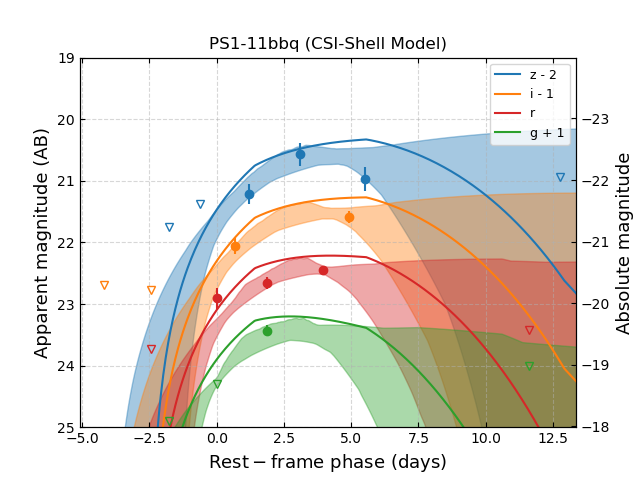}
\includegraphics[width=0.32\textwidth,angle=0]{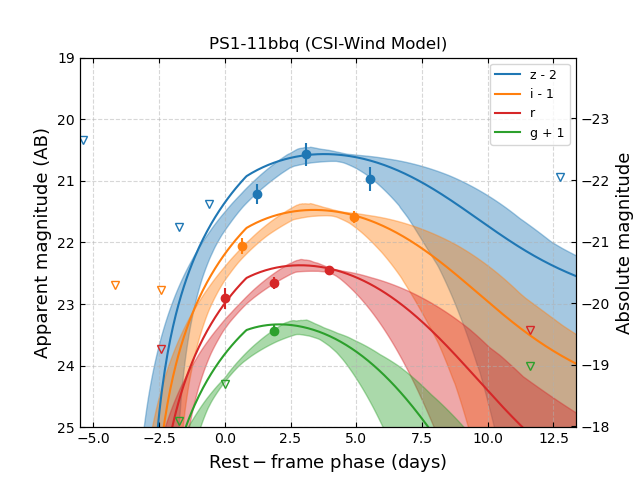}
\includegraphics[width=0.32\textwidth,angle=0]{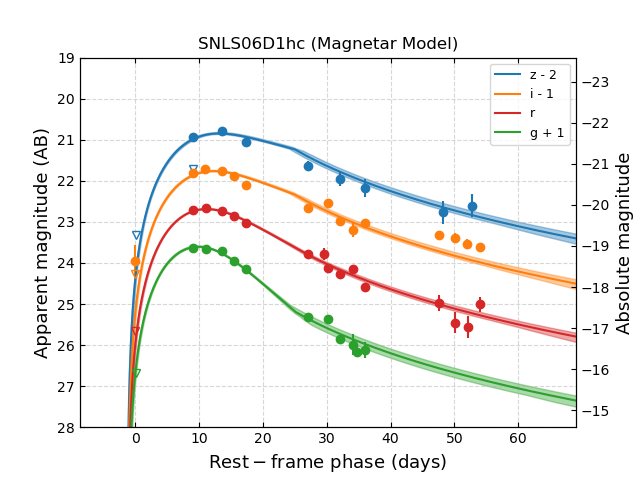}
\includegraphics[width=0.32\textwidth,angle=0]{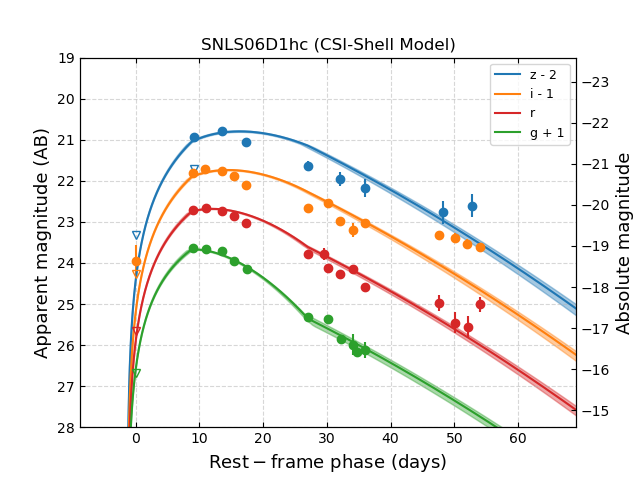}
\includegraphics[width=0.32\textwidth,angle=0]{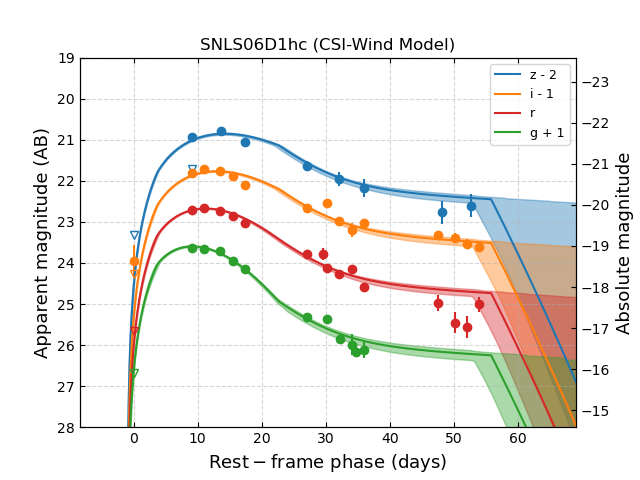}
\caption{(Continued).}
\end{figure}

\clearpage

\begin{figure}
\centering
\includegraphics[width=0.32\textwidth,angle=0]{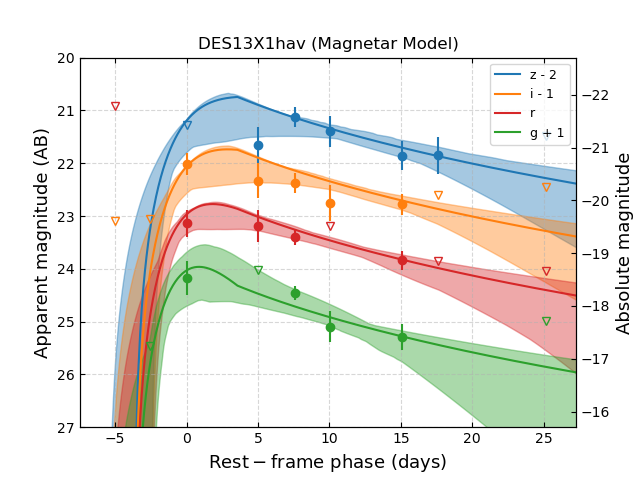}
\includegraphics[width=0.32\textwidth,angle=0]{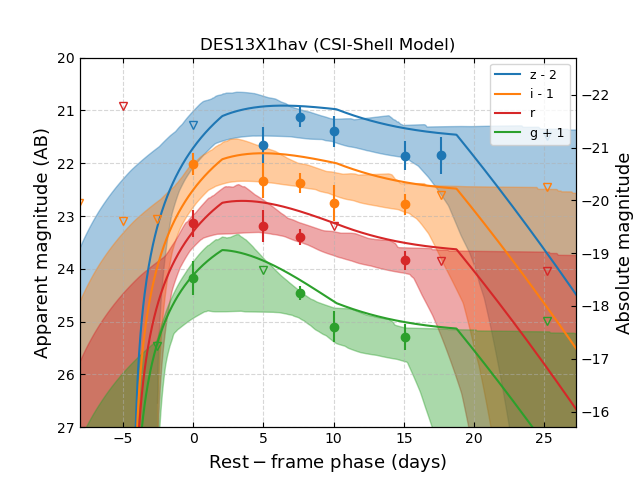}
\includegraphics[width=0.32\textwidth,angle=0]{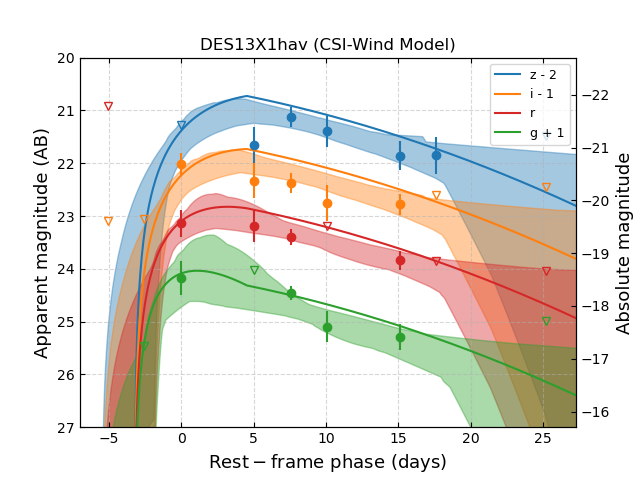}
\includegraphics[width=0.32\textwidth,angle=0]{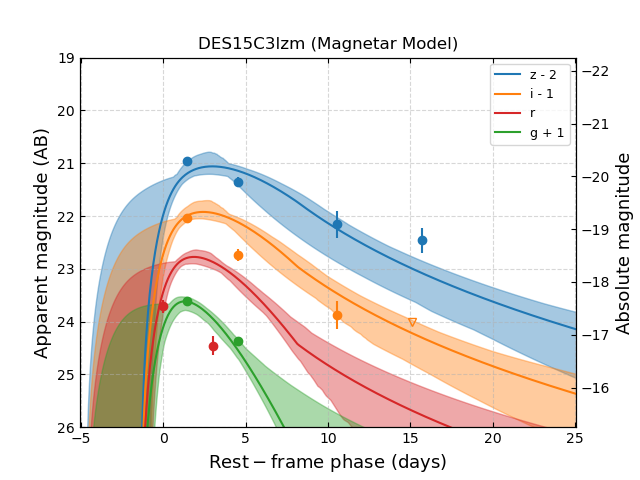}
\includegraphics[width=0.32\textwidth,angle=0]{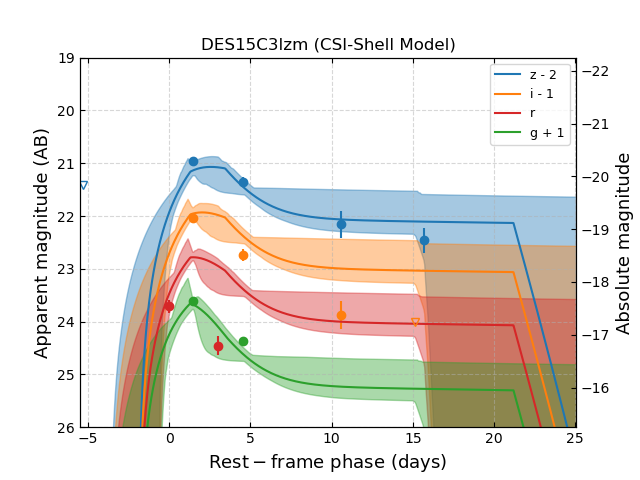}
\includegraphics[width=0.32\textwidth,angle=0]{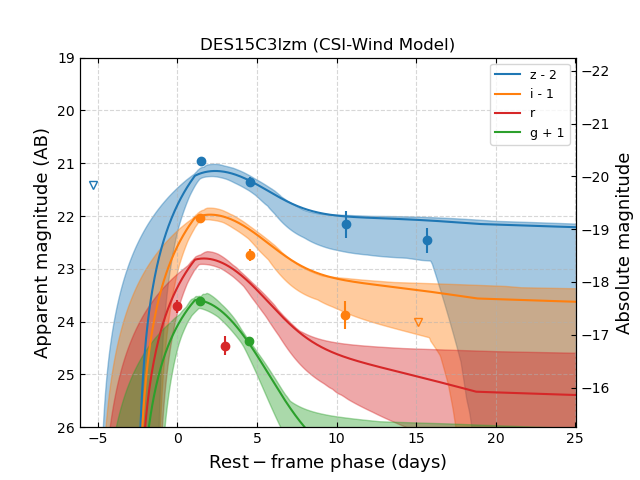}
\includegraphics[width=0.32\textwidth,angle=0]{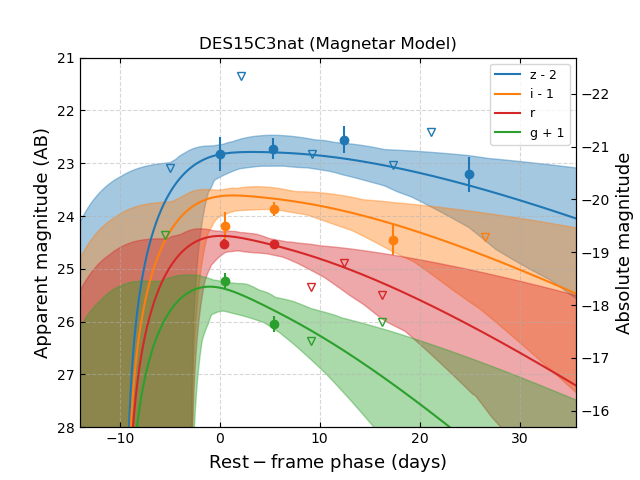}
\includegraphics[width=0.32\textwidth,angle=0]{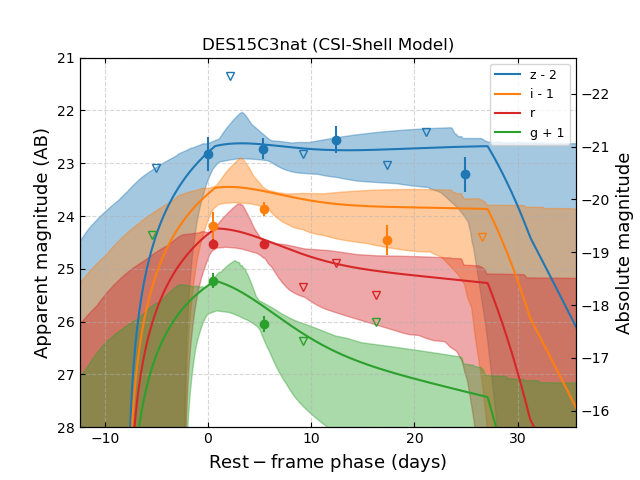}
\includegraphics[width=0.32\textwidth,angle=0]{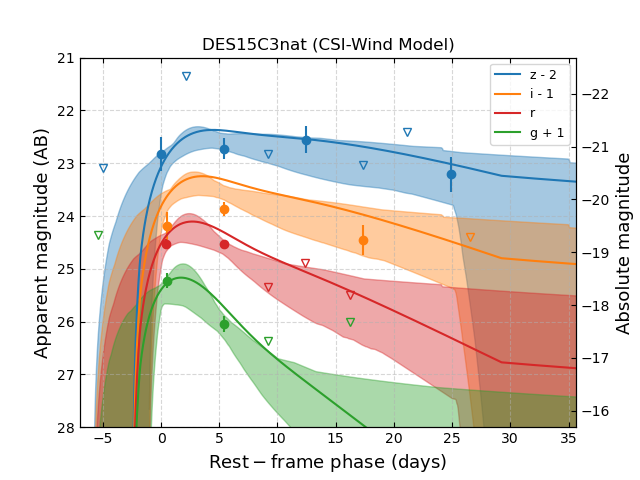}
\caption{{Same as Fig. \ref{fig:LCfits_excellent}, but for the 3 REOTs that cannot be fitted by the models.}}
\label{fig:LCfits_bad}
\end{figure}

%
%

\clearpage

\clearpage

\end{document}